\pdfoutput=1
\documentclass[3p,nopreprintline]{elsarticle}
\usepackage[T1]{fontenc}

\usepackage{amsmath}
\usepackage{amssymb}
\usepackage{amsthm}
\usepackage{mathtools}
\usepackage{newtxtext}
\usepackage{newtxmath}
\usepackage[hidelinks]{hyperref}
\usepackage[nameinlink,capitalize]{cleveref}
\usepackage{multirow}
\usepackage{tikz}
\usepackage{pgfplots}
\pgfplotsset{compat=1.18}
\usepackage{subcaption}
\usepackage{placeins}
\usepackage{enumitem}

\usepackage[textwidth=1.3cm,textsize=footnotesize]{todonotes}
\makeatletter\@mparswitchfalse\makeatother\normalmarginpar

\newtheorem*{theorem*}{Remark}
\newtheorem*{remark*}{Remark}

\crefname{appendix}{}{}

\providecommand{\doi}[1]{\href{https://doi.org/#1}{\path{#1}}}

\makeatletter
\newcommand{\slope}[6]{

  \pgfplotsextra{
    \pgfkeys{/pgf/fpu}
    \pgfmathparse{#5*(#4/#3)^#2}
    \edef\by{\pgfmathresult}
    \pgfkeys{/pgf/fpu=false}

    \pgfmathprintnumberto{\by}\tmp
    \typeout{Calculated \\by for the slope: \tmp}

    \coordinate (a) at (axis direction cs:#3,#5);
    \coordinate (b) at (axis direction cs:#4,\by);
    \csname slope@draw@#1\endcsname{$1$}{#6}
  }
}
\newcommand\slope@draw@bl[2]{%
  \draw (a) -- (b) -| (a)
        node [pos=0.25,anchor=north] {\small #1}
        node [pos=0.75,anchor=east] {\small #2};
}
\newcommand\slope@draw@tr[2]{%
  \draw (b) -- (a) -| (b)
        node [pos=0.25,anchor=south] {\small #1}
        node [pos=0.75,anchor=west] {\small #2};
}
\makeatother
 
\title{On bifurcations and traction forces on an obstacle in incompressible flow\tnoteref{t1}}
\tnotetext[t1]{Accepted for publication in revised form in the \href{https://www.cambridge.org/core/journals/journal-of-fluid-mechanics}{Journal of Fluid Mechanics}. This version has not undergone copyediting or typesetting by Cambridge University Press.}

\author[3]{J.~Cach}
\author[3]{K.~T\r{u}ma\corref{cor1}}
\author[3]{J.~Blechta}
\author[3,4]{S.~Schwarzacher}
\address[3]{Charles University, Faculty of Mathematics and Physics, Sokolovsk\'{a} 83, 186 75 Prague, Czechia}
\address[4]{Uppsala Universitet, Ångströmlaboratoriet, Lägerhyddsvägen 1, 751 06 Uppsala, Sweden}

\begin{document}

\begin{abstract}
A systematic numerical investigation of flow-regime transitions in
the two-dimensional incompressible Navier--Stokes flow past a confined circular
cylinder is presented. For a fixed benchmark geometry, we observe a clear empirical
correspondence between qualitative changes in steady traction profiles,
understood here as the pointwise force density given by the Cauchy stress tensor
on the obstacle boundary, and bifurcations in the long-time behavior of the
unsteady Navier--Stokes equations. The observed transitions include onset of time-periodic oscillations, the appearance of multiple steady solutions and loss of effective symmetry.

The well-known planar Schäfer--Turek benchmark is considered for Reynolds
numbers up to~500. Several numerical techniques are employed to compute steady
solutions, boundary traction profiles, and linear stability spectra such as a
duality-based approach for traction evaluation, deflation methods for detecting
multiple steady states, and both two- and three-dimensional linear stability analysis.

The results suggest that steady boundary traction profiles can serve as a
sensitive diagnostic indicator of critical Reynolds numbers at which qualitative
changes in flow dynamics occur. This suggests a computationally inexpensive,
complementary approach for detecting flow-regime transitions within this
benchmark configuration.
\end{abstract}


\maketitle

\section{Introduction}

Understanding the nature of steady and unsteady solutions to the Navier--Stokes equations is crucial for both theoretical analysis and practical applications. This study is motivated by the need to bridge the gap between experimental observations and rigorous mathematical analysis regarding the questions of bifurcation. We focus on numerical methods to study the behavior of solutions to the Navier--Stokes equations. Besides its relevance for applications, our investigation is also relevant for the mathematical analysis of fluid mechanics, where stability and uniqueness still remain substantial open problems.

We performed vast numerical experiments based on the Sch\"{a}fer--Turek
benchmark \citep*{Schafer1996}, which concerns the 2D flow around an asymmetrically
placed cylinder in a~narrow channel; see \cref{fig:turek_geometry}. Our key observation is that changes in the dynamics of the flow are observable
upstream on the surface of the obstacle itself. This relates to the common descriptions in much of the engineering and
fluid-mechanics literature, where the onset of vortex shedding and related
transitions is often associated with the separated shear layers issuing from the
obstacle \citep{Williamson1996,PrasadWilliamson1997,Mittal2008}. A different perspective is the global wake-stability viewpoint, in which the change of dynamics
is interpreted in terms of the instability and nonlinear saturation of the wake
downstream of the body
\citep{NoackEckelmann1994,Barkley2006,SippLebedev2007}. While these insights do not contradict each other, they motivate a~more detailed study of how changes in
the flow regime are reflected in quantities measured on the obstacle.

Many viscosity‑dominated fluid flows observed in everyday life are either stable steady states or time‑periodic motions. In such flows, regardless of their initial conditions, the motion typically undergoes an initial adjustment period and then converges to a persistent regime, such as the mentioned time-periodic oscillations or a~steady state (trivial time-periodic motion). We refer to this persistent regime as the long‑time behavior of the flow. 
This behavior changes drastically with increasing Reynolds number, and has been studied in the literature for a very long time. For example, as the Reynolds number $\mathrm{Re}$ increases, the system transitions from a~stable, steady, and unique attractor to a~regime where a~non-trivial time-periodic attractor appears and it differs from the steady state. Further bifurcations can be observed with increasing Reynolds numbers, such as the coexistence of multiple steady-state solutions together with qualitative changes of the time-periodic attractor.

This work is dedicated to the numerical investigation of these changes, with a focus on their relation to observable changes of the fluid forces acting on the obstacle. Specifically, we examine the pointwise traction profiles of the fluid along the obstacle surface, that is the pointwise force density given by the Cauchy stress tensor of the fluid. It turned out that for the {\em steady} Navier-Stokes equations, the traction profiles change drastically at certain Reynolds numbers. This observation led us to formulate our central objective: To analyze how the {\em steady} traction profiles relate to bifurcations of the long-time behavior of the {\em unsteady} Navier–Stokes equations. Thus, we investigate to what extent steady traction profiles can serve as an indicator quantity to detect critical Reynolds numbers.

The physical idea behind this objective is that the traction field, which represents the transfer of momentum in between the fluid and the obstacle, contains, in compressed form, information about the processes that shape the wake dynamics. As the Reynolds number increases, the pressure and viscous stresses change their distribution, reflecting changes in separation, recirculation and shear-layer development. The first Hopf bifurcation provides the clearest example of this connection: A~qualitative change in the steady traction field occurs at the onset of vortex shedding. As mentioned above, our objective goes beyond this expected relation: It is to clarify whether and to what extent the steady momentum transfer exhibits any dynamical transitions including higher Reynolds numbers far beyond the first Hopf bifurcation.

Our approach refines classical indicators for flow transitions (e.\ g.\ the drag-crisis phenomenon) as the total drag and lift coefficients are integral means of the traction profiles in the direction of flow and orthogonal to the direction of flow, cf.\ \eqref{eq:drag-lift-coeffs}. In contrast to integral quantities, traction profiles retain local spatial information while incorporating global effects through their dependence on pressure and convected velocity gradients. The decomposition into drag and lift components of the traction is particularly natural from the viewpoint of flow symmetry. The drag component reflects the streamwise force balance and is primarily associated with symmetric modifications of the flow, whereas the lift component is sensitive to asymmetric structures and symmetry-breaking effects. We found that {\em steady} traction profiles undergo characteristic qualitative changes at precisely such Reynolds numbers, where the long‑time solutions of the {\em unsteady} Navier--Stokes equations change their dynamical regime. See \cref{fig:teaser_result}, where the synchronized behavior between the traction of the steady solution and its relation to unsteady solutions is summarized.

This work focuses on flow‑regime transitions for Reynolds numbers up to 500, which includes a broad set of steady and unsteady transition scenarios, see \cref{fig:teaser_result}. Most prominent are canonical wake-transition phenomena that are observed both experimentally and numerically in bluff-body flows, such as vortex shedding, wake asymmetries and three-dimensional instabilities \citep*{thompson2021bluff}. In this range of Reynolds numbers the dimensions of the numerical setting of the Sch\"afer--Turek benchmark are computationally stable with regard to numerous perturbations; see \cref{ap:C}.

To present a comprehensive study of flow-regime transitions, several complementary numerical approaches are developed and used. The definitive reference for identifying long‑time flow behavior remains direct numerical integration of the unsteady Navier--Stokes equations. As mentioned above, the computation of pointwise traction is used throughout as a tool of field observation; it is recorded from steady solutions to document how its behavior varies across different flow regimes. As it is the central computational reference quantity, great care is given to guarantee its numerical accuracy. In particular, we use variational methods; see \cref{ap:A,ap:B} for their convergence analysis. In parallel, we explore the structure of steady solutions through \emph{deflated continuation}, which allows us to uncover multiple steady solution branches. Another approach is two-dimensional linear stability analysis (LSA) based on classical linear stability analysis to assess how its stability predictions, when applied to the computed steady states, are consistent with the observed transition scenarios. In particular, we later show that LSA accurately captures the onset of the first Hopf bifurcation, both in terms of the critical Reynolds number and the associated global velocity field of the emerging time‑periodic solution.
The last numerical approach we use is three-dimensional linear stability analysis (3D LSA), which allows us to identify the onset of the steady three-dimensional instability of the steady two-dimensional base flow, referred to as Mode~E by~\citet*{Thompson2013}. This mode manifests at lower Reynolds numbers, if the centerline mirror symmetry is sufficiently disrupted \citep*{Thompson2016, Thompson2017}.

The choice of the Schäfer--Turek benchmark configuration as our primary computational setting is pragmatic. While its widespread use and standardization enable validation, reproducibility, and comparison across studies, the benchmark also offers several methodological and conceptual advantages that are useful for the present investigation. First, it provides a controlled environment in which confinement and wall effects give rise to additional flow phenomena that are particularly convenient for observation, while simultaneously restricting effects that are computationally expensive to resolve. In particular, the large blockage ratio (ratio of obstacle diameter to channel width) and the short distance between the cylinder and the outflow boundary suppress large‑scale wake phenomena that in unconfined domains typically emerge far downstream, such as secondary vortex streets \citep*{Jiang_Cheng_2017} or delayed three‑dimensional transitions \citep*{williamson-1996}. As a result, changes in the flow regime manifest in a small domain, making the time‑periodic nature of the long‑time solution easier to identify and enabling a clearer classification of steady and unsteady solution branches. This is further amplified by the slightly asymmetric placement of the obstacle, see \citep*{Lu_Aljubaili_Zahtila_Chan_Ooi_2023}, where similar confinement-induced asymmetries in unsteady wakes have been reported.  As we focus on the Schäfer--Turek benchmark as a single canonical configuration, all numerical investigations performed here are restricted to a case study. Nevertheless, the obtained observations retain their validity under rather general perturbations. See \cref{ap:C}, where further experiments show that our conclusions are not specific to the particular choices of the benchmark, by assessing the influence of the {\em symmetric obstacle placement}, a~{\em perturbed obstacle shape}, and the {\em channel of double width}.

Even though the focus of this work is on the dynamical‑systems aspects of the Navier--Stokes equations and on the classification of flow regimes, we use the setting for a systematic numerical study on the computation of the key diagnostic quantity, the pointwise traction along the obstacle boundary. We show numerically the advantage of variational approaches for its computation in comparison with direct evaluation of the traction. The pointwise traction, as well as the integral drag and lift, are compared numerically in \cref{ap:B}.

The paper is structured as follows: In \cref{sec:problem_description}, we introduce the
Navier--Stokes equations, define branching and multiplicity, introduce the linear
stability and the deflated continuation, and provide an overview of the theoretical
understanding of existence, uniqueness, and stability of Navier--Stokes
solutions. We conclude this section by summarizing our findings and outlining
their implications (see \cref{fig:teaser_result}). In \cref{sec:numerical_methods}, we briefly
describe the numerical methods used to solve the Navier--Stokes equations in the Sch\"{a}fer--Turek benchmark configuration and
a~related linear eigenvalue problem, which are standard in the field.
\Cref{sec:res_intro} presents our results in the order of increasing Reynolds numbers, starting
with vortex-free flow, $\mathrm{Re}\in(0,7)$, followed by unique steady
flow with a~small vortex wake behind the cylinder for
$\mathrm{Re}\in(7,48)$, which is also a~stable solution of the unsteady problem.
The main results focus on vortex shedding and the resulting vortex street in the range $\mathrm{Re} \in (48, 500)$.
In~\cref{ap:A}, we outline, in the context of the Poisson equation, both the
direct and the variational computation of the normal derivative. \Cref{ap:B}
extends this derivation to the unsteady Navier--Stokes equations and presents
computational experiments in the Sch\"{a}fer--Turek geometry. Finally, \cref{ap:C} showcases results for variations of the Sch\"{a}fer--Turek benchmark geometry.

\section{Problem description}
\label{sec:problem_description}

\subsection{Sch\"{a}fer--Turek benchmark}
\label{sec:turek_benchmark}
\begin{figure}
  \centering
  \begin{tikzpicture}[scale=4.5]

    \def\length{2.2}  
    \def\width{0.41}   
    \def\scaledLength{8.8 / 4}  
    \def\scaledWidth{\width}   

    \def\originalCircleX{0.2}
    \def\originalCircleY{0.2}
    \def\originalRadius{0.05}

    \def\scaledCircleX{\originalCircleX}
    \def\scaledCircleY{\originalCircleY}
    \def\scaledRadius{\originalRadius}

    \draw (0,0) rectangle (\scaledLength,\scaledWidth);

    \draw[<->] (0,\scaledWidth + 0.05) -- (\scaledLength,\scaledWidth + 0.05) node[midway, above] {$2.2\,\mathrm{m}$};
    \draw[<->] (\scaledLength+0.05,0) -- (\scaledLength+0.05,\scaledWidth) node[midway, right] {$0.41\,\mathrm{m}$};

    \node at (\scaledLength/2, \scaledWidth) [below] {$\mathbf{v}=\mathbf{0}$};
    \node at (\scaledLength/2, 0) [above] {$\mathbf{v}=\mathbf{0}$};
    \node at (\scaledLength, \scaledWidth/2) [left] {outflow};
    \node at (0, \scaledWidth/2) [left] {$\mathbf{v} =\mathbf{v}_\text{in}$};

    \draw (\scaledCircleX,\scaledCircleY) circle (\scaledRadius);

    \node at (\scaledCircleX + \scaledRadius-0.05,\scaledCircleY + \scaledRadius + 0.05) {$\mathbf{v}=\mathbf{0}$};
  \end{tikzpicture}
  \caption{%
    The Sch\"{a}fer--Turek benchmark. The lower left corner is at
    $(0\,\mathrm{m}, 0\,\mathrm{m})$ and the cylinder center is located at
    $(0.2\,\mathrm{m}, 0.2\,\mathrm{m})$ with radius $R = 0.05\,\mathrm{m}$.
    The domain is slightly asymmetric. The benchmark does not require use
    of a~specific outflow condition.
  }
  \label{fig:turek_geometry}
\end{figure}
We consider the flow around a~slightly asymmetrically placed circular cylinder
in a~channel; see \cref{fig:turek_geometry}. This setup is known as the
Sch\"{a}fer--Turek benchmark \cite{Schafer1996} and has been widely used to
validate numerical methods for incompressible flow. The governing equations for
velocity $\mathbf{v}$ and pressure $p$ are the incompressible Navier--Stokes
equations, with emphasizing the Cauchy stress tensor $\mathbb{T}$:
\begin{equation}
  \label{eq:navier-stokes}
  \begin{alignedat}{2}
    \operatorname{div} \mathbf{v} &= 0
    &\qquad &\text{in $\Omega$,} \\
    \rho\Bigl(\frac{\partial\mathbf{v}}{\partial t}+\mathbf{v}\cdot\nabla\mathbf{v}\Bigr) &= \operatorname{div} \mathbb{T}
    &\qquad &\text{in $\Omega$,} \\
    \mathbb{T} &= -p\mathbb{I} + \mu\bigl(\nabla\mathbf{v}+(\nabla\mathbf{v})^\top\bigr)
    &\qquad &\text{in $\Omega$,} \\
    \mathbf{v} &= \mathbf{v}_\mathrm{D}
    &\qquad &\text{on $\Gamma_\mathrm{D}$,} \\
    \mathbb{T}\mathbf{n} = \boldsymbol{0}
    \quad&\mathbin{\text{or}}\quad
    \left\{\begin{aligned}
      (\mathbb{T}\mathbf{n})\cdot\mathbf{n} &= 0,\\
      \mathbf{v} - (\mathbf{v}\cdot\mathbf{n})\mathbf{n} &= 0
    \end{aligned}\right.
    &\qquad &\text{on $\partial\Omega \setminus \Gamma_\mathrm{D}$.}
  \end{alignedat}
\end{equation}
In the sequel, $\Omega\subset\mathbb{R}^2$ represents the domain in~\cref{fig:turek_geometry}
(as an exception, slightly modified domains in~$\mathbb{R}^2$ will be considered in \cref{ap:C}). It resembles a three-dimensional domain of the form $\Omega\times\mathbb{R}$, which will explicitly be used in the context of 3D LSA.
Here, $\Gamma_\mathrm{D}\subset\partial\Omega$ collects all the Dirichlet boundaries; see~\cref{fig:turek_geometry}.
The surface of the obstacle is denoted by $\Gamma\subset\Gamma_\mathrm{D}$.
At the inflow, a parabolic velocity profile $\mathbf{v}_\text{in} = (4Uy(H
- y)/H^2, 0)$ is prescribed with peak velocity $U~[\mathrm{m}\cdot
\mathrm{s}^{-1}]$ and the channel width is $H = 0.41\,\mathrm{m}$.
The channel walls and the cylinder surface are equipped with the no-slip
boundary condition $\mathbf{v} = \mathbf{0}$.
At the outflow $\partial\Omega\setminus\Gamma_\mathrm{D}$, the condition
is not specified by the benchmark. We employ either the ``do nothing''
condition, which constrains the \emph{traction}, defined as $\mathbf{t} \coloneqq
\mathbb{T} \mathbf{n}$, to be zero, or we force the velocity to be perpendicular
to the outflow boundary and constrain the normal traction
$\mathbf{t}\cdot\mathbf{n}=(\mathbb{T}\mathbf{n})\cdot\mathbf{n}$ to be zero.
We note that either considered outflow condition is useful and natural in the
weak formulation, but it also has a~well-known limitation that the backflow is
not a~priori controlled by the energy estimate. General well-posedness results
are available only under smallness assumptions on the data, even in two
dimensions; see \citep*{Galdi2008}, also \citep*{LanzendorferHron2020} for
numerical investigation.

The Reynolds number is defined as $\mathrm{Re} \coloneqq V\!L \rho / \mu$, where $V =
2U/3$ is the mean inflow velocity, $L = 2R = 0.1\,\mathrm{m}$ is the cylinder
diameter, $\rho = 1\,\mathrm{kg}\cdot \mathrm{m}^{-3}$ is the fluid density,
and $\mu = 0.001\,\mathrm{Pa}\cdot \mathrm{s}$ is the dynamic viscosity.
Frequency $f~[\mathrm{Hz}]$ of a~periodic solution is characterized
by the Strouhal number $\mathrm{St} \coloneqq f L / V$.
The important output quantities are the drag and lift forces acting at the
cylinder~$\Gamma$, i.e., the horizontal and vertical components of $\int_\Gamma
\mathbf{t}$. In the two-dimensional setting, the forces have physical dimension of
$\mathrm{N}\cdot\mathrm{m}^{-1}$, i.e., force per unit length in the out-of-plane direction.
The dimensionless drag and lift coefficients are defined as
\begin{equation}
  \label{eq:drag-lift-coeffs}
  C_D = \frac{1}{\rho V^2 R}\int_\Gamma \mathbf{t}\cdot\mathbf{e}_x,
  \qquad
  C_L = \frac{1}{\rho V^2 R}\int_\Gamma \mathbf{t}\cdot\mathbf{e}_y.
\end{equation}
For the details on numerical evaluation of traction $\mathbf{t}$,
see~\cref{ap:B}.

\subsection{Branching, multiplicity, and linear stability analysis}
\label{sec:branching}
By \emph{branching} we refer to the bifurcation of solutions from a steady state. When such a state loses stability, perturbations may exhibit nontrivial temporal evolution. Branching of solutions to the unsteady Navier--Stokes equations may
be studied analytically using linear stability analysis; see
\citep*{Babenko1980}. Since we consider a bounded domain only, the spectrum of the linearized Navier--Stokes operator around the steady state consists of isolated eigenvalues of finite multiplicity. It may be proved that branching occurs at critical Reynolds number $\mathrm{Re}_\text{crit}$ if the linearized operator associated with the perturbation equations around the steady state does not involve a zero eigenvalue and a simple pair of nonzero purely imaginary conjugate eigenvalues is present, with each eigenvalue having a one-dimensional subspace over the complex field~\citep*{galdi2016bifurcating}. Under these conditions on eigenvalues, a~periodic
solution branching off from the stationary one may be constructed. This
phenomenon is commonly referred to as a~Hopf bifurcation. However, the rigorous proof of the existence of such eigenvalues for the linearized Navier--Stokes operator remains an open problem.

To study branching we will employ, beside the direct time integration and the
steady traction profiles, spectral analysis of the linearized system for the
perturbation $(\hat{\mathbf{v}}, \hat{p})$,
\begin{equation}
  \label{EQ:LinearPerturbation}
  \begin{alignedat}{2}
    \operatorname{div} \hat{\mathbf{v}} &= 0
    &\qquad &\text{in $\Omega$,} \\
    \lambda\rho\hat{\mathbf{v}} &= \operatorname{div} \hat{\mathbb{T}} - \rho(\hat{\mathbf{v}}\cdot\nabla\mathbf{v}_\text{st} + \mathbf{v}_\text{st}\cdot\nabla\hat{\mathbf{v}})
    &\qquad &\text{in $\Omega$,} \\
    \hat{\mathbb{T}} &= -\hat{p}\mathbb{I} + \mu\bigl(\nabla\hat{\mathbf{v}}+(\nabla\hat{\mathbf{v}})^\top\bigr)
    &\qquad &\text{in $\Omega$,} \\
    \vphantom{\hat{\mathbb{T}}}  
    \hat{\mathbf{v}} &= \boldsymbol{0}
    &\qquad &\text{on $\Gamma_\mathrm{D}$,} \\
    \vphantom{\hat{\mathbb{T}}}  
    \hat{\mathbf{v}} - (\hat{\mathbf{v}}\cdot\mathbf{n})\mathbf{n} &= 0
    &\qquad &\text{on $\partial\Omega \setminus \Gamma_\mathrm{D}$,} \\
    (\hat{\mathbb{T}}\mathbf{n})\cdot\mathbf{n} &= 0
    &\qquad &\text{on $\partial\Omega \setminus \Gamma_\mathrm{D}$,}
  \end{alignedat}
\end{equation}
which has been obtained by expanding $\mathbf{v}$ in~\eqref{eq:navier-stokes}
around a~steady solution $(\mathbf{v}_\text{st}, p_\text{st})$ as
\begin{equation}
  \label{EQ:2DLinearPerturbation}
  \begin{alignedat}{4}
    \mathbf{v} &= \mathbf{v}_\text{st} &{}+{}& \varepsilon e^{\lambda t}\hat{\mathbf{v}}, \\
    p          &= p_\text{st} &{}+{}& \varepsilon e^{\lambda t}\hat{p},
  \end{alignedat}
\end{equation}
with $(\hat{\mathbf{v}}, \hat{p})$ independent of time,
neglecting the $O(\varepsilon^2)$ terms.
The homogeneous boundary conditions are prescribed for the perturbation $\hat{\mathbf{v}}$,
as $\mathbf{v}_\text{st}$ satisfies the inhomogeneous conditions in~\eqref{eq:navier-stokes},
and so does $\mathbf{v}$.
We refer to this formulation as the two‑dimensional linear stability analysis (2D LSA).

It is well established, both theoretically and computationally, that the
critical Reynolds number at which the Hopf bifurcation occurs, predicted by
an eigenmode crossing the imaginary axis, agrees well with direct numerical
simulations; see, e.g., \citep{2010IJNMF..63..533M}. If the LSA is applied
to analyze transitions beyond the first Hopf bifurcation,
however, the long-time solution of the unsteady system is a time-periodic
attractor rather than a steady state, and the steady solution around which
the LSA is performed is no longer dynamically realized. Furthermore, as LSA
relies on the linear superposition of eigenmodes, it is not expected to
capture nonlinear mode interactions that become increasingly important at
higher Reynolds numbers. This is an inherent limitation of LSA.

In order to investigate steady three-dimensional branching of the flow, we
consider a three-dimensional linear stability analysis (3D LSA) based on a
spanwise-periodic three-dimensional perturbation ansatz in domain
$\Omega\times\mathbb{R}$.
As before we have
$\mathbf{v}_\mathrm{st}=(v_{\mathrm{st},x},v_{\mathrm{st},y},0)$ and
$p_\mathrm{st}$ the steady two-dimensional base flow, independent of the
spanwise coordinate~$z$.
We consider infinitesimal three-dimensional perturbations of the form
\begin{equation}
  \label{EQ:3DLinearPerturbation}
  \begin{alignedat}{4}
v_x(x,y,z,t) &=
v_{\mathrm{st},x}(x,y)
&{}+{}& \varepsilon e^{\lambda t}\,\hat v_x(x,y)\cos(\kappa z), \\
v_y(x,y,z,t) &=
v_{\mathrm{st},y}(x,y)
&{}+{}& \varepsilon e^{\lambda t}\,\hat v_y(x,y)\cos(\kappa z), \\
v_z(x,y,z,t) &=
&& \varepsilon e^{\lambda t}\,\hat v_z(x,y)\sin(\kappa z), \\
p(x,y,z,t) &=
p_\mathrm{st}(x,y)
&{}+{}& \varepsilon e^{\lambda t}\,\hat p(x,y)\cos(\kappa z),
  \end{alignedat}
\end{equation}
where $\kappa>0$ is an unknown spanwise wavenumber and
$(\hat{\mathbf{v}},\hat{p})=\bigl((\hat v_x,\hat v_y,\hat v_z),\hat{p}\bigr)$
is the sought perturbation on the two-dimensional domain $\Omega$.
Further details can be found in \citep*{Barkley_Henderson_1996,
Thompson2016} and the references therein. Substituting this ansatz
in~\eqref{eq:navier-stokes} and neglecting terms of order
$O(\varepsilon^2)$, as before, yields the generalized eigenvalue problem
\begin{equation}
\label{EQ:3D_LSA_final_T}
\begin{aligned}
\operatorname{div} \hat{\mathbf v} + \kappa\,\hat v_z &= 0 &\qquad& \text{in $\Omega$,}
\\[0.3em]
\lambda \rho \hat{\mathbf v}
&=
\operatorname{div} \hat{\mathbb T}
- \rho(\hat{\mathbf v}\cdot\nabla\mathbf v_{\mathrm{st}}
+ \mathbf v_{\mathrm{st}}\cdot\nabla\hat{\mathbf v})
- \mu \kappa^2 \hat{\mathbf v}
- \kappa \hat p\,\mathbf e_z &\qquad& \text{in $\Omega$,} \\[0.3em]
\hat{\mathbb T}
&=
-\hat p\,\mathbb I
+ \mu\bigl(\nabla\hat{\mathbf v}
+ (\nabla\hat{\mathbf v})^\top\bigr)&\qquad& \text{in $\Omega$.}
\end{aligned}
\end{equation}
Here, $\nabla$ and $\operatorname{div}$ denote differential operators acting only on the
in-plane spatial variables $(x,y)$ as before, i.e., $\nabla = (\partial_x,
\partial_y, 0)$. Boundary conditions are identical to those in~\eqref{EQ:LinearPerturbation}.
Observe that \eqref{EQ:3D_LSA_final_T} reduces to~\eqref{EQ:LinearPerturbation}
when $\kappa=0$.

Observe that if $\lambda=0$ and $\kappa>0$,
ansatz $(\mathbf{v},p)$ in~\eqref{EQ:3DLinearPerturbation} is a~three-dimensional steady state,
which breaks the translational symmetry in the spanwise direction~$z$
and which has spanwise period $L_z=2\pi/\kappa$.
Recall that this has been obtained using the perturbation theory, thus valid for
small $\varepsilon>0$.
Mode~E is the first (in~$\mathrm{Re}$) three-dimensional mode,
which manifests as an instability sensitive to spanwise perturbations.
\citet*{Thompson2016} show that this is indeed the first three-dimensional
instability; for lower Reynolds numbers, spanwise perturbations decay.
This corresponds to $\lambda$ with negative real part in~\eqref{EQ:3DLinearPerturbation};
cf.\ \cref{fig:modeE}. Thus we will look for pairs $(\mathrm{Re},\kappa)$
with smallest $\mathrm{Re}>0$ such that $\lambda=0$ is an~eigenvalue in~\eqref{EQ:3D_LSA_final_T}.

To study the multiplicity of steady solutions, we employ the deflated continuation
and search for imperfect pitchfork bifurcations. Deflated continuation penalizes already
found solutions and searches the solution space for another one; yet, it does not guarantee that all the existing steady solutions are found \citep*{farrell2015}. According to
our results, the Reynolds number at which this bifurcation occurs is
significantly higher than the critical Reynolds number associated with
branching. Within the framework of two-dimensional linear stability analysis, this behavior is associated with the presence of a purely real eigenvalue that remains negative over a finite range of Reynolds numbers and gradually drifts toward zero as $\mathrm{Re}$ approaches the bifurcation point from below. At the bifurcation, this eigenvalue reaches zero, giving rise to new steady solution branches.
This mechanism is consistent with the hydrodynamic pseudospectral picture described in \citep*{Gerecht2012}, which demonstrates that, beyond the first Hopf bifurcation, the linearized Navier--Stokes operator typically exhibits several real eigenvalues that may trigger steady‑state bifurcations. Our observations provide a concrete realization of this scenario in a confined wake configuration and further establish a connection between the approach of a real eigenvalue to zero, the emergence of new steady branches, and qualitative changes in pointwise traction.

\subsection{Theoretical context of the Navier--Stokes equations}
\label{ssec:theory}
The mathematical theory about the Navier--Stokes equations is rich and diverse. Many proofs are available, even though the famous millennium problem about proving or disproving uniqueness for the initial value problem of the unsteady Navier--Stokes equations in three space dimensions is still unsolved. What is known with respect to uniqueness is that the initial value problem with two space dimensions is unique. For any initial data in $L^2$, all weak solutions are actually smooth and unique, which is a~celebrated result by \citet*{Lady}.

For the unsteady 3D Navier--Stokes equations in the last decade, several breakthrough results toward non-uniqueness of weak or very weak solution concepts were shown by \citet*{Sve14,JiaSve15,BucVic19}, and most recently, non-uniqueness for weak solutions to the forced Navier--Stokes equations \citep*{ABC}. Active discourse currently centers on whether these non-uniqueness techniques can be related to the millennium problem and whether they can be related to oscillating behavior observed of the Navier--Stokes equations after rigid obstacles.

The steady Navier--Stokes equations is known to be not unique in 2D and 3D, except for the case of small Reynolds numbers. The related question on the uniqueness of time-periodic solutions, which are one concern of this paper, is also not solved except for the case of small Reynolds numbers. Indeed, the steady solutions are known to be the only time-periodic solutions in the context of a single obstacle in a fluid channel, if the Reynolds number is sufficiently small \citep*{Galdi-Gazzola}.

Multiplicity of steady or time-periodic solutions is connected to what is arguably the most important task for the mathematical theory of Navier--Stokes equations: To explain, quantify, or certify what makes a~solution stable. Beyond small Reynolds numbers, rigorous mathematical results on global stability and on the uniqueness of long-time attractors or time-periodic regimes remain limited, even in two space dimensions. However, necessary conditions related to the spectrum of the linearized operator for a Hopf bifurcation to appear have been developed \citep*{galdi2016bifurcating,galdi2020problem}, and our computations as well as others indicate that these conditions are met. At the same time, the stability of flow behavior is very well studied through physical experiments, numerical simulations and hydrodynamic stability theory, especially for flows past obstacles. For physical
experiments see \citep*{Gerrard_1966,Bearman_Zdravkovich_1978,schewe1983force}; see
also \citep*{ZDRAVKOVICH199053,michalek2022influence} for a setting similar to the
Sch\"{a}fer--Turek benchmark.
Here, we study numerically how steady solutions, time-dependent attractors, stability properties and traction profiles change as the Reynolds number increases. In this way, the computations complement both the mathematical theory and experimental observations. Indeed, numerical predictions have the advantage that they represent purely the idealized partial differential equations, allowing, for example, the detection of multiple unstable steady solutions, which is not something a physical experiment can easily show.
The same also holds for the perspective of this paper, which analyses the traction of the fluid on the obstacle as a local boundary quantity that may reflect changes in the surrounding flow.

\subsection{Critical Reynolds numbers}
By the methods described above, we identify several critical Reynolds numbers associated with bifurcations and transitions in the long-time dynamics of the Navier--Stokes solutions. These include the emergence of solutions that oscillate in time, as well as distinct transitions in the pointwise traction along the obstacle boundary, features that are not captured by classical analyses but can be reliably detected from steady-state data. We further analyze how these changes in traction relate to classical concepts of bifurcation theory, such as the appearance of complex or unstable eigenvalues and the related approximations by linear perturbation, an approach that remains widely used and relevant. 
We also analyze the connection between these changes and the emergence of multiple (unstable) steady solutions revealed by deflated continuation.

At first, several critical Reynolds numbers were identified through direct observation of changes in the long-time behavior of the solutions to unsteady Navier--Stokes equations (see \cref{fig:teaser_result}). In each case, we then examined the corresponding steady-state traction profiles and found that they exhibit clear and systematic localized changes at the same Reynolds numbers in either the lift or drag component. Since these components are aligned with the principal flow directions, we expect the lift to be particularly sensitive to asymmetrical transitions, analogous to lift generation on bluff bodies, e.g., an airfoil, while changes in drag might reflect more global alterations in the flow regime, analogous to the drag-crisis phenomenon \citep*{Bot2016}. The first such transition occurs at $\mathrm{Re} = 7$, where the traction profile begins to deviate from the shape of traction in the Stokes regime ($\mathrm{Re} = 0$), which is similar to the traction profile of Stokes's law (in an open domain). Even though the steady solution remains unique and stable, two vortices appear in the formerly vortex-free flow. At $\mathrm{Re} = 48$, the steady solution loses stability and gives way to a stable, nontrivial time-periodic flow, which is again accompanied by distinct changes in the traction. Around $\mathrm{Re} = 315$, multiple steady states become observable, and the time-periodic solution exhibits pronounced asymmetry, with vortices beginning to drift toward the top wall. We concluded our exploration of the unsteady Navier--Stokes regime at $\mathrm{Re} = 500$. In all cases, the observed bifurcations in the unsteady flow relate to qualitative changes in the traction profile. A~schematic summary is shown in \cref{fig:teaser_result}.
\begin{figure}
    \centering
    \begin{tikzpicture}[
        every node/.style={align=center, font=\small},
        scale=1.15
    ]
        \def\yA{6}
        \def\yC{4.5}

        \draw[->] (0,\yA) -- (11.25,\yA) node[below] at (11.25, \yA-0.2) {\small$\mathrm{Re}$};
        \foreach \x/\val in {0/0, 1.5/7, 4.25/48, 7/315} {
            \draw[thick] (\x,\yA-0.1) -- (\x,\yA+0.1);
            \node[below] at (\x,\yA-0.2) {\val};
        }
        \node[above] at (0.75,\yA+0.15) {stationary,\\vortex-free};
        \node[above] at (2.895,\yA+0.15) {stationary,\\with vortex wake};
        \node[above] at (5.6125,\yA+0.15) {decaying\\vortex street};
        \node[above] at (9,\yA+0.1) {non-symmetric vortex street\\(also multiplicity of steady states)};

        \draw[->] (0,\yC) -- (11.25,\yC) node[below] at (11.25, \yC-0.2) {\small$\mathrm{Re}$};
        \foreach \x/\val in {0/0, 1.5/7, 4.25/48, 7/315} {
            \draw[thick] (\x,\yC-0.1) -- (\x,\yC+0.1);
            \node[below] at (\x,\yC-0.2) {\val};
        }
        \node[above] at (0.0,\yC+0.1) {};
        \node[above] at (1.5,\yC+0.15) {lift};
        \node[above] at (4.25,\yC+0.1) {drag};
        \node[above] at (7,\yC+0.1) {lift, drag};
    \end{tikzpicture}
    \caption{The upper row illustrates the evolution of long-time flow regimes in the
      Sch\"{a}fer--Turek benchmark as the Reynolds number increases, governed
      by the time-dependent Navier--Stokes equations. Dynamical transitions include the onset of periodic shedding with the formation of a vortex street, and the subsequent development of pronounced asymmetry toward the upper wall. The lower row marks the Reynolds numbers at which the traction
      profile of the steady solution undergoes characteristic change in
      either the lift or the drag on the upstream face of the cylinder. These
      qualitative changes in the steady profiles align strikingly with
      transitions in the time-dependent dynamics, suggesting that bifurcations
      in the (possibly unstable) steady solution correspond to global-in-time
      regime changes in the unsteady flow.}
    \label{fig:teaser_result}
\end{figure}
Most of the changes in the pointwise traction occur on the front face of the cylinder boundary, supporting our observation that changes in the flow regime have identifiable traction signatures on the upstream face of the obstacle.

\section{Numerical methods}
\label{sec:numerical_methods}
For spatial discretization, the lowest-order Hood--Taylor method is used.
The mesh results in approximately 200,000 degrees of freedom, with 300 nodes
along the cylinder boundary. It is a~priori refined around the cylinder, in the
near wake, and at the outflow.
For the deflated continuation, which is described below, we use a~coarser mesh
yielding only about 100,000 degrees of freedom to keep the compute expenses of
deflated continuation reasonable, but we keep the 300 boundary nodes on the
cylinder.
This resolution has been found sufficient for computing steady states.

Temporal discretization is carried out using the Crank--Nicolson scheme.
The time step is chosen to ensure at least 30 steps per oscillation period,
resulting in the CFL number below $0.5$.
To observe the long-time behavior, each simulation is run past the initial transient until the lift and drag histories exhibit the persistent regime. The required final time depends on the Reynolds number. Near the first Hopf bifurcation, the departure from the unstable steady state is slow and longer integrations are needed, whereas for larger Reynolds numbers the oscillations develop much faster. 
The nonlinear systems are solved using Newton's method, where we employ
lagged Jacobians to reduce computational costs; the LU factors computed
using MUMPS \citep*{MUMPS:1} are reused over multiple Newton/time steps.

For the convective term $(\mathbf{v} \cdot \nabla)\mathbf{v}$ in unsteady
simulations, we employ the EMAC (energy, momentum, and angular momentum
conserving) formulation \citep*{Schroeder2017}, which is given by $(\mathbf{v}
\cdot \nabla)\mathbf{v} = 2\bigl(\nabla \mathbf{v} + (\nabla
\mathbf{v})^\top\bigr)\mathbf{v} + (\operatorname{div} \mathbf{v})\mathbf{v}$.
We note that this is essential at especially higher Reynolds numbers, as it not
only preserves kinetic energy but also ensures the conservation of both linear
and angular momentum at the discrete level;
see \citep*[Theorem~2.1]{Schroeder2017}.

For the details on the computation of traction, see \cref{ap:B}.

For solution of the eigenvalue problem~\eqref{EQ:LinearPerturbation} we employ
the Krylov--Schur algorithm~\citep*{stewart-2001} as implemented in
SLEPc \citep*{Hernandez2005}, with shift-and-invert targeting for vanishing real
part.

To compute multiple solutions of the stationary Navier--Stokes equations, we
employ \emph{deflated continuation} \citep*{farrell2015,farrell2016}.
Continuation in the Reynolds number with step $\Delta\mathrm{Re}=4/3$ is combined with
a~certain deflation procedure, which penalizes already known solutions to
``motivate'' the Newton solver to converge to another solution if it exists.
Unlike many other bifurcation tracking techniques, deflated continuation does
not require eigenvalue computations or any explicit bifurcation detection. All
subproblems remain standard nonlinear solves as the Jacobian for the deflated
problem is obtained by the Sherman--Morrison formula, while the original
preconditioner ($LU$ factorization in our case) is reused. The cost of the method
consists of the need to perform many nonlinear solves, including failed ones. After a prescribed number of unsuccessful Newton iterations at a given Reynolds number, the search is terminated; additional solutions may exist, but are not accessible within the present computational procedure.
Therefore, the steady branches reported below should be understood as the branches detected by this procedure. Additional steady branches may exist and are not excluded by the present computations.

\section[Unsteady Schafer–Turek benchmark observations]{Unsteady Sch\"{a}fer--Turek benchmark observations}
\label{sec:res_intro}
We present our observations of the Sch\"{a}fer--Turek benchmark over the Reynolds number range $\mathrm{Re} \in (0, 500)$. All physical parameters, such as kinematic viscosity and characteristic length, are held fixed, and the Reynolds number is varied solely through changes in the peak of the prescribed parabolic inflow velocity profile.

To investigate multiplicity and branching behavior, we employ the numerical techniques described in the previous section. In particular, we perform time integration of the unsteady Navier--Stokes equations to resolve the long-time behavior of the flow. From the resulting velocity and pressure fields, $\mathbf{v}(x,t)$ and $p(x,t)$,
we examine velocity streamlines and magnitudes in the wake, time series of integral quantities (such as total lift, drag, and wall vorticity), and oscillations in pointwise traction. The time-dependent pointwise traction signal may also yield further insight; however, its analysis is beyond the scope of the present work. In contrast, the pointwise traction profile of the corresponding steady solutions revealed more than one would initially have expected: It consistently exhibited localized structural changes aligned with transitions observed in the unsteady simulations. To further assess the results, we apply deflated continuation to detect bifurcations in the steady-state solutions. We also compute the discrete spectrum of the operator linearized about each steady solution to analyze its linear stability.

Nevertheless, the focus is on how major qualitative changes in the long-time behavior of the flow relate to local features of the steady-state traction profile acting on the cylinder. Specifically, we focus on the lift and drag components of traction, aligned with the principal flow directions. Across the considered Reynolds number range, we observe several key transitions in the flow: From a~steady vortex-free regime (perfect laminar flow around a cylinder) to one with a recirculating wake, followed by the onset of time-periodicity in the form of a decaying vortex street (constant Strouhal number, with vortices in the wake not yet closed, or decaying after one shedding period). At slightly higher Reynolds numbers, symmetric vortices begin to advect along the walls. At even higher Reynolds numbers, multiple steady solutions with pronounced asymmetry appear, even though only one global-in-time attractor is observed in the unsteady simulations. Simultaneously, the time-periodic attractor also exhibits pronounced asymmetry, and wall vortices along the top wall become larger than those along the bottom.

\begin{figure}
    \centering
    \begin{subfigure}{0.49\textwidth}
        \includegraphics[width=\linewidth]{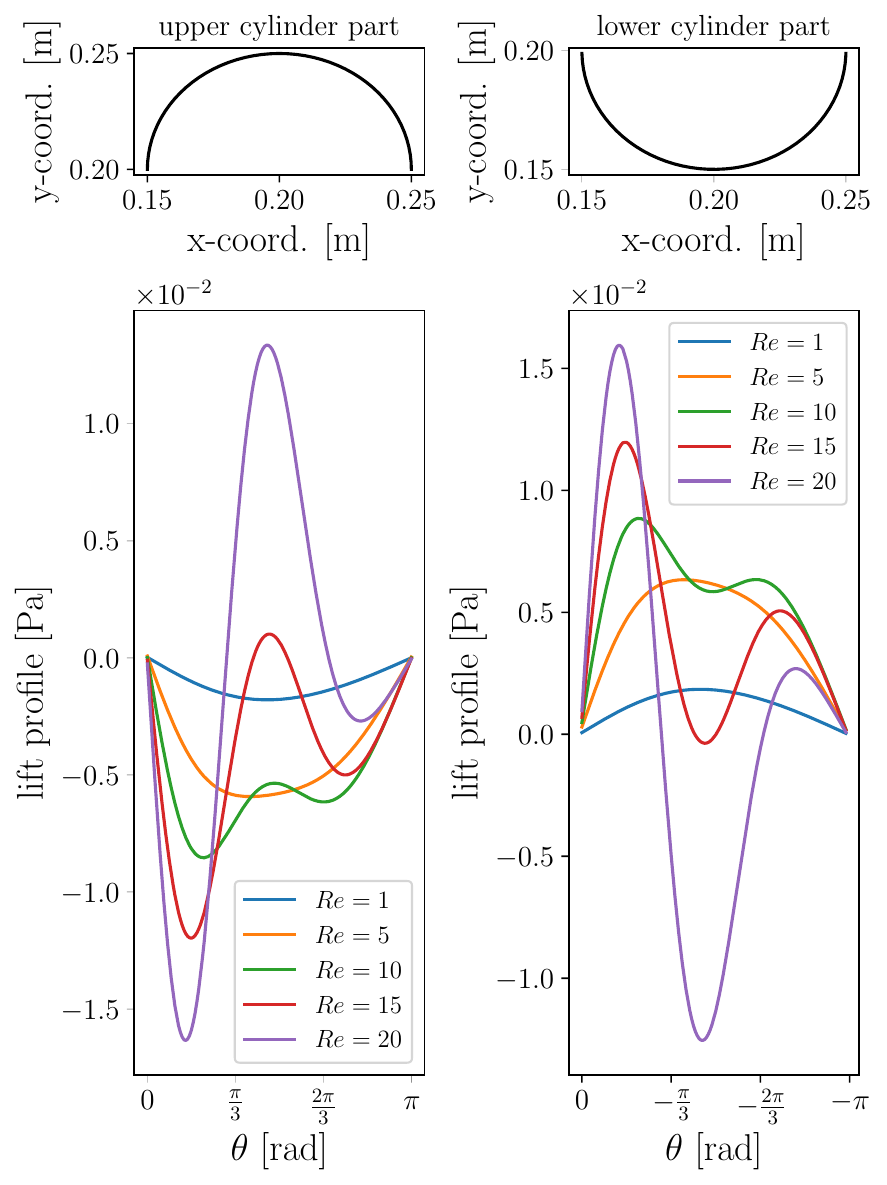}
        \caption{Pointwise lift profile}
    \end{subfigure}
    \begin{subfigure}{0.49\textwidth}
        \includegraphics[width=\linewidth]{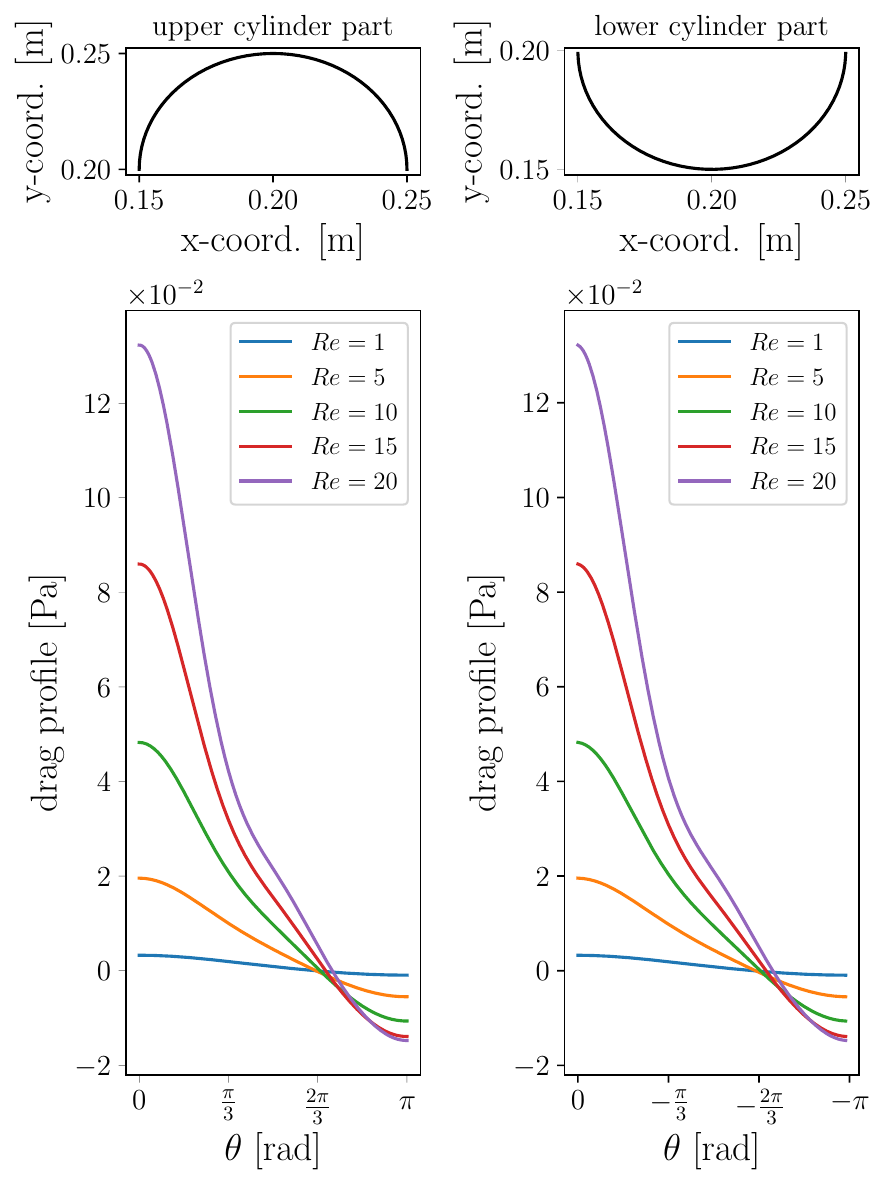}
        \caption{Pointwise drag profile}
    \end{subfigure}
\caption{Traction profiles across the range of Reynolds numbers where the transition from low Reynolds number flow to flow with a~vortex wake occurs}
\label{fig:profiles_Stokes}
\end{figure}

In what follows, we make empirical observation relating critical Reynolds
numbers and certain distinguished points of steady-state traction~$\mathbf{t}$.
Consider a~traction component $t_\text{c}(\theta,\mathrm{Re})$ as a~function of
Reynolds number~$\mathrm{Re}$ and position~$\theta$ on the cylinder~$\Gamma$, where $\theta = 0$ is the most upstream point on the cylinder (the stagnation point), i.e.,
either $t_\text{c}=t_\text{drag}$ or $t_\text{c}=t_\text{lift}$; in the present
case $t_\text{drag} = \mathbf{t}\cdot\mathbf{e}_x$ and $t_\text{lift} =
\mathbf{t}\cdot\mathbf{e}_y$.
For the traction component~$t_\text{c}$, we consider its stationary points with
respect to~$\mathrm{Re}$, i.e., the points $(\theta, \mathrm{Re})$ satisfying
$\partial_{\mathrm{Re}} t_\text{c}(\theta,\mathrm{Re}) = 0$; see \cref{fig:profiles_curves}.
We use this derivative as a local sensitivity measure of the steady traction
profile. It indicates where the local force response on the cylinder changes
direction as the Reynolds number is varied. This is useful because the traction
profiles vary smoothly with~$\mathrm{Re}$, and their absolute values do not always
locate the transition clearly.
Then we consider the following two types of \emph{turning points},

\begin{subequations}
  \label{eq:turning_points}
  \begin{alignat}{3}
    \label{eq:turning_points_a}
    \partial_{\mathrm{Re}} t_\text{c}(\theta,\mathrm{Re}) &= 0
    \quad &&\text{and}& \quad
    \partial_{\theta} \partial_{\mathrm{Re}} t_\text{c}(\theta,\mathrm{Re}) &= 0
    \shortintertext{or}
    \label{eq:turning_points_b}
    \partial_{\mathrm{Re}} t_\text{c}(\theta,\mathrm{Re}) &= 0
    \quad &&\text{and}& \quad
    \partial^2_{\mathrm{Re}} t_\text{c}(\theta,\mathrm{Re}) &= 0.
  \end{alignat}
\end{subequations}
The horizontal folds in \cref{fig:profiles_curves} correspond to
condition~\eqref{eq:turning_points_a} and are exhibited in the following way:
There is initially for lower values of~$\mathrm{Re}$ no stationary point anywhere on
the cylinder; increasing~$\mathrm{Re}$, \eqref{eq:turning_points_a} marks the point at
which a~stationary point first appears on the cylinder.
Similarly, the vertical folds in \cref{fig:profiles_curves} happen when
a~stationary point traveling the curve $\partial_{\mathrm{Re}}
t_\text{c}(\theta,\mathrm{Re})=0$ reverses its direction on the cylinder; these
satisfy condition~\eqref{eq:turning_points_b}.

The turning points consistently appear near the same Reynolds numbers at which
transitions are observed in the unsteady simulations. Depending on the nature
of the transition, turning points appear at different locations on the cylinder surface.
For each transition we have observed, there was at least one turning point on the
upwind side of the cylinder, i.e., for $\theta \in (-\pi/2, \pi/2)$.
A~comprehensive view of the turning points is presented in
\cref{sec:global_turning_points}.

The observed folds of the curves $\partial_{\mathrm{Re}} t_c(\theta,\mathrm{Re})=0$
resemble degeneracies known from bifurcation theory, but they should be interpreted at the level of a derived observable, not at the level of the Navier--Stokes solution branch itself. Classical bifurcation theory studies singularities, changes of stability, or branching of the parameter-to-solution map $(\mathbf v,p)(\mathrm{Re})$; see, e.g., \citep[Ch.~3]{Kuznetsov1998} and \cite{GuckenheimerHolmes1983}. Here, by contrast, we track the local boundary observable
$t_c(\theta,\mathrm{Re};(\mathbf v,p)(\mathrm{Re}))$, computed from steady solutions; see, e.g., \citep[Ch.~3]{Strogatz2015} and \citep[Ch.~I.2]{Temam2001}. The condition
$\partial_{\mathrm{Re}} t_c(\theta,\mathrm{Re})=0$
therefore marks a point where the local traction is stationary with respect to the Reynolds number, or equivalently where its first-order sensitivity with respect to $\mathrm{Re}$ vanishes.

Physically, the pointwise traction is the local force density exerted by the fluid on the obstacle. A turning point of the traction profile therefore identifies a place on the cylinder where the local force response to increasing Reynolds number changes direction. Condition~\eqref{eq:turning_points_a} marks a critical point of this zero-sensitivity set with respect to the boundary coordinate~$\theta$, whereas condition~\eqref{eq:turning_points_b} marks a higher-order degeneracy in the Reynolds-number dependence of the same observable. We use these folds as empirical indicators of flow-regime transitions, not as rigorous bifurcation conditions for the full Navier--Stokes system.

\subsection{\texorpdfstring{Flow at low Reynolds number ($\mathrm{Re} < 7$)}{Flow at low Reynolds number (Re < 7)}}
\label{sec:low_re}
\begin{figure}
\centering
  \centering
  \includegraphics[width=0.6\linewidth]{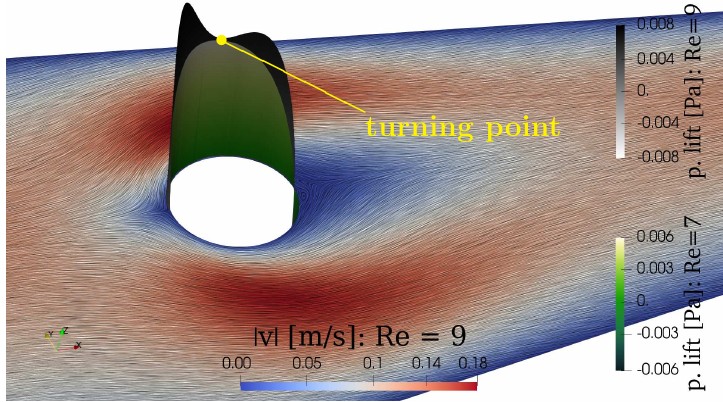}
  \caption{%
    Velocity field at $\mathrm{Re}=9$.
    Lift profiles (shown using warp-by-scalar plot, with reversed sign for
    plotting convenience) at the transition from stationary vortex-free flow ($\mathrm{Re}=7$,
    in shades of green) to stationary flow with a~vortex wake ($\mathrm{Re}=9$, in shades of gray);
    mind the pair of small vortices past the cylinder downstream.
    The turning point ($\theta = 85^\circ$, $\mathrm{Re}=7$, in yellow),
    where the lift is stationary with respect to $\mathrm{Re}$.
  }
  \label{fig:turning_point_Stokes_Wake}
\end{figure}
\begin{figure}
    \centering
    \begin{subfigure}{0.24\textwidth}
        \includegraphics[width=\linewidth]{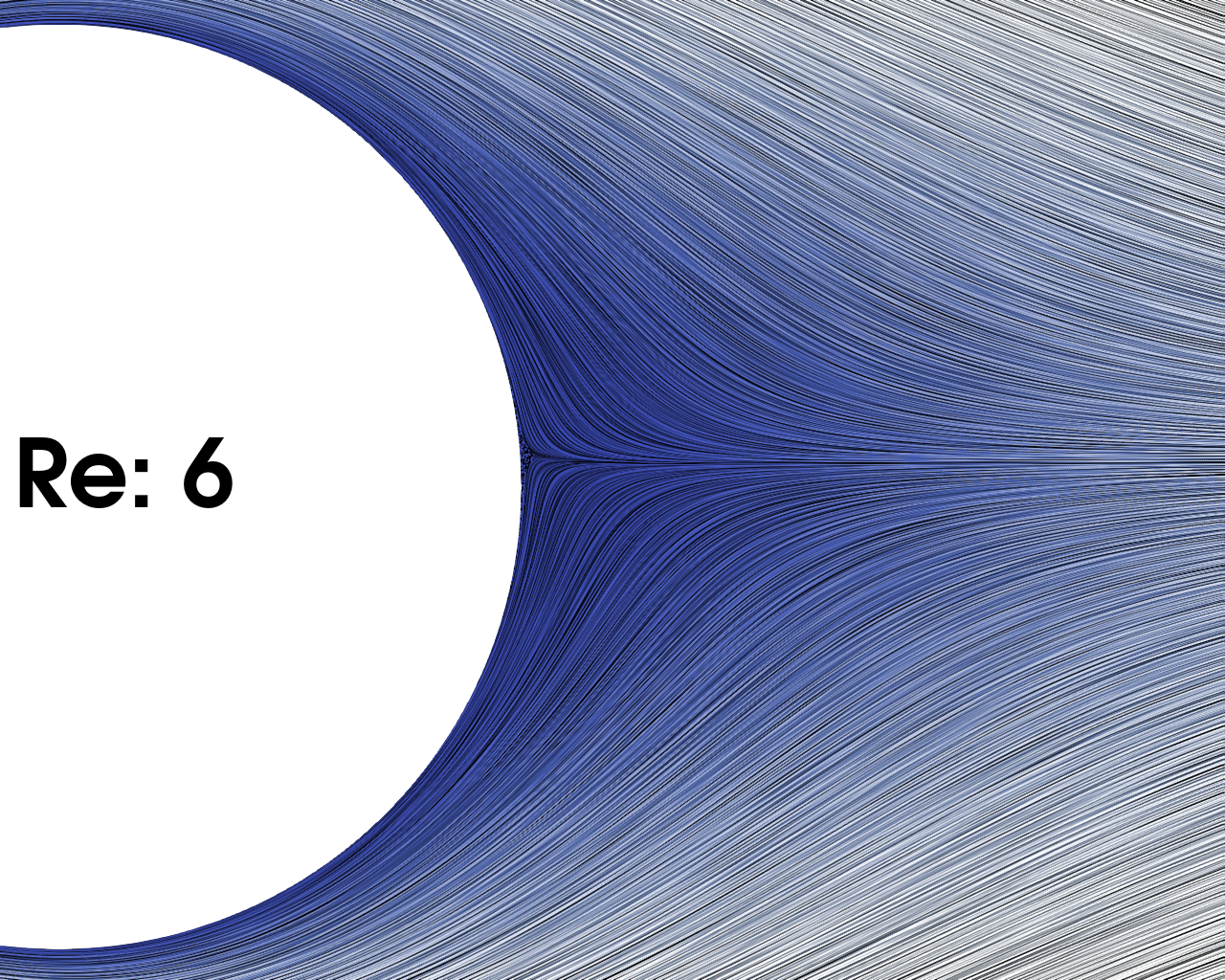}
        \captionsetup{width=\linewidth}
        \caption{$\mathrm{Re} = 6$}
    \end{subfigure}\hfill
    \begin{subfigure}{0.24\textwidth}
        \includegraphics[width=\linewidth]{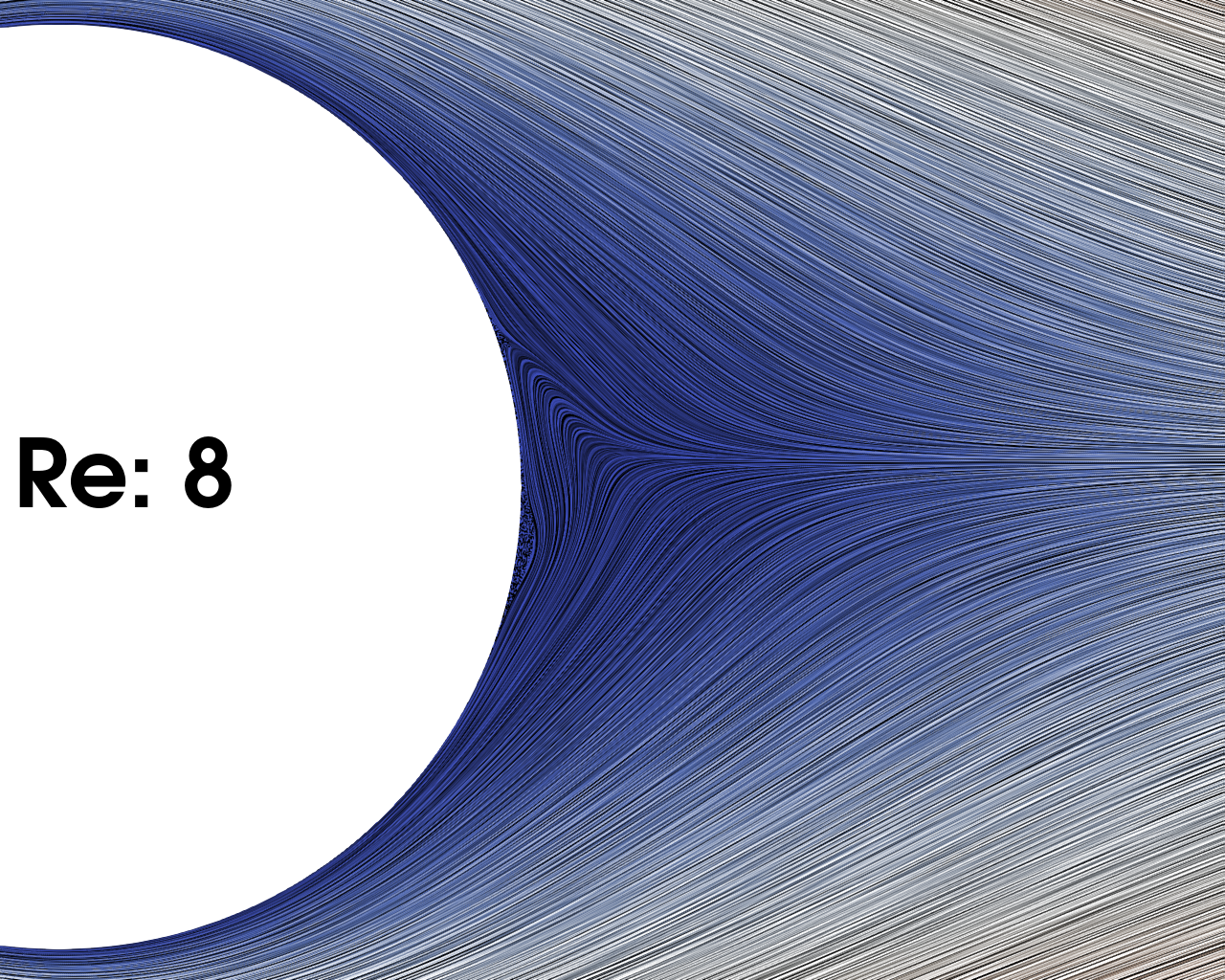}
        \captionsetup{width=\linewidth}
        \caption{$\mathrm{Re} = 8$}
    \end{subfigure}\hfill
    \begin{subfigure}{0.24\textwidth}
        \includegraphics[width=\linewidth]{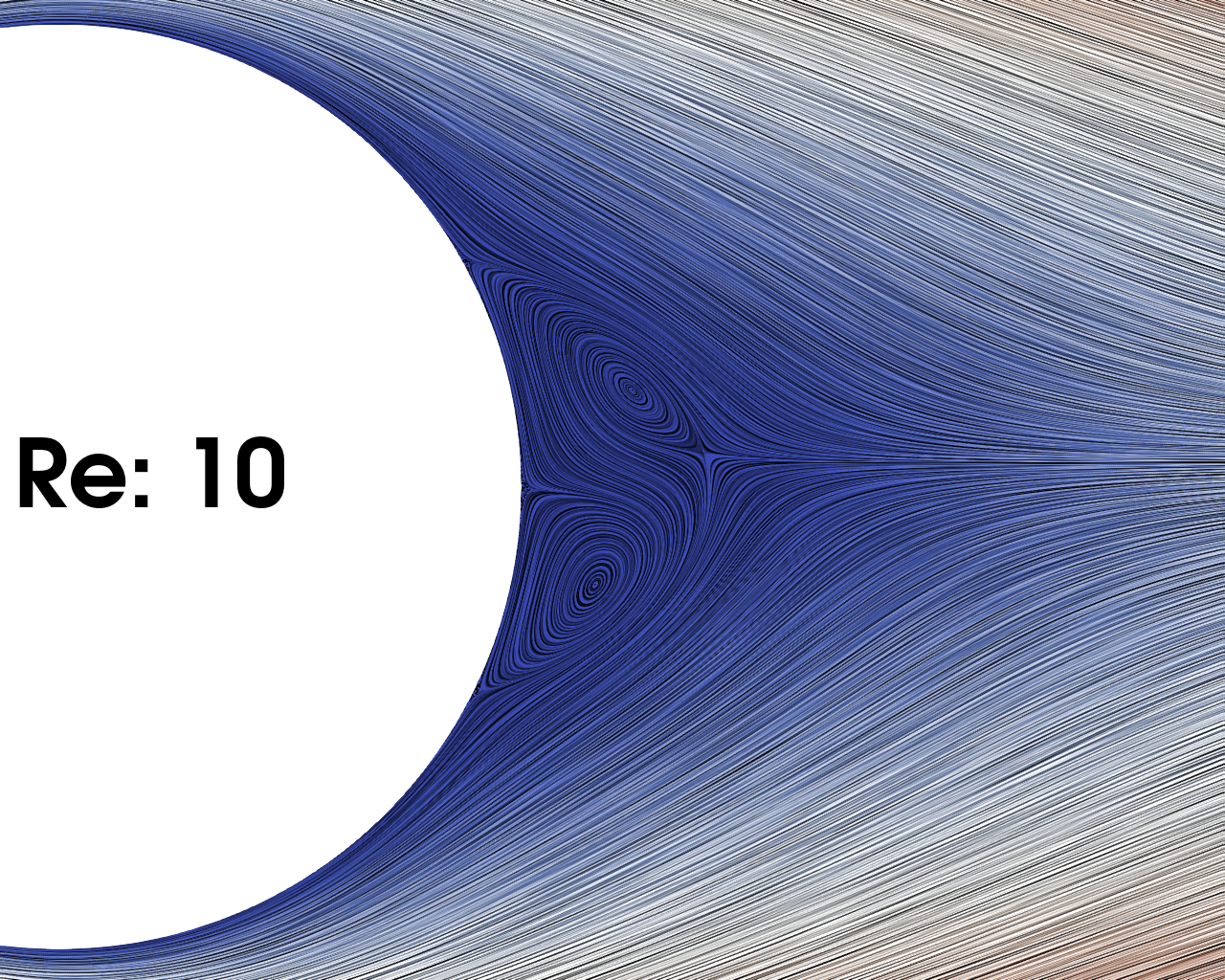}
        \captionsetup{width=\linewidth}
        \caption{$\mathrm{Re} = 10$}
    \end{subfigure}\hfill
    \begin{subfigure}{0.24\textwidth}
        \includegraphics[width=\linewidth]{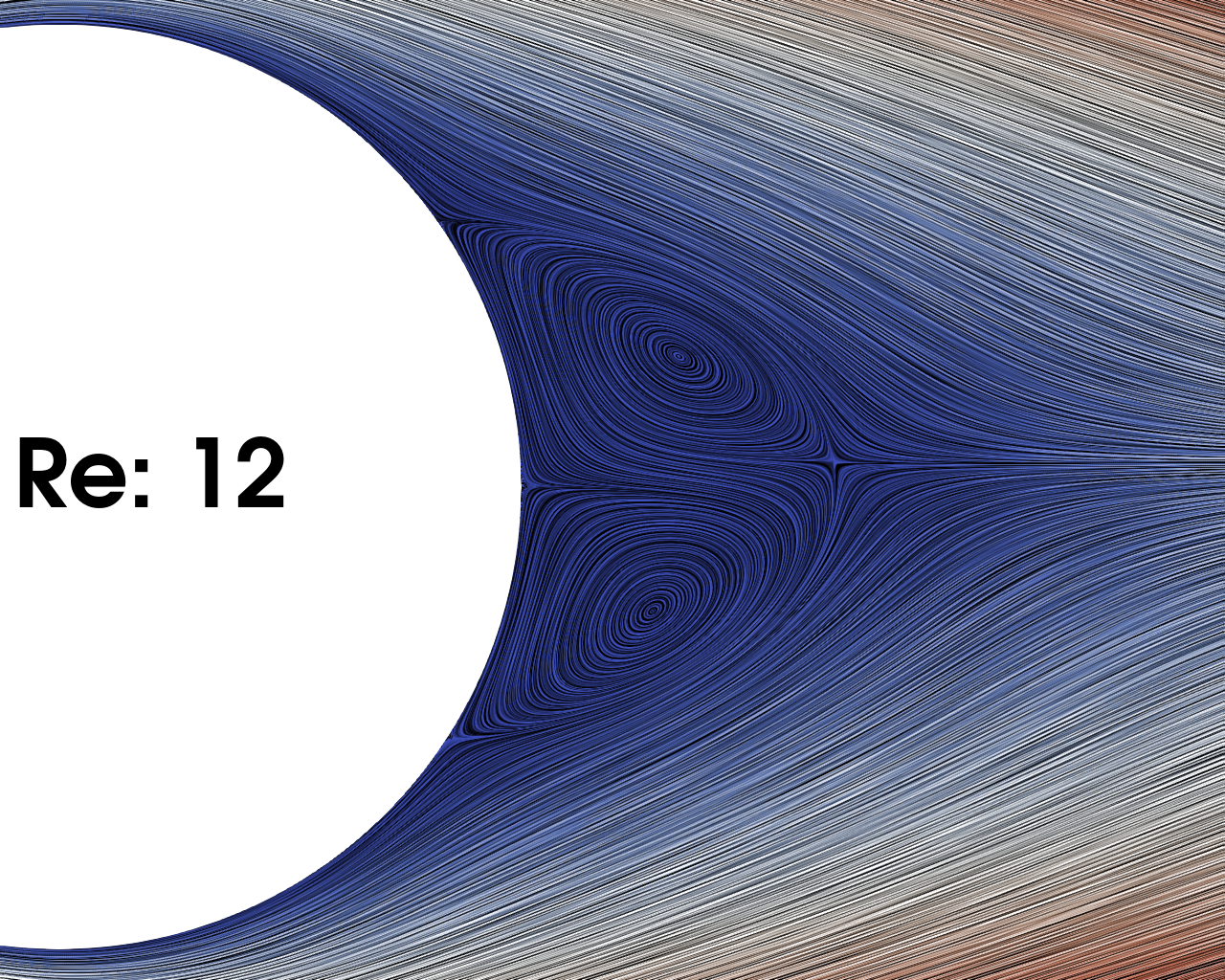}
        \captionsetup{width=\linewidth}
        \caption{$\mathrm{Re} = 12$}
    \end{subfigure}\hfill

    \caption{%
      Velocity field on the downstream side of the cylinder in line integral
      convolution plot. A~pair of stationary vortices appears at $\mathrm{Re}=7$ and
      grows in size with increasing $\mathrm{Re}$.
    }
    \label{fig:turek_vortex_reynolds}
\end{figure}

Perfectly laminar (creeping) vortex-free flow is disrupted by the emergence of convective effects. Specifically, a~pair of recirculation regions begins to form behind the cylinder; see \cref{fig:turek_vortex_reynolds}. This transition is also reflected in the continuous deviation from the characteristic sine-like lift profile, which is analytically computable in the Stokes regime in an open domain; see \cref{fig:profiles_Stokes}. The previously monotonic growth of the profile with increasing inflow velocity is suddenly interrupted by a turning point at the topmost and bottommost points on the cylinder boundary, as shown in \cref{fig:turning_point_Stokes_Wake}.

\begin{figure}
\centering
  \centering
  \includegraphics[width=0.6\linewidth]{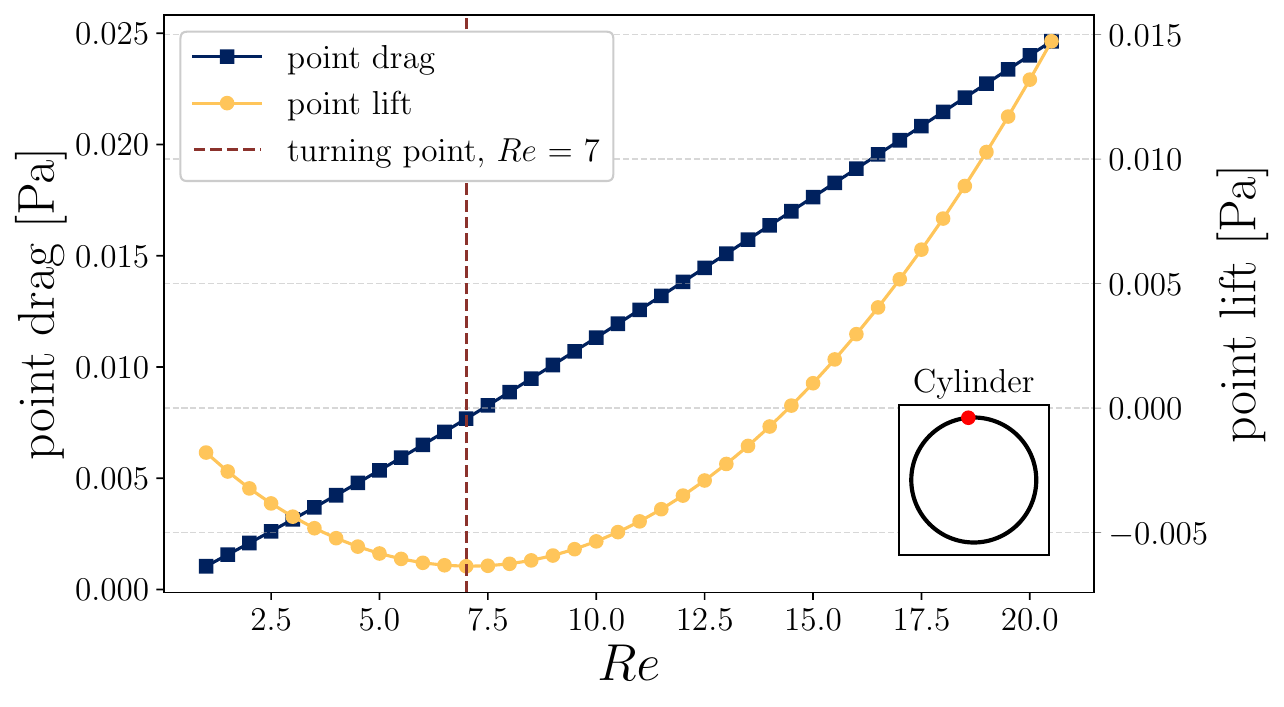}
\caption{Turning point at $(\theta = 85^\circ, \mathrm{Re} = 7)$ in the lift profile corresponding to the bifurcation from low-Reynolds-number vortex-free flow to steady flow with a~vortex wake}
  \label{fig:turning_point_Stokes}
\end{figure}
This is, in fact, the most important initial observation we made: The pointwise lift and drag values at this location are shown in \cref{fig:turning_point_Stokes}. It marks a critical point of the steady traction with respect to $\mathrm{Re}$, occurring precisely at the Reynolds number where the pair of vortices appears, and notably, its location does not coincide with the location at which the separation layer emerges. This motivated us to systematically track critical points where $\partial_{\mathrm{Re}} t_{\text{drag,lift}} = 0$ to determine whether bifurcations in the unsteady system are imprinted in the steady-state traction profile. Since we later observe that these zero-derivative points trace out a smooth curve, and this is the first such point in $\mathrm{Re}$, it must be the location where the curve changes direction in $\mathrm{Re}$. This constitutes one type of turning point, and in fact, the very next one exhibits the same behavior, but in the drag profile. The point itself, along with its coordinates $(\theta = 85^\circ, \mathrm{Re} = 7)$, is clearly visible in the global overview in \cref{fig:profiles_curves_lift} at the end of this section.

\subsection{\texorpdfstring{Steady unique flow ($\mathrm{Re} < 48$)}{Steady unique flow (Re < 48)}}
Even after the recirculating wake appears (see \cref{fig:turek_vortex_reynolds}) behind the cylinder, the long-time unsteady flow agrees with the steady flow and remains unique. With increasing Reynolds number, the transition into vortex shedding occurs and a decaying vortex street emerges. This corresponds to the branching; see \cref{sec:branching}. This phenomenon has been extensively studied in the context of flow past a~cylinder in an infinite domain using both analytical and computational methods \citep*{Benard1908, Karman1911, Zebib1987, Morzynski1991, Jackson1987, Provansal1987}. In such settings, the critical Reynolds number is approximately $\mathrm{Re} = 48$, with a~corresponding Strouhal number $\mathrm{St} = 0.12$.

In contrast, the Sch\"{a}fer--Turek benchmark features a narrow channel, which undoubtedly tends to affect the flow at higher $\mathrm{Re}$. However, at low Reynolds numbers, the influence of the walls appears negligible, as does the asymmetric placement of the obstacle---since the first Hopf bifurcation occurs at a similar $\mathrm{Re}$, as confirmed by \citet*{Jiang_Cheng_2019} in an aspect-ratio study. Consequently, the asymmetrical placement of the obstacle plays no significant role, and the flow quickly becomes symmetric downstream; see the baseline branch values in \cref{fig:bifurcation_diagram}.

\begin{figure}
    \centering
    \begin{subfigure}{0.49\textwidth}
        \includegraphics[width=\linewidth]{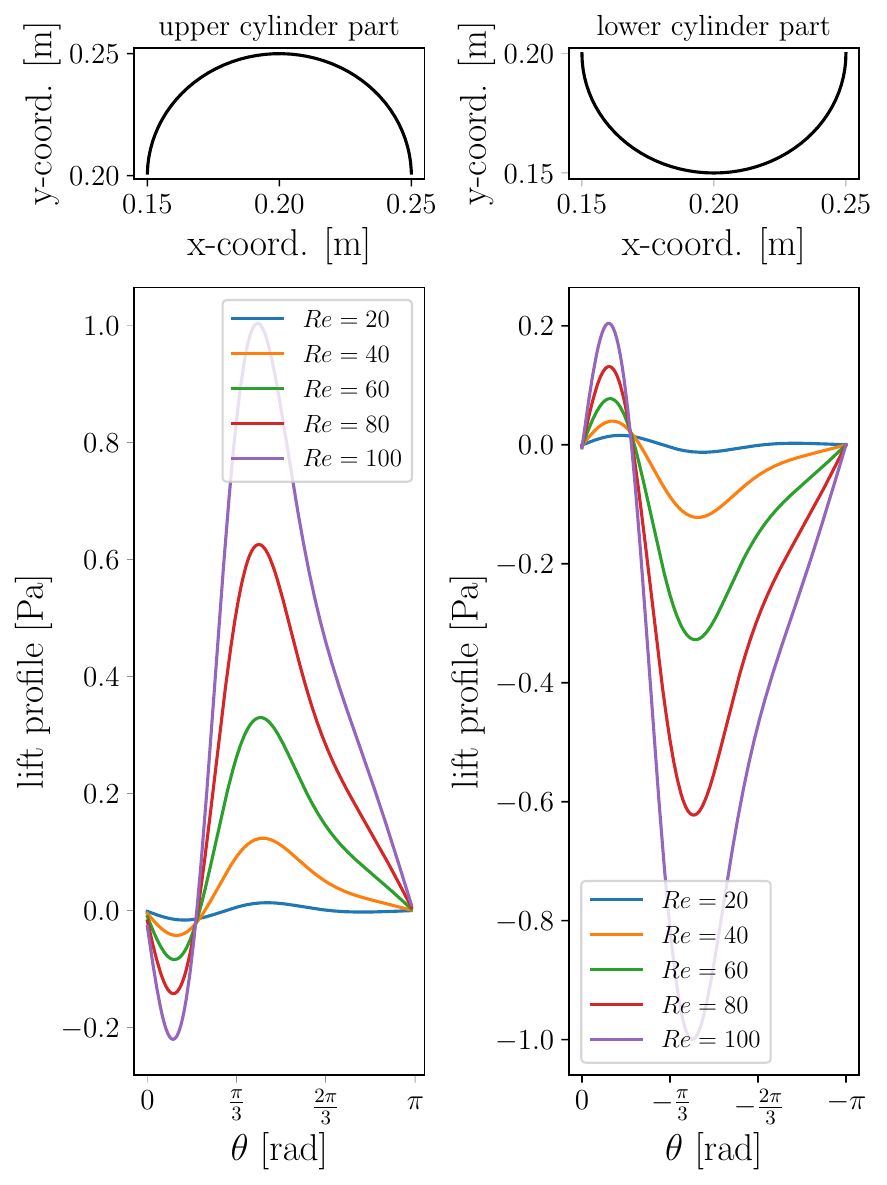}
        \caption{Pointwise lift profile}
    \end{subfigure}
    \begin{subfigure}{0.49\textwidth}
        \includegraphics[width=\linewidth]{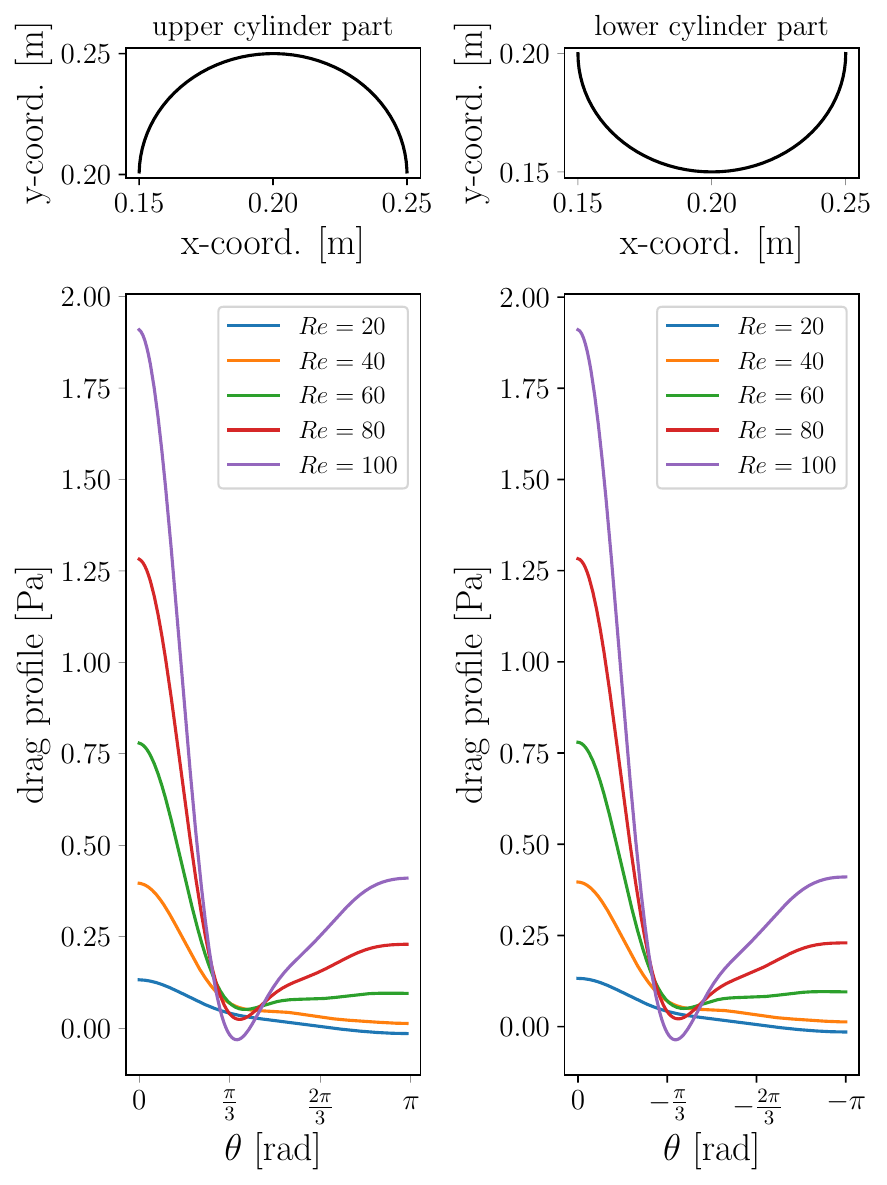}
        \caption{Pointwise drag profile}
    \end{subfigure}
\caption{Traction profiles across the range of Reynolds numbers where the steady flow with a vortex wake transitions to a regime in which the long-time unsteady solution no longer matches the steady one}
\label{fig:profiles_branching}
\end{figure}
For this critical Reynolds number, we were also able to identify a turning point in the steady traction profiles, see \cref{fig:profiles_branching}. The point itself and its coordinates $(\theta = 67.5^\circ, \mathrm{Re} = 48)$ are clearly visible in \cref{fig:profiles_curves_drag}. We observe that, relative to the previous turning point at $\mathrm{Re} = 7$, this one is shifted toward the stagnation point ($\theta = 0$), not toward the separation layer. 

\begin{figure}
  \begin{subfigure}{\textwidth}
    \centering
    \includegraphics[width=.9\linewidth]{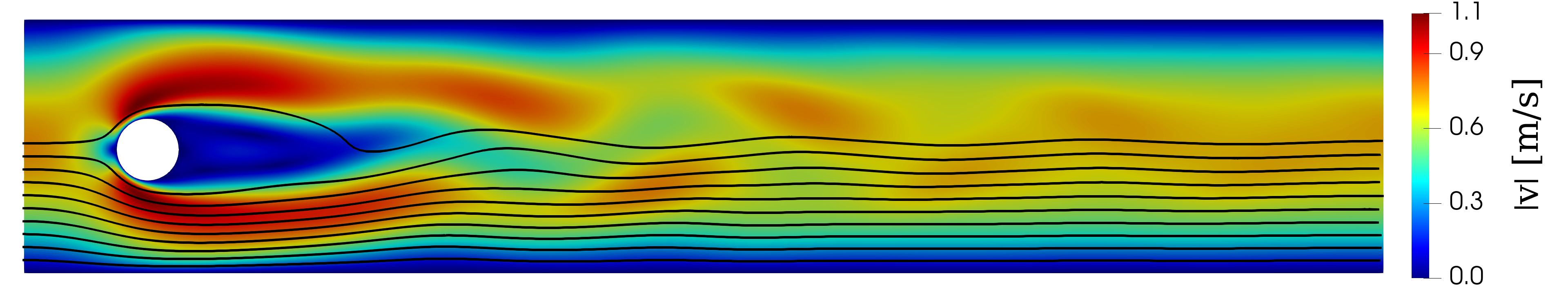}
    \caption{Snapshot of velocity magnitude and streamlines of unsteady solution}
    \label{fig:sfig3a:flow}
  \end{subfigure}
  \begin{subfigure}{\textwidth}
    \centering
    \includegraphics[width=.9\linewidth]{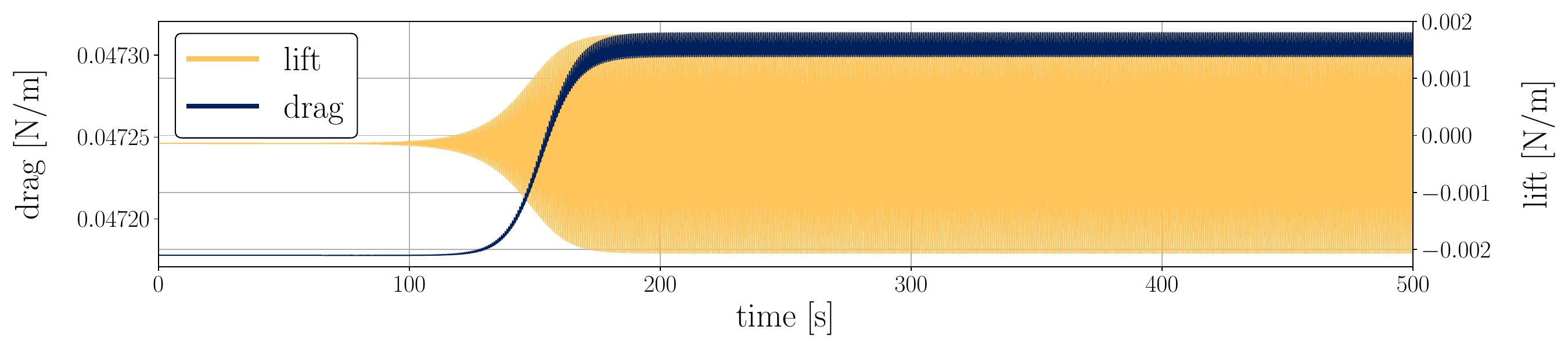}
    \caption{Lift and drag oscillations}
    \label{fig:sfig3b:oscillations_drag_lift}
  \end{subfigure}
  \caption{Results from direct numerical time integration of the unsteady equations at $\mathrm{Re} = 50$}
  \label{fig:sfig3:critical_re_direct}
\end{figure}
The critical Reynolds number is, of course, confirmed by visualization of the velocity field in the domain. By numerically integrating the unsteady equations starting from the steady state, perturbed only by numerical errors associated with the steady-solver residual tolerance, we conclude that the flow indeed diverges from the steady state and becomes unstable. The long-time solution (the attractor) becomes time-periodic at the critical value $\mathrm{Re}_\text{crit} = 48$, as illustrated by a snapshot of the velocity field in \cref{fig:sfig3a:flow}. Since this case is close to the first Hopf bifurcation, the initial growth away from the unstable steady state is slow, and the time integration was continued until the periodic regime was clearly visible. The most appropriate quantity for observing and measuring the frequency of oscillations in the flow appears to be the total lift; see \cref{fig:sfig3b:oscillations_drag_lift}.

\begin{figure}
  \centering
  \begin{subfigure}{\textwidth}
    \centering
    \includegraphics[width=.9\linewidth]{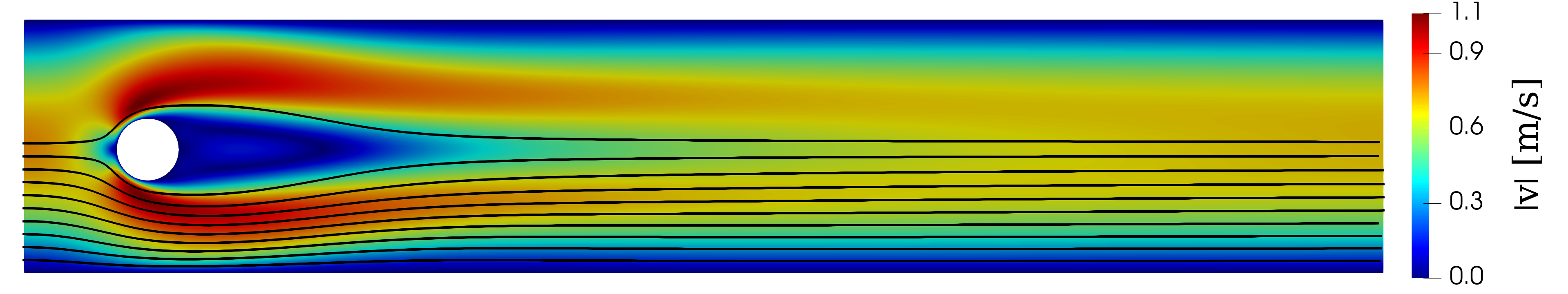}
    \caption{Velocity magnitude and streamlines of steady solution $\mathbf{v}_\text{s}$}
    \label{fig:sfig3a:steady_state}
  \end{subfigure}
  \begin{subfigure}{\textwidth}
    \centering
    \includegraphics[width=.9\linewidth]{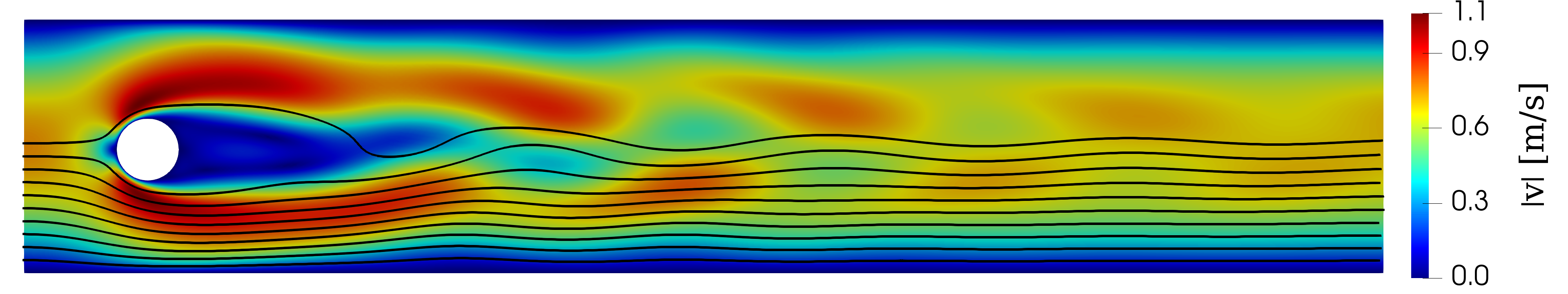}
    \caption{%
      Snapshot of the perturbation solution
      \eqref{EQ:2DLinearPerturbationCritical} for a~suitable choice
      of~$\varepsilon$, selected to provide a~close visual match of the
      velocity field with the full-system solution shown in
      \cref{fig:sfig3a:flow}
    }
    \label{fig:sfig3b:eigenmode_reconstruction}
  \end{subfigure}
  \caption{Construction of perturbation solution using 2D LSA for $\mathrm{Re} = 50$}
  \label{fig:sfig3:critical_re_eigenmode}
\end{figure}
\begin{figure}
  \centering
  \includegraphics[width=.8\linewidth,trim={0 0.75cm 0 0},clip]{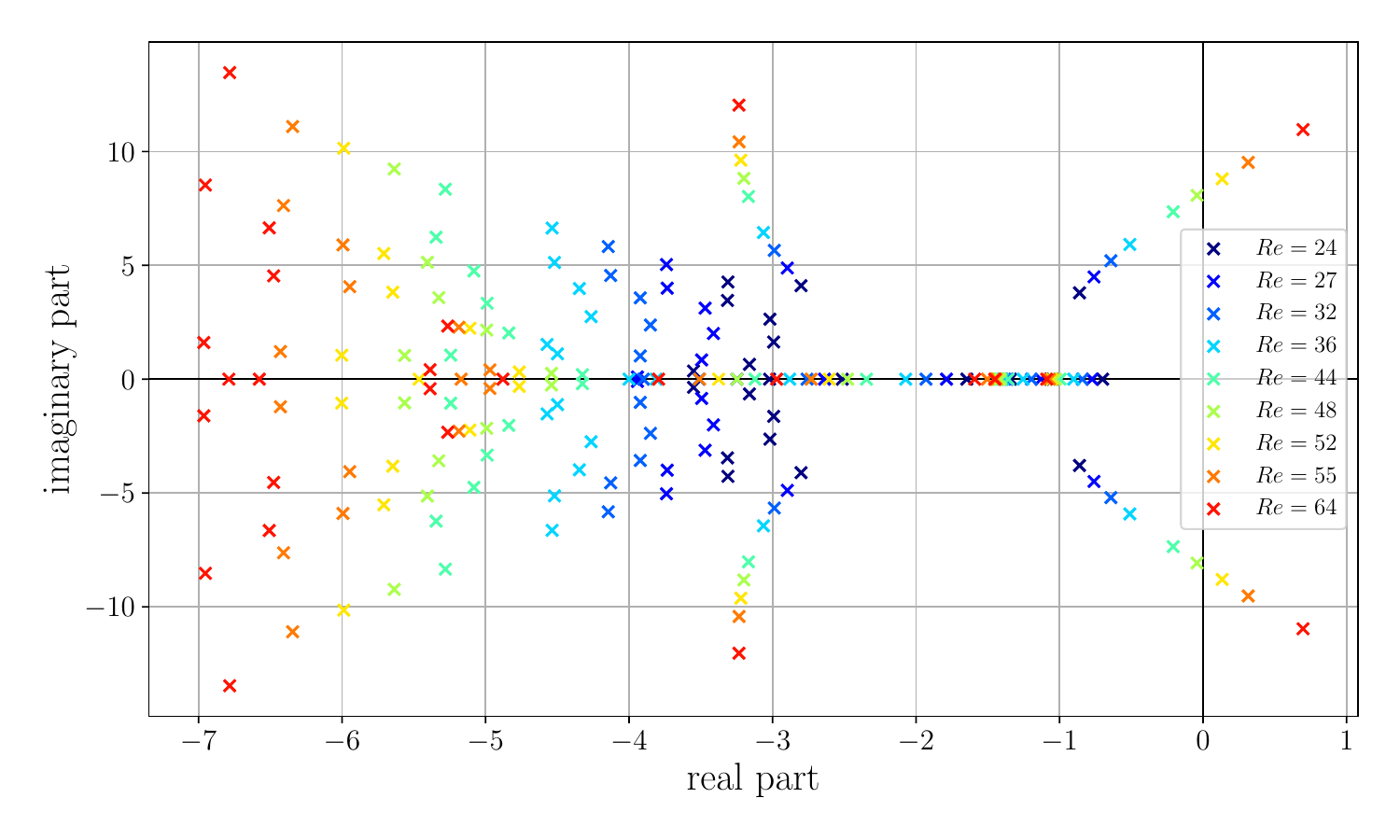}
  \caption{Plot of eigenvalues in the complex plane of the 2D LSA operator around critical $\mathrm{Re}_\text{crit} = 48$. A pair of complex conjugate eigenvalues crosses the imaginary axis at this value, marking the onset of instability.}
  \label{fig:sfig2:critical_re_eigenvalues}
\end{figure}
The critical value at $\mathrm{Re}_\text{crit} \approx 48$ corresponds well to the 2D~LSA
spectrum crossing the imaginary axis; see \cref{fig:sfig2:critical_re_eigenvalues}.
Suppose first that the rightmost eigenvalue pair of~\eqref{EQ:LinearPerturbation}
is $\lambda_{\pm} = \pm i \omega$, $\omega>0$, i.e., is pure imaginary, and the
corresponding eigenvectors are
$\hat{\mathbf{v}}_{\pm} = \texttt{Re}(\hat{\mathbf{v}}) \pm i \texttt{Im}(\hat{\mathbf{v}})$.
Analogously to \eqref{EQ:2DLinearPerturbation}, we can express a~perturbation solution
of the full unsteady system \eqref{eq:navier-stokes} as
\begin{equation}
  \label{EQ:2DLinearPerturbationCritical}
  \mathbf{v} =
  \mathbf{v}_\text{st} + \frac{\varepsilon}{2} \bigl(
      e^{ i\omega t} \hat{\mathbf{v}}_+
    + e^{-i\omega t} \hat{\mathbf{v}}_-
  \bigr)=
  \mathbf{v}_\text{st} + \varepsilon \bigl(
      \texttt{Re}(\hat{\mathbf{v}}) \, \cos(\omega t)
    - \texttt{Im}(\hat{\mathbf{v}}) \, \sin(\omega t)
  \bigr).
\end{equation}
All other modes would feature negative real parts, i.e., the transient
$\exp(-\sigma t)$, $\sigma>0$, in \eqref{EQ:2DLinearPerturbation} and thus
would decay over time.

For Reynolds numbers above the critical value $\mathrm{Re}_\text{crit} = 48$,
there are eigenvalues of~\eqref{EQ:LinearPerturbation} that have positive real
part; see \cref{fig:sfig2:critical_re_eigenvalues}.
Here, the perturbation solution~\eqref{EQ:2DLinearPerturbation} would predict
the exponentially growing, non-physical, transient $\exp(\sigma t)$,
$\sigma>0$.
But such a~solution is only valid for small times, as long as the perturbation
remains small, before the nonlinear terms in~\eqref{eq:navier-stokes} become
significant.
In practice, one observes that, for Reynolds numbers not too far above the
critical value, the exponential growth saturates and the flow eventually
settles into a~time-periodic solution; see
\cref{fig:sfig3b:oscillations_drag_lift}.
If $\lambda_{\pm}=\sigma\pm i\omega$, with $\sigma,\,\omega>0$,
is the only eigenvalue pair of~\eqref{EQ:LinearPerturbation} with non-negative
real part and
$\hat{\mathbf{v}}_{\pm} = \texttt{Re}(\hat{\mathbf{v}}) \pm i \texttt{Im}(\hat{\mathbf{v}})$
are the corresponding eigenvectors, the perturbation
solution~\eqref{EQ:2DLinearPerturbationCritical} with a~suitable
constant~$\varepsilon$ approximates the limiting time-periodic solution;
see \cref{fig:sfig3:critical_re_direct,fig:sfig3:critical_re_eigenmode}.
LSA does not predict $\varepsilon$; the amplitude of the limit cycle
oscillations is determined by nonlinear effects in the full
system~\eqref{eq:navier-stokes}.
This can be modeled by the Stuart--Landau equation; see, e.g.,
\citep*{Provansal1987,NoackEckelmann1994,SippLebedev2007}.

It is worth noting that, at the first Hopf bifurcation at $\mathrm{Re}_{\mathrm{crit}} = 48$, the oscillation frequency predicted by LSA,
$\omega / 2\pi = 1.39$ Hz ($\mathrm{St} = 0.278$), matches closely the
frequency of the time-periodic solution, $f = 1.37$ Hz ($\mathrm{St} = 0.274$),
as measured from the total lift oscillations using Fast Fourier Transform
(FFT). This value is notably more than twice the Strouhal number observed in
the open channel configuration, which is approximately $\mathrm{St} = 0.12$.
The first Hopf bifurcation also provides a~useful link to the spectral study by
\citet*{Gerecht2012}. Their analysis shows that the spectrum beyond the first
Hopf bifurcation may contain real eigenvalues associated with later
steady-state bifurcations. This gives spectral context to the additional steady
branches detected by deflated continuation in \cref{sec:multiplicity}.

\subsection{\texorpdfstring{The flow with different long-time and steady solutions ($\mathrm{Re} \geq 48$)}{The flow with different long-time and steady solutions (Re >= 48)}}
An important finding of our study is that, in the decaying vortex street regime, the system consistently converges to a unique time-periodic attractor, in the numerical sense that all initial conditions tested in our simulations converge to the same long-time state. This attractor appears to be highly stable, as none of our perturbation strategies were able to deflect the solution toward an alternative state. We tested a variety of initial conditions, such as multiple steady states (found in the next section), a zero velocity field, and a halved-in-magnitude snapshot of the periodic solution at twice the Reynolds number. The last case was an attempt to induce an overfrequency effect. Furthermore, we investigated the impact of following a hysteretic path in Reynolds numbers during evolution. We also temporarily modified the inflow profile at the shedding frequency to induce resonance. In all cases, the solution ultimately returned to the same time-periodic attractor, demonstrating its stability.

We now examine the nature of the time-periodic solution through the structure of the velocity field, focusing on how it evolves with increasing Reynolds number. As $\mathrm{Re}$ increases, the wake width gradually increases and eventually saturates at $\mathrm{Re} = 78$ with the formation of vortices along the walls, which are advected downstream. Interestingly, this minor transition is also reflected as a turning point $(\theta = 33.3^\circ,\ \mathrm{Re} = 78)$ in the traction profile. It is visible as a fold in the curve corresponding to a change in direction with respect to $\theta$; see \cref{fig:profiles_curves_lift}. From that point onward, the periodic solution remains structurally consistent across Reynolds numbers in the range $\mathrm{Re} \in (78,315)$, with only minor variations. In this regime, the spatial wavelength $\chi$ of the vortex street appears approximately unchanged in the computed flow fields; see \cref{fig:turek_unsteady_solution} for the definition of~$\chi$. This observation can be related to the Strouhal numbers of solutions at $\mathrm{Re}_1$ and $\mathrm{Re}_2$. Assuming that the convection speed of the vortex pattern is approximately equal to the mean inflow velocity, we have
\begin{equation*} \frac{1}{\mathrm{St}_{\mathrm{Re}_1}}-\frac{1}{\mathrm{St}_{\mathrm{Re}_2}} \approx \frac{1}{L}\bigl(\chi_{\mathrm{Re}_1}-\chi_{\mathrm{Re}_2}\bigr) \approx 0. \end{equation*}
Thus, the observed near-invariance of the vortex wavelength is consistent with an approximately constant $\mathrm{St}$ within this regime. Consequently, the shedding frequency increases approximately linearly with the mean inflow velocity, and hence with $\mathrm{Re}$. This scaling is confirmed by direct frequency measurements obtained from the FFT of the total lift, yielding a mean Strouhal number of $\overline{\mathrm{St}}=0.316$ over the specified interval, with a standard error of the mean of $0.003$. The steepest growth of the Strouhal number occurred in the interval $\mathrm{Re} \in (48,78)$, reaching $\mathrm{St} = 0.297$ by the end, while the wake width was still increasing.

\begin{figure}
  \centering
  \begin{subfigure}{\textwidth}
    \centering
    \includegraphics[width=.9\linewidth]{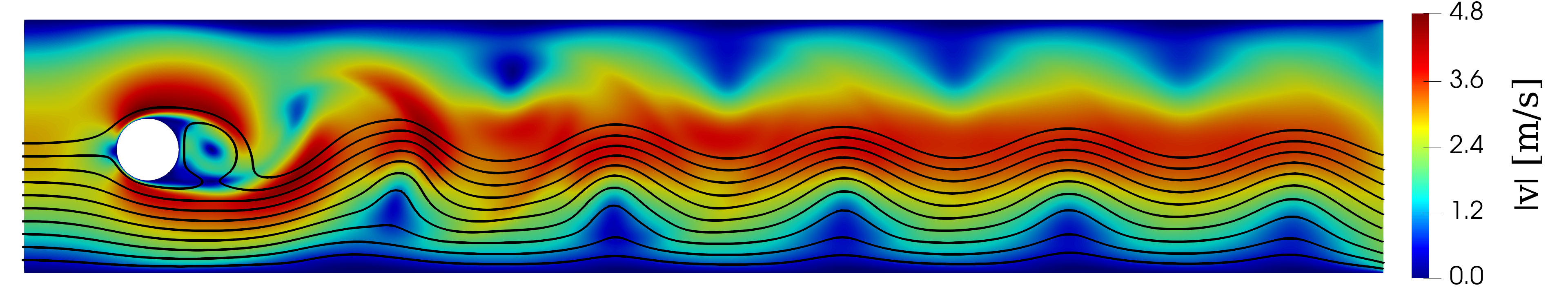}
  \end{subfigure}
  \begin{subfigure}{\textwidth}
    \centering
    \includegraphics[width=.9\linewidth]{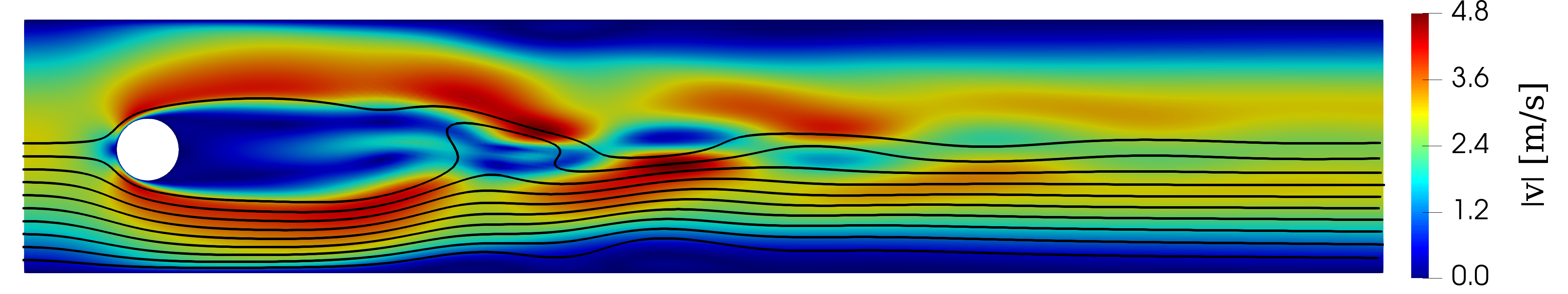}
  \end{subfigure}
  \caption{Comparison of snapshot of unsteady solution (with plotted streamlines over half of the domain) with reconstruction of unsteady solution from steady state and unstable eigenmode at $\mathrm{Re} = 200$}
  \label{fig:sfig3:reconstructionRe200}
\end{figure}
Another notable trend is the behavior of the stationary vortex wake length over time. In the periodic shedding regime, increasing $\mathrm{Re}$ leads to a shortening of the residual steady wake behind the cylinder. Ultimately, the vortex street begins immediately behind the cylinder. This behavior contrasts with the steady-state case, where increasing $\mathrm{Re}$ results in an elongated wake; see \cref{fig:sfig3:reconstructionRe200}. As discussed in \cref{sec:branching}, LSA has to be interpreted with care in this regime. After the first Hopf bifurcation, the long-time solution of the unsteady system is a time-periodic attractor, whereas LSA is performed around the corresponding steady solution. Thus, LSA describes the stability of the steady branch, but it is not expected to describe the nonlinear time-periodic attractor or nonlinear interactions between modes.

Overall, our results indicate that in the decaying vortex street regime, the time-periodic flow pattern remains structurally unchanged across different Reynolds numbers. While minor adjustments occur, such as those described above, the approximately constant Strouhal number and spatial pattern prevail. These findings suggest an inherent self-similarity in the attractor of the shedding dynamics in the Sch\"{a}fer--Turek benchmark, which persists across a broad range of Reynolds numbers.

\begin{figure}
\centering
\begin{subfigure}{.33\textwidth}  \includegraphics[width=\linewidth]{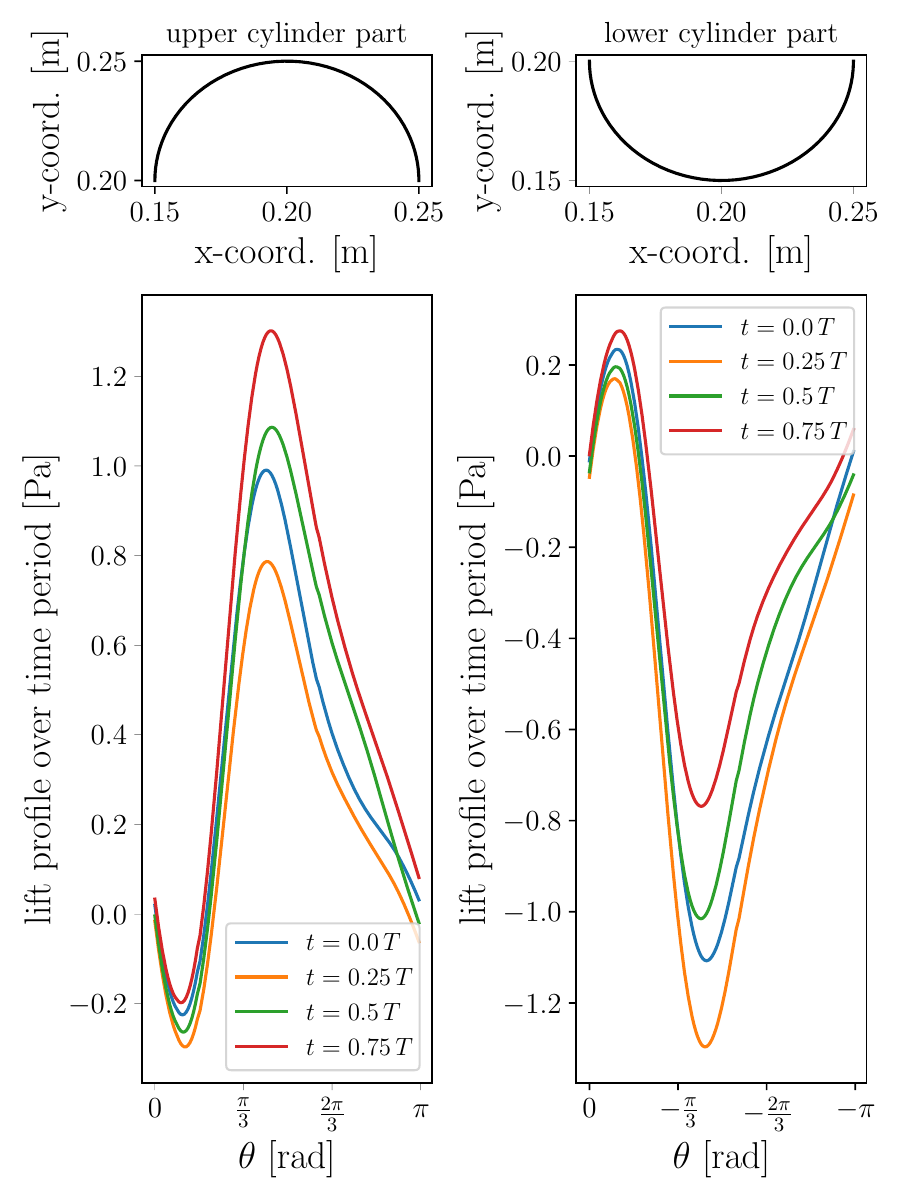}
  \caption{Pointwise lift}
  \label{fig:sfig3}
\end{subfigure}
\begin{subfigure}{.33\textwidth}  \includegraphics[width=\linewidth]{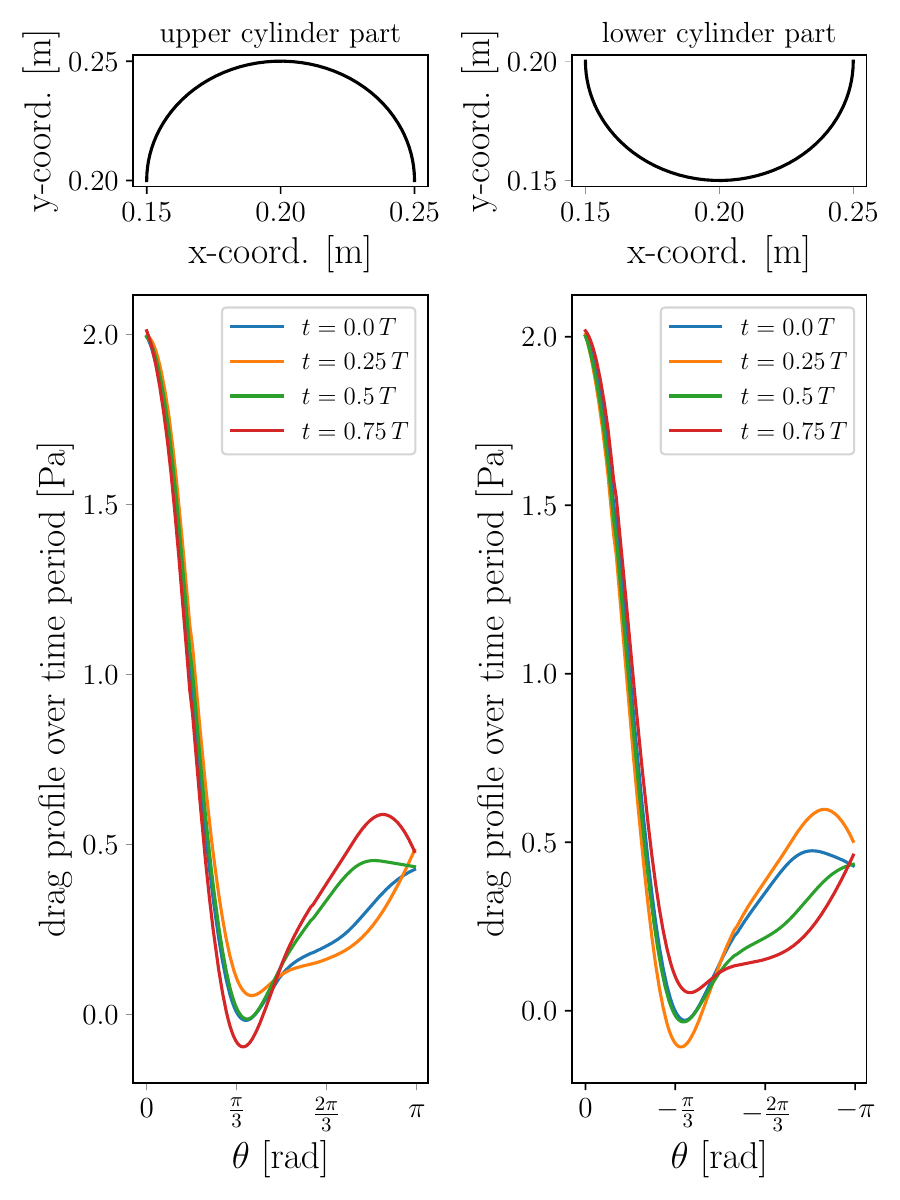}
  \caption{Pointwise drag}
  \label{fig:sfig1}
\end{subfigure}%
\begin{subfigure}{.33\textwidth}  \includegraphics[width=\linewidth]{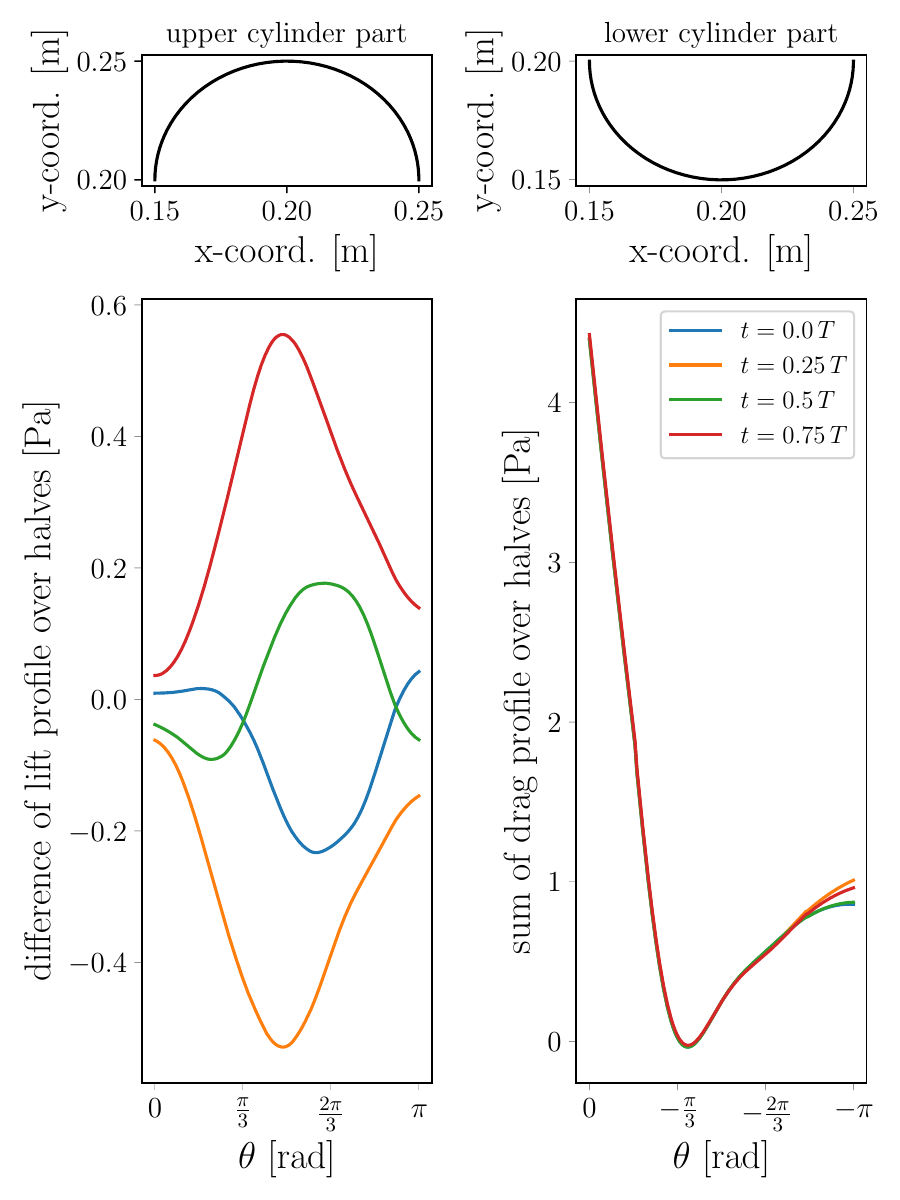}
  \captionsetup{width=\linewidth}
  \caption{Pointwise sum over both halves}
  \label{fig:sfig2}
\end{subfigure}
\caption{Traction profiles throughout the time period $T$ of the time evolution of the periodic solution of the Sch\"{a}fer--Turek benchmark at $\mathrm{Re}=100$}
\label{fig:time_pointwise_traction}
\end{figure}
Hence, it is natural to investigate this attractor more closely. A movie showing the simulation of one time period (movie~1) is available as supplementary material accompanying this paper. We examine one period of the pointwise traction of the attractor at $\mathrm{Re} = 100$, as this corresponds to the original setting of the evolutionary Sch\"{a}fer--Turek benchmark; see \cref{fig:time_pointwise_traction}. The drag profile exhibits noticeable oscillations in the region of the vortex wake; however, their influence on the total drag is almost negligible. In contrast, the lift profile oscillates on each side of the cylinder in opposite directions, resulting in a large temporal variation in the total lift. Interestingly, the pointwise lift difference between the upper and lower halves of the cylinder approximately follows a superposition of two modes: $\sin(\theta) \cos(f t)$ and $\sin(\theta) \cos(f t + \phi)$. Here, $\theta$ is the angular parameter along the cylinder boundary, and the attractor oscillates with frequency $f = 3.01$ Hz ($\mathrm{St} = 0.301$), matching the observation by \citet{Schafer1996}.

\begin{figure}
    \centering
    \begin{subfigure}{0.49\textwidth}
        \includegraphics[width=\linewidth]{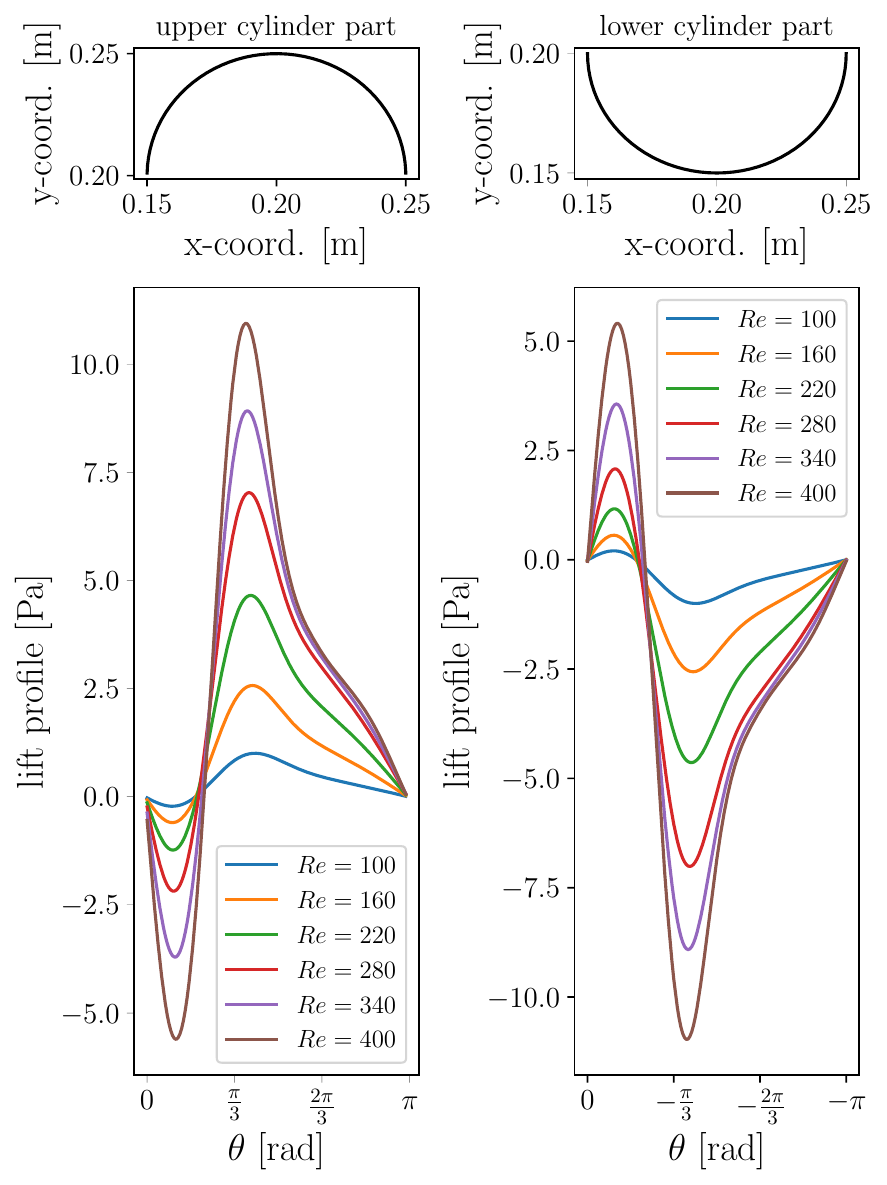}
        \caption{Pointwise lift profile}
    \end{subfigure}
    \begin{subfigure}{0.49\textwidth}
        \includegraphics[width=\linewidth]{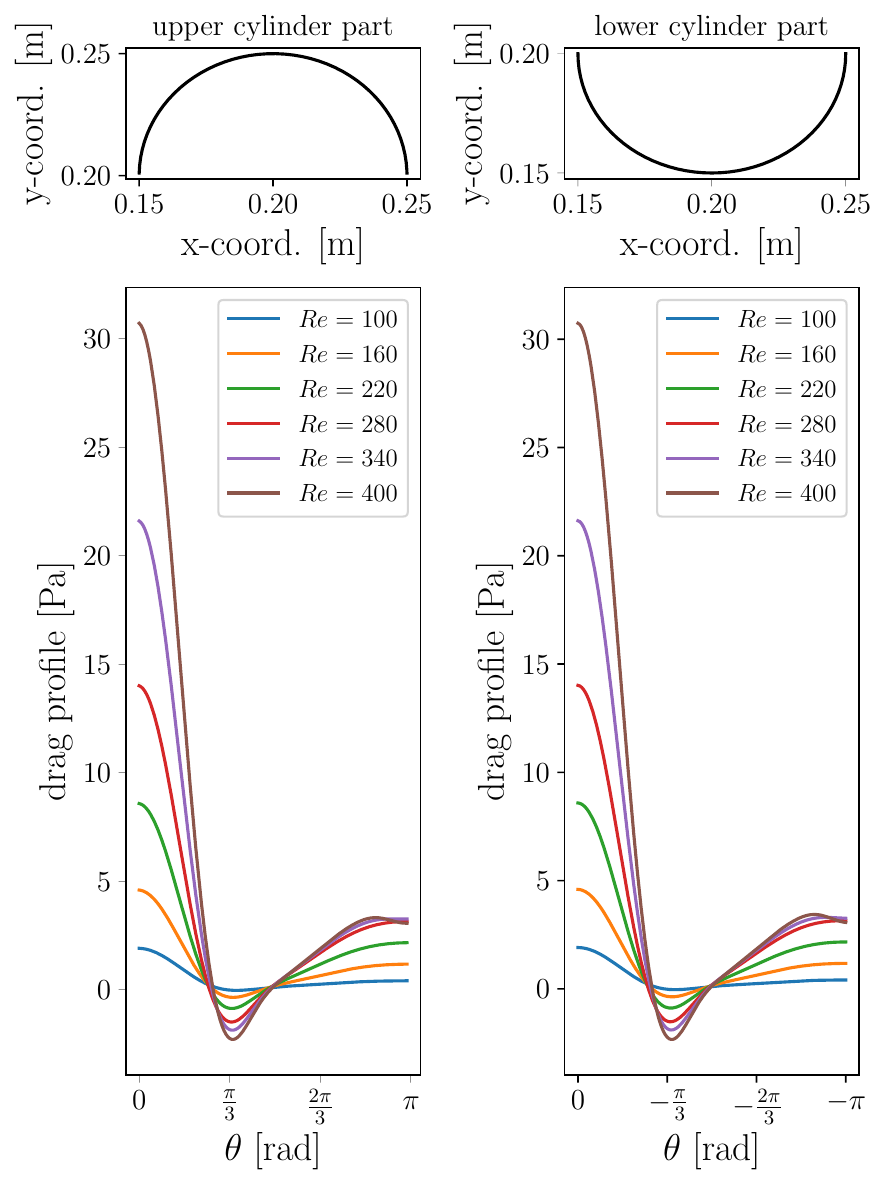}
        \caption{Pointwise drag profile}
    \end{subfigure}
\caption{Traction profiles across the range of Reynolds numbers where the long-time solution differs from steady solution and transition to the flow regime with multiple steady solutions occurs}
\label{fig:profiles_bifurcation}
\end{figure}
While time-dependent pointwise traction offers rich insight into the attractor; however, its analysis across a wide range of Reynolds numbers is beyond the scope of the present work. We therefore return to steady-state traction profiles as a more tractable tool for analysis. \Cref{fig:profiles_bifurcation} presents a selection of steady lift profiles for $\mathrm{Re} \in (100, 400)$. As shown later, a turning point in this range is linked to the bifurcations discussed in the next section.

\begin{figure}
\centering

\begin{minipage}[t]{0.43\textwidth}
    \vspace{0pt}
    \centering
    \begin{subfigure}{\linewidth}
        \centering
        \includegraphics[width=\linewidth,trim={8mm 8mm 8mm 6mm},clip]{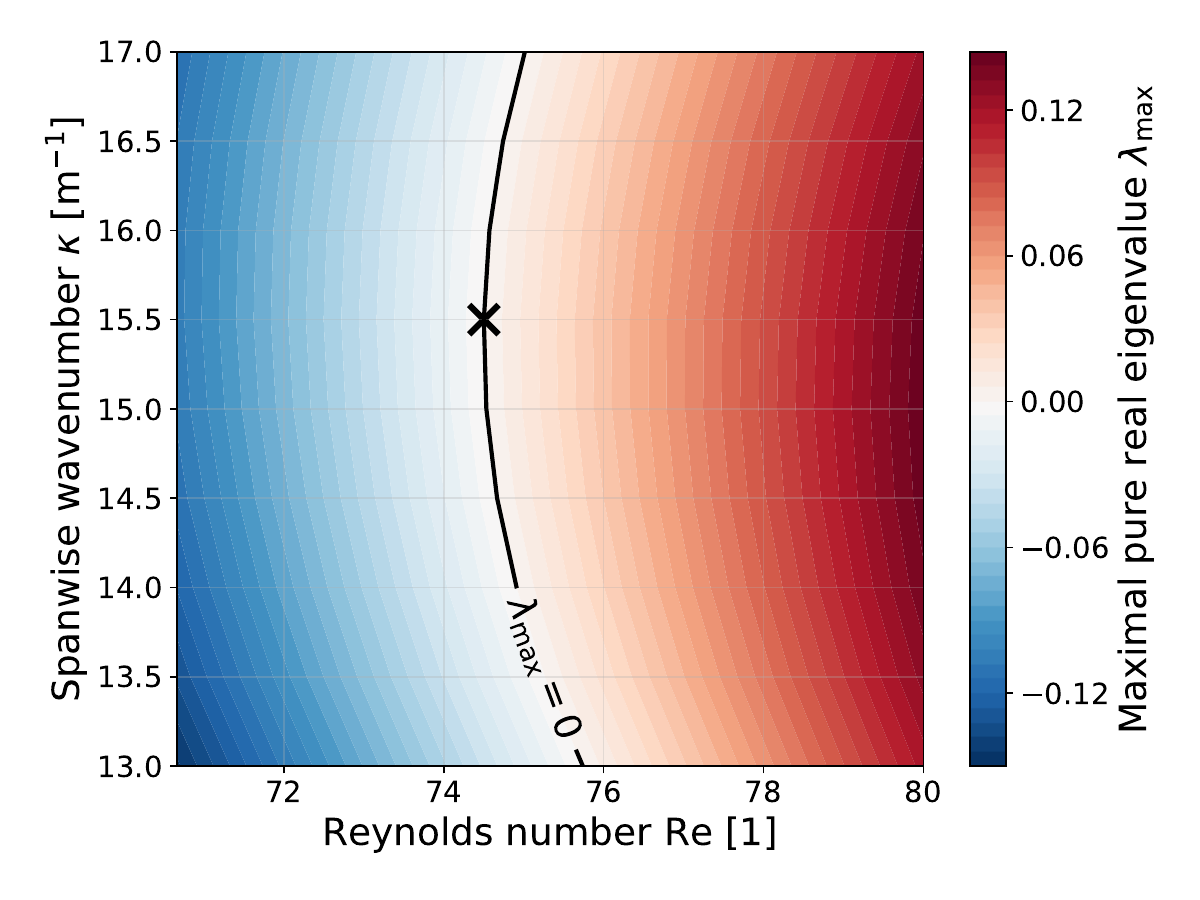}
        \caption{Maximal real eigenvalue}
    \end{subfigure}
\end{minipage}
\hfill
\begin{minipage}[t]{0.55\textwidth}
    \vspace{10pt}
    \centering
    
    \begin{subfigure}{\linewidth}
        \centering
        \includegraphics[width=\linewidth]{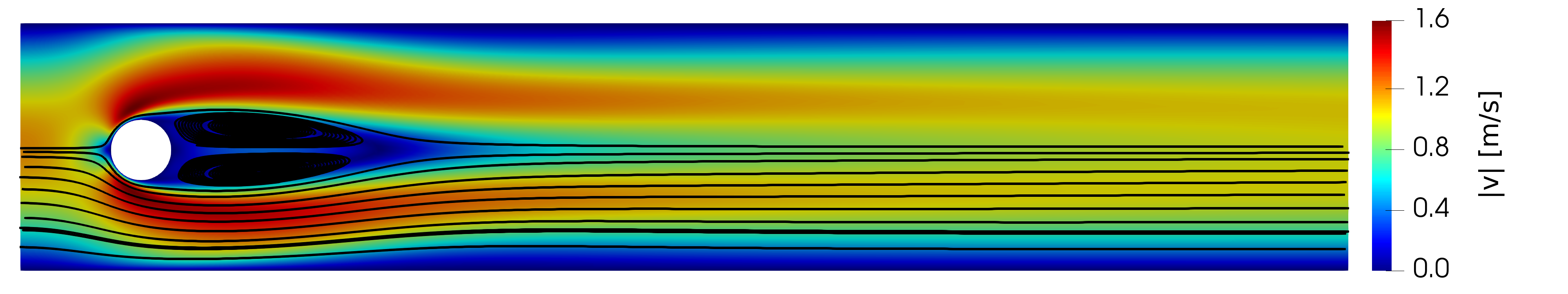}
        \caption{Slice at $z = L_z/2$}
        \label{fig:sfig3b:eigen1}
    \end{subfigure}
    
    \vspace{0.5em}
    
    \begin{subfigure}{\linewidth}
        \centering
        \includegraphics[width=\linewidth]{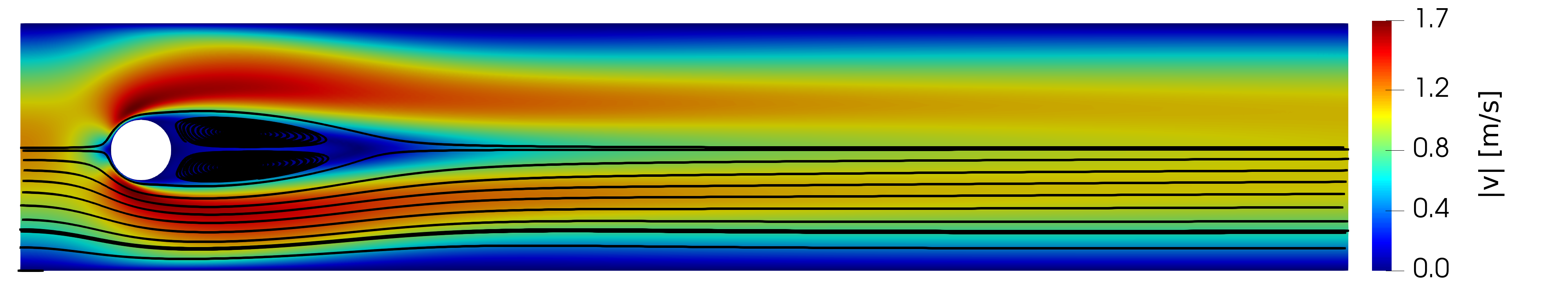}
        \caption{Slice at $z = L_z$}
        \label{fig:sfig3c:eigen2}
    \end{subfigure}

\end{minipage}

\caption{3D Mode~E identification. 
(a) Value of a maximum pure real eigenvalue in the $(\mathrm{Re},k)$ plane, and its zero-crossing. (b–c) Eigenmode reconstruction for $\mathrm{Re} = 75$ with small perturbation value $\varepsilon$ chosen for visualization, shown at two spanwise locations over one period~$L_z$. Mode~E manifests with periodic variation i.a., in recirculation region length.}
\label{fig:modeE}
\end{figure}

Using three-dimensional linear stability analysis (3D LSA), we detected the onset of a genuine three-dimensional steady instability (Mode~E) at $\mathrm{Re} = 74.5$. We identified this condition in \cref{sec:branching} and reconstructed slices of the solution (base flow plus a small perturbation) at two spanwise locations; see \cref{fig:modeE}. Notably, this occurs in close proximity to the onset of wall-induced effects observed in the unsteady two-dimensional solution.

Furthermore, the naturally selected spanwise wavenumber $\kappa \approx 15.5\ \mathrm{m}^{-1}$ corresponds to spanwise period $L_z = 2\pi/\kappa \approx 0.41\,\mathrm{m}$, which matches the channel width $H=0.41\,\mathrm{m}$. This suggests that the dominant spanwise wavelength of the instability is strongly influenced by the wall confinement.

\subsection{\texorpdfstring{Multiple steady solutions ($\mathrm{Re} > 315$)}{Multiple steady solutions (Re > 315)}}\label{sec:multiplicity}
\begin{figure}
  \begin{subfigure}{\textwidth}
    \centering
    \includegraphics[width=.9\linewidth]{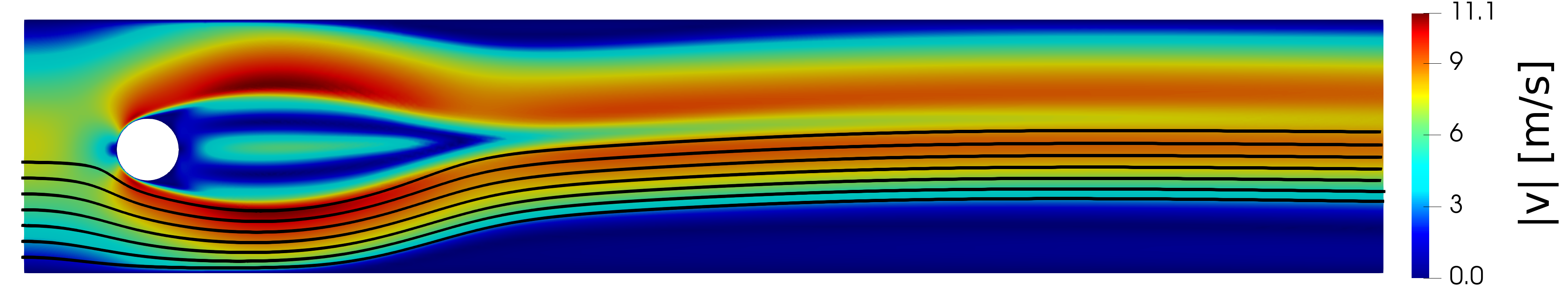}
    \caption{Baseline branch}
    \label{fig:sfig3a:flow_branch0}
  \end{subfigure}
  \begin{subfigure}{\textwidth}
    \centering
    \includegraphics[width=.9\linewidth]{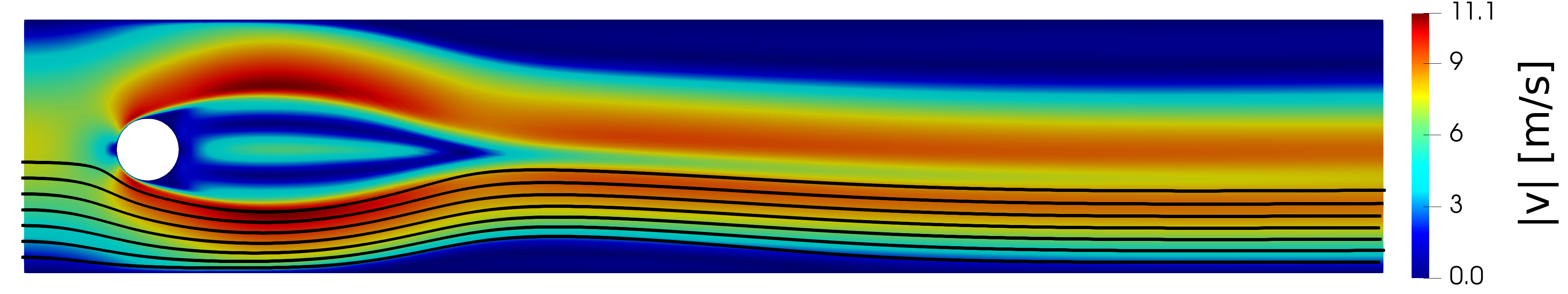}
    \caption{Branch A}
    \label{fig:sfig3a:flow_branch2}
  \end{subfigure}
  \begin{subfigure}{\textwidth}
    \centering
    \includegraphics[width=.9\linewidth]{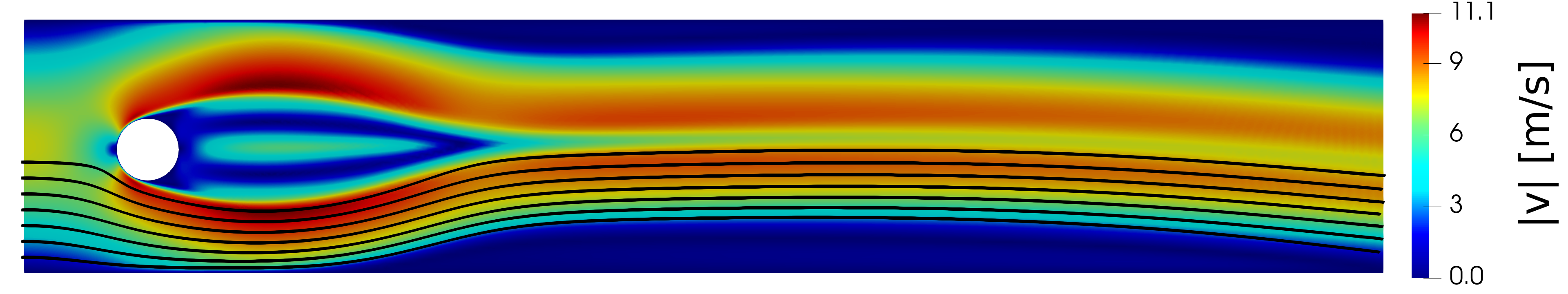}
    \caption{Branch B}
    \label{fig:sfig3b:flow_branch4}
  \end{subfigure}
  \caption{Plot of flow pattern of multiple steady states at $\mathrm{Re}=500$. By \emph{baseline} branch we denote the one that was found by plain continuation in $\mathrm{Re}$ (without using deflation).}
  \label{fig:flow_branches}
\end{figure}
\begin{figure}
    \centering
    \includegraphics[width=0.95\linewidth,trim={10mm 10mm 6mm 6mm},clip]{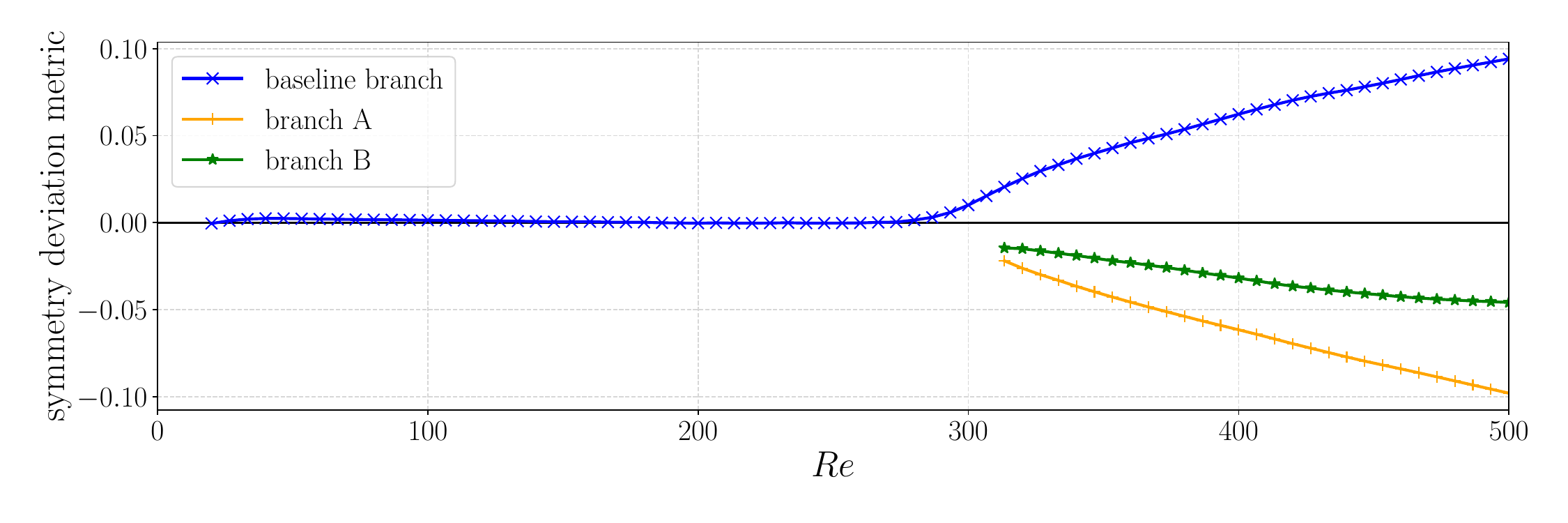}
    \caption{Bifurcation diagram based on (signed) symmetry deviation metric, defined as $\int_\Omega (\mathbf{v} - \mathbf{v}_{\text{sym}}) \cdot \mathbf{e}_y$, where $\mathbf{v}_{\text{sym}}$ is the velocity field reflected across the horizontal axis $y = 0.2$, and $\mathbf{e}_y = (0, 1)^T$}
    \label{fig:bifurcation_diagram}
\end{figure}

As we continue to increase the Reynolds number beyond $\mathrm{Re} > 315$, we observe a significant change in the structure of the steady flow field. Although the geometry lacks exact symmetry, the observed steady‑state structure is consistent with an imperfect (unfolded) pitchfork bifurcation associated with symmetry breaking in the symmetric limit. In \cref{ap:C}, we also present the case of a perfectly symmetric geometry, which yields the same qualitative behavior. At $\mathrm{Re} = 315$, using the deflation, we find multiple steady-state solutions, which we refer to as the baseline, A, and B branches. Each exhibits distinct flow trajectories downstream and a similar vortex structure (see \cref{fig:flow_branches}). Notably, the baseline solution at this point becomes markedly asymmetric. This marks a bifurcation that closely resembles a pitchfork type, as illustrated in the bifurcation diagram in \cref{fig:bifurcation_diagram}. The diagram plots a~signed symmetry deviation metric, defined as $\int_\Omega (\mathbf{v} - \mathbf{v}_{\text{sym}}) \cdot \mathbf{e}_y$, where $\mathbf{v}_{\text{sym}}$ is the velocity field reflected across the horizontal axis $y = 0.2$, and $\mathbf{e}_y = (0, 1)^T$ (extended by zero). Beyond the bifurcation Reynolds number, the streamlines of the steady solutions remain largely unchanged with further increases in $\mathrm{Re}$. A movie showing the complete steady continuation (movie~2) is available as supplementary material accompanying this paper.

This bifurcation is also reflected in the steady lift traction profile of the baseline branch, where a~turning point $(\theta = 45^\circ, \mathrm{Re} = 315)$ is observed, see \cref{fig:profiles_curves}. This turning point is observable in both pointwise lift and drag, and as the only one, in multiple places on the cylinder. For particular traction profiles in the range $\mathrm{Re} \in (100,400)$; see \cref{fig:profiles_bifurcation}. We pre-expose that this turning point also reflects transition in the time-periodic attractor, but more on that later. Notably, the location of this turning point on the cylinder is once again included in the front face of the cylinder, showing that the emerging dynamics have a detectable traction signature on the upstream face of the cylinder.

\begin{figure}
    \centering
    \begin{subfigure}{0.24\textwidth}
        \includegraphics[width=\linewidth]{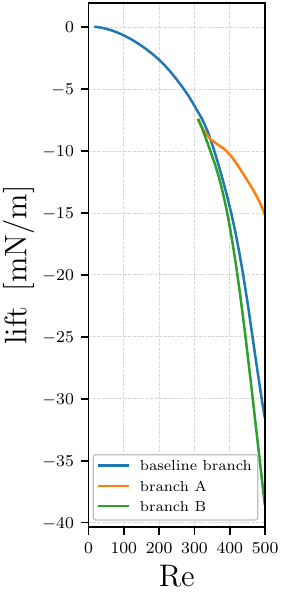}
        \caption{Lift}
        \label{fig:sfig1:bifurcation_lift}
    \end{subfigure}
    \begin{subfigure}{0.24\textwidth}
        \includegraphics[width=\linewidth]{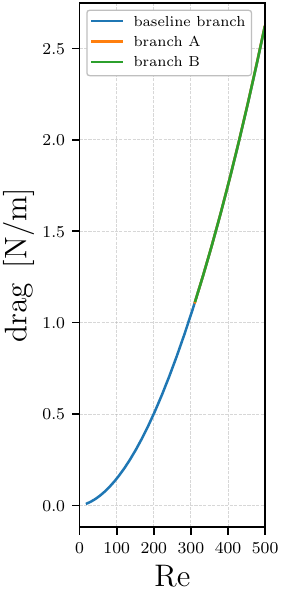}
        \caption{Drag}
        \label{fig:sfig1:bifurcation_drag}
    \end{subfigure}
    \begin{subfigure}{0.24\textwidth}
        \includegraphics[width=\linewidth]{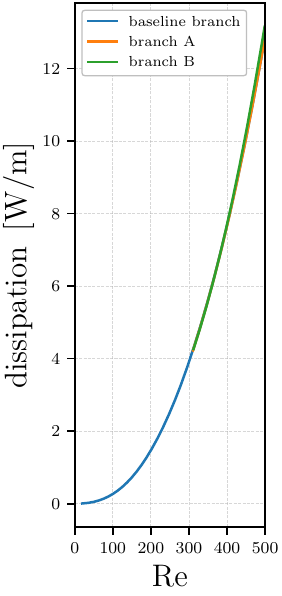}
        \captionsetup{width=\linewidth}
        \caption{Dissipation}
        \label{fig:sfig1:bifurcation_diss}
    \end{subfigure}
    \begin{subfigure}{0.24\textwidth}
        \includegraphics[width=\linewidth]{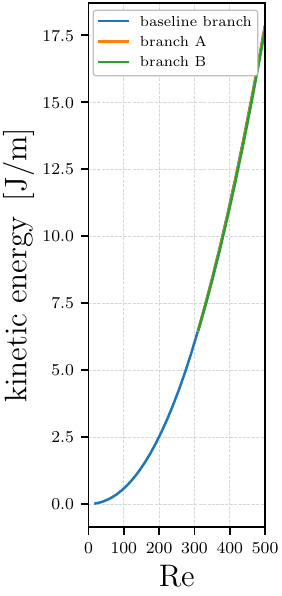}
        \captionsetup{width=\linewidth}
        \caption{Kinetic energy}
        \label{fig:sfig1:bifurcation_sqL2}
    \end{subfigure}
\caption{Sch\"{a}fer--Turek benchmark bifurcation point viewed by different quantities: Drag, lift, dissipation $\int_\Omega 2\mu|\nabla_\text{sym}\mathbf{v}|^2$, and kinetic energy $\int_\Omega\rho|\mathbf{v}|^2/2$}
\end{figure}

Again, the best physical quantity to distinguish between different solutions seems to be the total lift, see \cref{fig:sfig1:bifurcation_lift}. In other quantities, such as the total drag, see \cref{fig:sfig1:bifurcation_drag}, the dissipation $\int_\Omega 2\mu|\nabla_\text{sym}\mathbf{v}|^2$, see \cref{fig:sfig1:bifurcation_diss}, or the kinetic energy $\int_\Omega\rho|\mathbf{v}|^2/2$, see \cref{fig:sfig1:bifurcation_sqL2}, the branches basically coincide.

\begin{figure}
    \centering
    \begin{subfigure}{0.49\textwidth}
        \includegraphics[width=\linewidth]{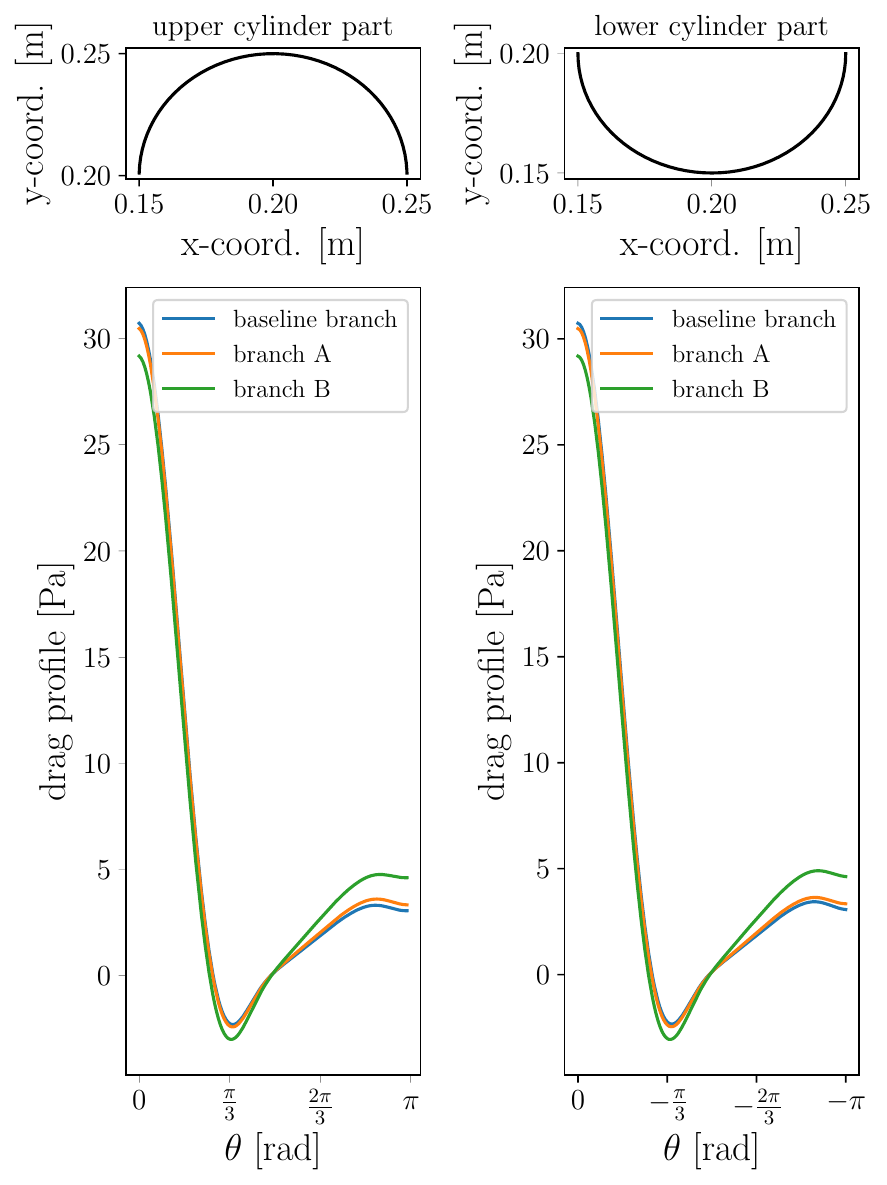}
        \caption{Drag}
        \label{fig:sfig1:profiles_3branch_drag_400}
    \end{subfigure}
    \begin{subfigure}{0.49\textwidth}
        \includegraphics[width=\linewidth]{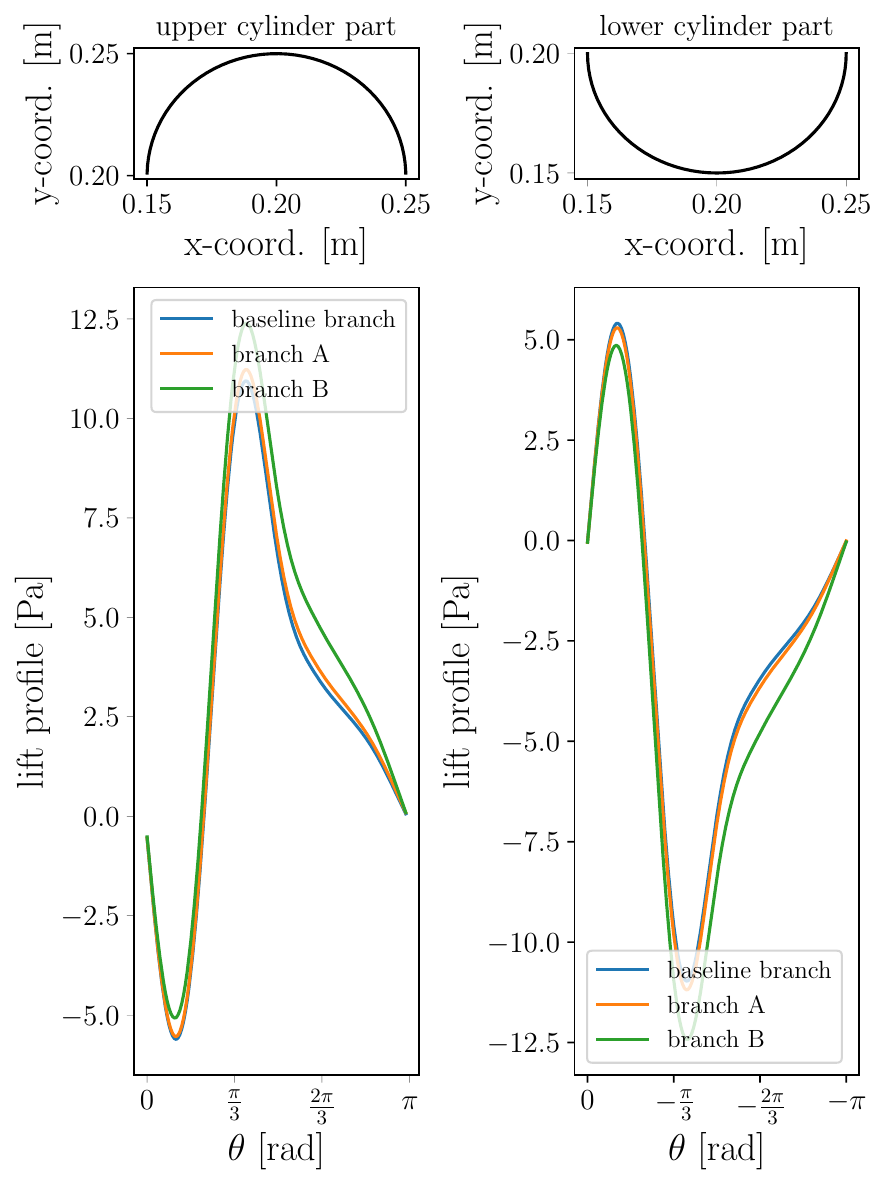}
        \caption{Lift}
    \end{subfigure}
\caption{Comparison of drag and lift profiles for different branches of the multiple steady solution at $\mathrm{Re} = 400$}
\label{fig:profiles_3branch}
\end{figure}

To investigate why the drag is comparable across different branches while the lift differs more significantly, we compare pointwise traction profiles at fixed $\mathrm{Re}$, see the pointwise drag and lift in \cref{fig:profiles_3branch} and pointwise. The similarity of the drag profiles across branches explains the close agreement in the drag values on the front face of the cylinder. In contrast, the lift is obtained from a near-cancellation of contributions from the two halves of the cylinder, where the pointwise lift has comparable magnitude but opposite sign. As a result, even small differences in the spatial structure of the traction lead to pronounced differences in the net lift. This highlights the strong sensitivity of the lift to asymmetrical effects, in contrast to the more robust drag response.

\begin{figure}
    \centering
    \includegraphics[width=0.95\linewidth,trim={3mm 3mm 2.5mm 3mm},clip]{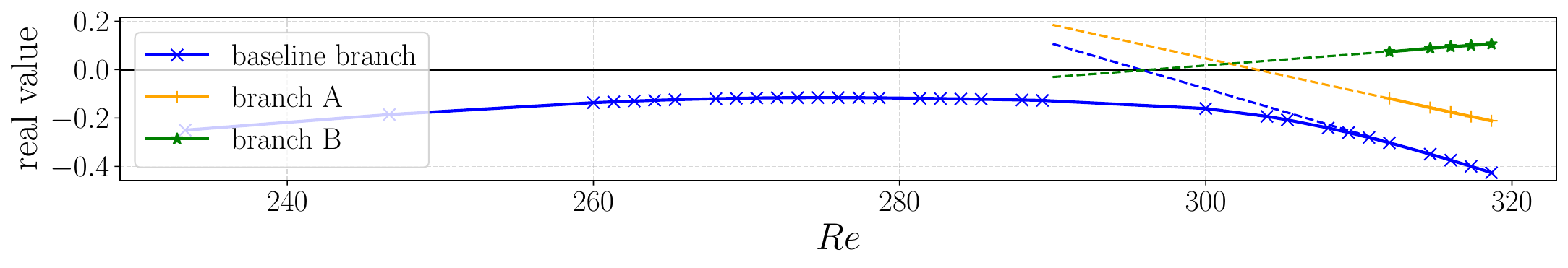}
    \caption{Evolution of the most right-hand-side purely real eigenvalue in the vicinity of the bifurcation point}
    \label{fig:bifurcation_eigenvalue}
\end{figure}

The identified bifurcation point is consistent with the spectrum of the LSA operator. In the theory of pseudospectra, see \cite{Gerecht2012}, it is shown that, beyond the first Hopf bifurcation induced by a pair of purely imaginary eigenvalues, additional real eigenvalues may play a role in subsequent steady-state bifurcations. This is reflected in our results: As $\mathrm{Re}$ approaches the bifurcation point, a real negative eigenvalue moves toward the imaginary axis. At the bifurcation point, it approaches zero real part, coinciding with the emergence of additional solution branches. For increasing $\mathrm{Re}$, this eigenvalue returns to the stable (negative) side on the baseline branch and on branch A, whereas on branch B it continues to drift toward positive real values, see \cref{fig:bifurcation_eigenvalue}. Note that we might not get an exact zero eigenvalue because of the discrepancy between the infinite-dimensional operator and its finite-dimensional approximation.

\begin{figure}
    \centering
    \includegraphics[width=0.9\linewidth]{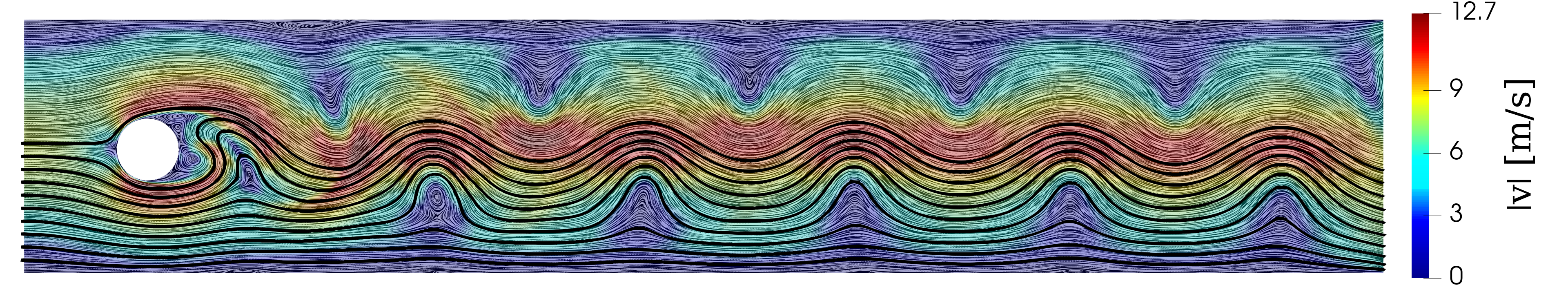}
    \caption{Snapshot of the time-periodic attractor corresponding to decaying vortex street at $\mathrm{Re} = 500$, shown using line integral convolution. Larger wall vortices are advected along the top wall. In the main flow, after one shedding period, the Kármán vortices in the street still fully decay, and the flow returns to the pattern observed at lower $\mathrm{Re}$. This underlines the fact that wall influence dominates the overall behavior.}
    \label{fig:snapshot_almost_turbulent}
\end{figure}
In addition to the imperfect symmetry breaking observed in the steady solutions at the critical Reynolds number $ \mathrm{Re} = 315$ ($\mathrm{St} = 0.334$), a pronounced change is also observed in the structure of the time‑periodic attractor as the Reynolds number is further increased. Although the underlying geometry lacks exact symmetry, the time‑periodic flow initially for lower $\mathrm{Re}$ exhibits an effectively symmetric vortex‑shedding pattern downstream of the cylinder, consistent with a regime in which viscous effects remain significant. At higher Reynolds numbers, however, this effective symmetry progressively deteriorates, indicating a shift toward inertia‑dominated dynamics in the wake. In particular, the decaying vortex street develops a marked asymmetry between vortices advected along the upper and lower channel walls, as illustrated in \cref{fig:snapshot_almost_turbulent} for $ \mathrm{Re} = 500 $.

\begin{figure}
    \centering
    \includegraphics[width=0.7\linewidth]{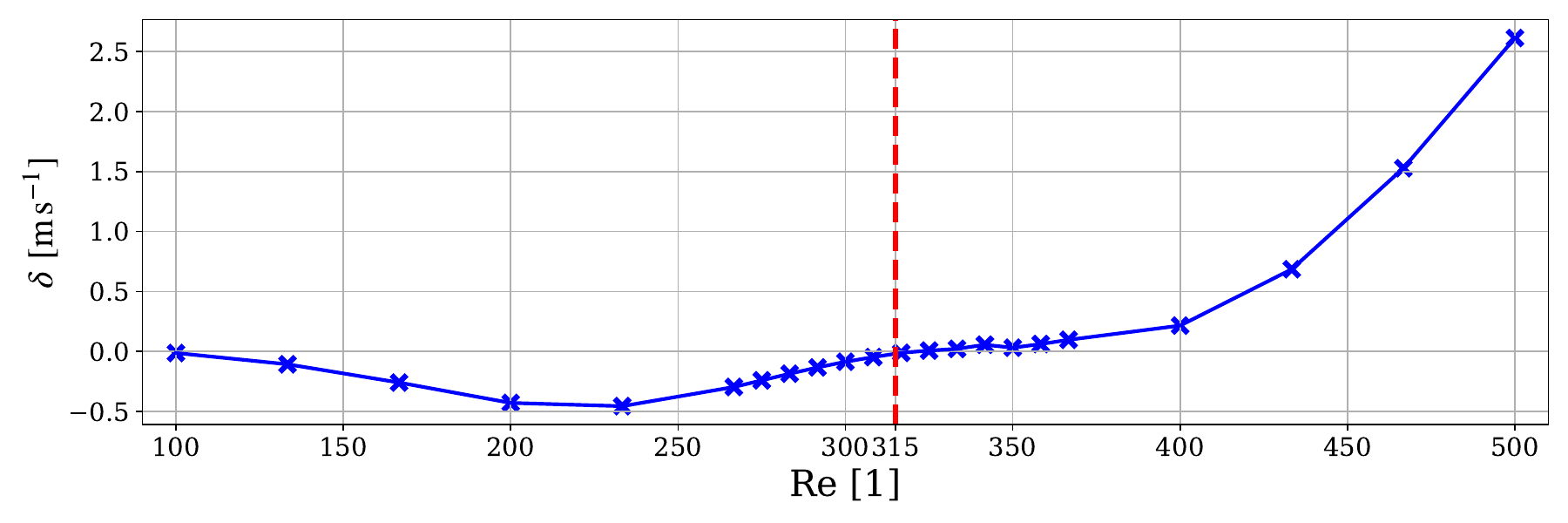}
    \caption{Time-averaged wall vorticity imbalance $\delta$ as a function of the Reynolds number. A zero crossing near $\mathrm{Re} = 315$ is followed by rapid growth of the signed imbalance, accompanying the development of pronounced asymmetry toward the upper wall.}
    \label{fig:vorticity_plot}
\end{figure}

To quantify the emergence of pronounced asymmetry in the flow response, we compute
a scalar diagnostic based on the time‑averaged wall vorticity imbalance,
defined as
\begin{equation*}
\delta = \frac{1}{T} \int_{\tau}^{\tau+T} \left( \int_{\text{top} \cup \text{bot}} (\partial_x v_y - \partial_y v_x) \,\mathrm{d}s \right) \,\mathrm{d}t,
\end{equation*}
where the integration is performed along both channel walls and accounts for the opposite signs of vorticity (representing clockwise and counterclockwise rotation). Although the geometry itself is weakly asymmetric, viscous effects at
low Reynolds numbers suppress the dynamical amplification of this imperfection,
resulting in an effectively symmetric flow response. The signed quantity $\delta$ provides a scalar description of the imbalance between the wall-vorticity contributions. Notably, this diagnostic exhibits a zero crossing near $\mathrm{Re} = 315$ with rapid growth afterwards,
accompanying the development of pronounced asymmetry toward the upper wall; see \cref{fig:vorticity_plot}.
The rapid growth of $\delta$ beyond this point indicates the transition, consistent with the observed steady‑state behavior.

However, this transition does not immediately produce the Kármán vortex street. Instead of being advected along the centerline, the vortical structures are transported along the channel walls. Wake structures along the centerline decay after approximately one shedding period, and the wake retains a qualitative similarity to the flow patterns observed at lower Reynolds numbers. This behavior is clearly illustrated in \cref{fig:snapshot_almost_turbulent} at $\mathrm{Re} = 500$, where the vortex street remains decaying. 

Finally, beyond the $\mathrm{Re} = 315$, in all our simulations, regardless of initial condition (steady states, zero field, or ramped inflow), solutions consistently converge to the same non‑symmetric periodic attractor. This robust selection indicates that the long‑time dynamics strongly favor a single asymmetric state. While the geometry contains a small inherent asymmetry, and such biases may influence branch selection, the present results only demonstrate the observed consistency of the selected attractor. Instead of allowing the spontaneous emergence of twin symmetry-related attractors (as would occur via a pitchfork bifurcation in a perfectly symmetric setup), the system displays an unfolded bifurcation structure, with the dynamics biased toward a single outcome: The attractor aligned with the side of the displaced obstacle. Preliminary results in the symmetric configuration, where a pitchfork bifurcation is confirmed in the steady-state branch at a similar critical Reynolds number, suggest that, in principle, the system could support both attractors. But in practice, even small imperfections, such as geometrical or numerical, tend to select one of them. Future work with strictly enforced symmetry or symmetry-breaking perturbations could help clarify the existence and structure of these twin attractors and the role of basin geometry in their selection.

\subsection{Global visualization of turning points} \label{sec:global_turning_points}
\begin{figure}
    \centering
    \begin{subfigure}{0.440\textwidth}
        \includegraphics[width=\linewidth]{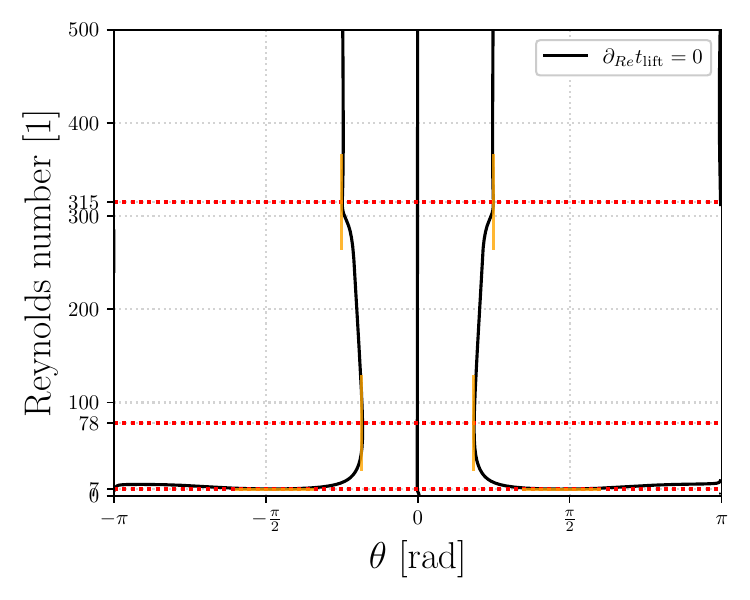}
        \captionsetup{width=0.9\linewidth}
    \caption{Pointwise lift: $\partial_{\mathrm{Re}} t_\text{lift}(\theta, \mathrm{Re}) = 0$ shown~in black}
    \label{fig:profiles_curves_lift}
    \end{subfigure}
    \begin{subfigure}{0.440\textwidth}
        \includegraphics[width=\linewidth]{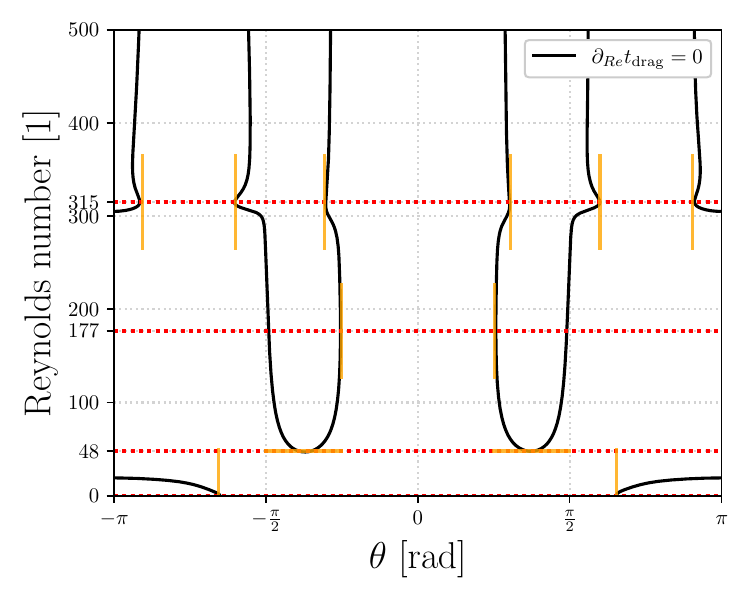}
    \captionsetup{width=0.9\linewidth}
    \caption{Pointwise drag: $\partial_{\mathrm{Re}} t_\text{drag}(\theta, \mathrm{Re}) = 0$ shown in black}
    \label{fig:profiles_curves_drag}
    \end{subfigure}
\caption{Implicit curves defined by $\partial_{\mathrm{Re}} t_{\text{drag,lift}}(\theta, \mathrm{Re}) = 0$ (black) along the cylinder boundary $\theta \in (-\pi, \pi)$, computed from the traction profiles of steady solutions of the baseline branch over the interval $\mathrm{Re} \in (0, 500)$. The folds of the black curve, highlighted by short orange tangent lines, correspond to turning points, whose positions are marked by red horizontal lines at $\mathrm{Re} = {0, 7, 48, 78, 177, 315}$.}
\label{fig:profiles_curves}
\end{figure}
To consolidate the local observations from the previous subsections, we now present a~global view of the critical points in the steady-state traction profile. \Cref{fig:profiles_curves} shows in black the implicit curves defined by $\partial_{\mathrm{Re}} t_{\text{drag/lift}}(\theta, \mathrm{Re}) = 0$ over the full parameter space $\theta \in (-\pi, \pi)$, $\mathrm{Re} \in (0, 500)$.

The critical Reynolds numbers identified a posteriori from the flow simulations across \crefrange{sec:low_re}{sec:multiplicity} match the Reynolds values at visible folds in these curves. We refer to these as turning points, as introduced in the introductory part of \cref{sec:res_intro}. This connection also reveals where each transition leaves its footprint on the obstacle boundary, typically on the front face ($|\theta|< \pi/2$). 

In our parametrization, $\mathrm{Re}=0$ corresponds to $V=0$ (no flow). For $\mathrm{Re}>0$, the flow enters a steady, vortex-free (creeping) regime, with inertia gradually becoming relevant as $\mathrm{Re}$ increases. This transition is also reflected in the drag by a turning point near $\mathrm{Re}=0$.

Next, the turning point in the lift at $\mathrm{Re} = 7$ is not sharply identifiable, as the zero-derivative curve continues almost horizontally toward the rear of the cylinder. We associate this behavior with the onset of recirculation regions, which itself is gradual: The transition from their initial appearance to a~clearly developed pair spans a~range of~$\mathrm{Re}$; see \cref{fig:turek_vortex_reynolds}.

Finally, we note that the drag turning point at $\mathrm{Re} = (177 \pm 60)$ lies in the range where a~secondary unsteady three-dimensional instability of the circular cylinder wake (Mode~A) has been reported in the literature, e.g., \citep*{Williamson1996, thompson2001physical,jiang-2016} and the references therein. We emphasize that this connection is only suggestive. The relatively large uncertainty arises from both the flatness of the turning-point curve in this region and the finite resolution in~$\theta$ (here $\Delta\theta = 0.01\,\text{rad}$) and~$\mathrm{Re}$. The estimate is obtained via a sensitivity analysis based on a second-order Taylor expansion of the spline fit. Although the present analysis is strictly two-dimensional, it is conceivable that weak signatures of three-dimensional instabilities may be reflected in two-dimensional observables, as, for example, in the steady spanwise-periodic bifurcation identified here (Mode~E).

Among all observed features, the most pronounced transition occurs at $\mathrm{Re} \approx 315$, where multiple turning points are present, notably in both drag and lift. In contrast to the broad and uncertain turning point at $\mathrm{Re} = (177 \pm 60)$, the turning point structure near $|\theta| \approx 1.9$ exhibits significantly smaller uncertainty, yielding an estimate of $\mathrm{Re} = (315 \pm 2.5)$.

For comparison with the other fold-type feature, we now provide an uncertainty estimate using the same methodology. In the case of the first Hopf bifurcation, the local description of the zero-derivative curve takes the form $\mathrm{Re}(\theta)$; thus, the primary uncertainty is associated with the critical position $\theta$. By propagating this uncertainty, we obtain an estimate of $\mathrm{Re} = (48 \pm 0.5)$ for the onset of the Hopf bifurcation, which is clearly dominated by the grid spacing in $\mathrm{Re}$.

In summary, with this interpretation, we are able to classify all turning points in the pointwise traction response, except for one observed in the drag at $\mathrm{Re}=20$ at the trailing edge ($\theta = \pi$).

We now propose a unified interpretation and classification of the observed turning points. Empirically, the folds of the zero-derivative traction curves, see \eqref{eq:turning_points}, resemble degeneracies known from bifurcation theory, albeit for a derived observable rather than the solution map itself. Two types of turning points can be distinguished. The first type corresponds to a simple critical point, characterized by $\partial_{\mathrm{Re}} t_c = 0$ together with $\partial_{\theta}\partial_{\mathrm{Re}} t_c = 0$, and is typically associated with the first appearance of a new flow feature. The second type corresponds to a higher-order degeneracy, where in addition $\partial_{\mathrm{Re}}^2 t_c = 0$, and it appears to be linked to a reorganization of already established flow structures. This classification is consistent with the sequence of transitions identified above.

Furthermore, the traction components exhibit a clear separation in their sensitivity to flow features. The drag primarily reflects (overall) symmetric, streamwise modifications of the flow and captures global force redistribution, while the lift is more sensitive to asymmetric flow structures and symmetry-breaking effects. Accordingly, transitions associated with symmetry breaking tend to manifest more prominently in the lift, whereas changes in the overall flow organization are more clearly reflected in the drag. Cases where both components are involved indicate increased dynamical complexity and possible interaction of multiple flow mechanisms.

\section{Conclusion}
\begin{figure}
  \centering
\begin{tikzpicture}[scale=5.5]
\newcommand{\vortex}[3]{%
  \begin{scope}
    \def\vortexCenterX{#1}
    \def\vortexCenterY{#2}
    \def\vortexScale{#3}

    \def\vortexA{0.175/8.0/\vortexScale}  
    \def\vortexB{0.175/4.0/\vortexScale} 

    \foreach \factor in {1, 0.8, 0.6, 0.4} {
      \draw[thick]
        ({\vortexCenterX - \factor*\vortexA}, {\vortexCenterY})
        arc[start angle=180, end angle=-180,
            x radius=\factor*\vortexA, y radius=\factor*\vortexB];
    }
  \end{scope}
}
\newcommand{\wallvortex}[3]{%
  \begin{scope}
    \def\vortexCenterX{#1}
    \def\vortexCenterY{#2}
    \def\vortexScale{#3}

    \def\vortexA{0.175/\vortexScale}  
    \def\vortexB{0.0875/\vortexScale} 

    \foreach \factor in {1, 0.8, 0.6, 0.4} {
      \draw[thick]
        ({\vortexCenterX - \factor*\vortexA}, {\vortexCenterY})
        arc[start angle=-180, end angle=180,
            x radius=\factor*\vortexA, y radius=\factor*\vortexB];
    }
  \end{scope}
}

  \def\length{2.2}
  \def\width{0.41}
  \def\scaledLength{2.2}
  \def\scaledWidth{\width}

  \def\originalCircleX{0.2}
  \def\originalCircleY{0.2}
  \def\originalRadius{0.05}

  \def\scaledCircleX{\originalCircleX}
  \def\scaledCircleY{\originalCircleY}
  \def\scaledRadius{\originalRadius}

  \draw (0,0) rectangle (\scaledLength,\scaledWidth);

  \def\ellipseA{0.25} 
  \def\ellipseB{0.05} 
  \def\ellipseCenterX{\scaledCircleX + \scaledRadius - 0.10 + \ellipseA/2}
  \def\ellipseCenterY{\scaledCircleY}
  \draw[fill=gray!40, draw=gray] (\ellipseCenterX,\ellipseCenterY) ellipse [x radius=\ellipseA/2, y radius=\ellipseB];

  \def\waveStartX{\ellipseCenterX + \ellipseA/2}
  \def\waveEndX{\scaledLength}
  \def\waveY{\scaledCircleY}
  \def\waveAmplitude{0.06}
  \def\wavePeriod{0.35}

  \draw[thick, blue, domain=\waveStartX:\waveEndX, samples=100, smooth, variable=\x]
    plot ({\x}, {\waveY + \waveAmplitude * sin(360*(\x - \waveStartX)/\wavePeriod)});

  \draw[dotted, thick] (0,0.2) -- (\scaledLength,0.2);

  \draw[fill=white, thick] (\scaledCircleX,\scaledCircleY) circle (\scaledRadius);

  \draw[|-|]
    (\scaledLength/2.0 + 0.04 - 0.35/2.0, \scaledWidth - 0.3) --
    ({\scaledLength/2.0 + 0.04 + 0.35/2.0}, \scaledWidth - 0.3)
    node[midway, below] {$\chi \approx \text{const.}$};

  \draw[<->]
    ({\scaledCircleX + 2.5*\scaledRadius + \ellipseA/2}, {\scaledCircleY - 0.105}) --
    ({\scaledCircleX + 2.5*\scaledRadius + \ellipseA/2}, {\scaledCircleY + 0.105});

  \draw[<->]
    ({\scaledCircleX - \scaledRadius - 0.015}, {\scaledCircleY - 0.035}) --
    ({\scaledCircleX - \scaledRadius - 0.015}, {\scaledCircleY + 0.035});

  \draw[<->]
    ({\scaledLength - 0.05}, {\scaledCircleY + 0.035}) --
    ({\scaledLength - 0.05}, {\scaledCircleY - 0.035});
  \node at ({\scaledLength - 0.41}, {0.035}) [left] {$\mathrm{Re} \geq 78$};

    \draw[|-<]
  ({\ellipseCenterX - \ellipseA/4}, {\ellipseCenterY - \ellipseB - 0.05}) --
  ({\ellipseCenterX + \ellipseA/2}, {\ellipseCenterY - \ellipseB - 0.05})
  node[midway, below] {$\mathrm{Re} \geq 48$};

    \def\eps{0.0075}

    \def\blueLineStartX{0}  
    \def\blueLineEndX{\scaledCircleX - \scaledRadius -\eps}  
    \def\blueLineY{\scaledCircleY + \eps}  

    \draw[thick, blue]
      (\blueLineStartX, \blueLineY) --
      (\blueLineEndX, \blueLineY);

    \draw[thick, blue]
      ({\scaledCircleX - \scaledRadius -\eps}, {\scaledCircleY + \eps})
      arc[start angle=180, end angle=90, radius=\scaledRadius];

    \wallvortex{\scaledLength/2.0 + 0.04 - 0.35 -0.35/2}{\scaledWidth-0.03}{4.5}
    \wallvortex{\scaledLength/2.0 + 0.04-0.35/2}{\scaledWidth-0.03}{4.5}
    \wallvortex{\scaledLength/2.0 + 0.04 + 0.35-0.35/2}{\scaledWidth-0.03}{4.5}
    \wallvortex{\scaledLength/2.0 + 0.04 + 0.35 + 0.35-0.35/2.0}{\scaledWidth-0.03}{4.5}
    \wallvortex{\scaledLength/2.0 + 0.04 + 0.35 +0.35}{+0.0161}{9.0}
    \wallvortex{\scaledLength/2.0 + 0.04-0.35}{+0.0161}{9.0}
    \wallvortex{\scaledLength/2.0 + 0.04 + 0.35-0.35}{+0.0161}{9.0}
    \wallvortex{\scaledLength/2.0 + 0.04 + 0.35 + 0.35-0.35}{+0.0161}{9.0}
    \vortex{\scaledLength/2.0 + 0.04 - 0.35}{\scaledWidth/2.0-0.015}{2}
    \vortex{\scaledLength/2.0 + 0.04 -0.35 - 0.35/2.0}{\scaledWidth/2.0+0.015}{1}
    \vortex{\scaledLength/2.0 + 0.04 - 0.35/2.0}{\scaledWidth/2.0+0.015}{4}
    \node at (1.875, \scaledWidth-0.08) [above] {$\mathrm{Re} \geq 315$};

\draw[|-|]
  ({-0.0}, -0.001) -- ({-0.0}, 0.2);
\node[rotate=90] at ({-0.05}, 0.1) {$0.2$};

\draw[|-|]
  ({-0.0}, 0.2) -- ({-0.0}, \scaledWidth+0.001);
\node[rotate=90] at ({-0.05}, {0.2 + (\scaledWidth - 0.2)/2}) {$0.21$};
\end{tikzpicture}
\caption{%
  Sch\"{a}fer--Turek benchmark. Sketch of the evolution of the observed unique
  time-periodic global-in-time attractor with increasing Reynolds number up to
  500. The arrows indicate trends with increasing Reynolds number. The onset
  Reynolds numbers of the key flow regimes correspond closely to characteristic
  changes in the pointwise traction profiles:
  (i) decaying vortex street at $\mathrm{Re} = 48$,
  (ii) saturation of wake width and onset of wall vortices at $\mathrm{Re} = 78$, and
  (iii) non-symmetric time-periodic flow with larger vortices advected along
  the top wall (and still with decaying vortices along the centerline) at $\mathrm{Re}
  = 315$. Here, $\chi$ denotes the streamwise wavelength associated with one shedding period. Its approximately unchanged value is used only as a qualitative description of the depicted flow pattern.
}
\label{fig:turek_unsteady_solution}
\end{figure}
We have presented a comprehensive numerical study of incompressible Navier--Stokes flow past a slightly off-centered cylinder in the Sch\"{a}fer--Turek benchmark configuration, exploring unsteady solutions for Reynolds numbers up to 500. The evolution of the unsteady long-time periodic velocity field with increasing Reynolds number is depicted in \cref{fig:turek_unsteady_solution}, which illustrates the emergence and progression of a unique time-periodic global-in-time attractor across the unsteady regime for all initial conditions tested. For Reynolds numbers up to $\mathrm{Re} \approx 7$, the flow remains vortex-free, and up to $\mathrm{Re} \approx 48$ it stays steady. At $\mathrm{Re} = 48$, the long-time solution becomes time-periodic, distinct from the corresponding steady solution (it becomes unstable), while retaining a spatiotemporal symmetry downstream with respect to the centerline. This marks the onset of a decaying vortex street. At $\mathrm{Re} \approx 78$, the appearance of wall-attached vortical structures is observed, possibly related to the saturation of the wake width across the channel. Around $\mathrm{Re} \approx 315$, the steady solution branch develops an imperfect (unfolded) symmetry-breaking bifurcation, and multiple (unstable) steady solutions coexist. At the same Reynolds number, the already time-periodic attractor changes character: In the downstream wake it develops into a pronounced asymmetric, decaying vortex street, with larger vortices advected along the upper channel wall.
Notably, across all unsteady regimes, we consistently observe convergence to a single periodic attractor from various initial conditions, e.g., multiple steady states.

\begin{figure}
    \centering
    \begin{tikzpicture}[
        every node/.style={align=center, font=\small},
        scale=1.13
    ]
        \def\yA{6}
        \def\yC{4.5}
        \def\yD{3}

        \draw[->] (0,\yA) -- (11.25,\yA) node[below] at (11.25, \yA-0.2) {\small$\mathrm{Re}$};
        \foreach \x/\val in {0/0, 1.5/7, 4.25/48, 7/315} {
            \draw[thick] (\x,\yA-0.1) -- (\x,\yA+0.1);
            \node[below] at (\x,\yA-0.2) {\val};
        }
        \node[above] at (0.75,\yA+0.15) {stationary,\\vortex-free};
        \node[above] at (2.895,\yA+0.15) {stationary,\\with vortex wake};
        \node[above] at (5.6125,\yA+0.15) {decaying\\vortex street};
        \node[above] at (9,\yA+0.1) {non-symmetric vortex street\\(also multiplicity of steady states)};

        \draw[->] (0,\yC) -- (11.25,\yC) node[below] at (11.25, \yC-0.2) {\small$\mathrm{Re}$};
        \foreach \x/\val in {0/0, 1.5/7, 4.25/48, 4.75/78, 6.0/177, 7/315} {
            \draw[thick] (\x,\yC-0.1) -- (\x,\yC+0.1);
            \node[below] at (\x,\yC-0.2) {\val};
        }
        \node[above] at (0.0,\yC+0.1) {drag};
        \node[above] at (1.5,\yC+0.15) {lift};
        \node[above] at (4.25,\yC+0.1) {drag};
        \node[above] at (4.75,\yC+0.15) {lift};
        \node[above] at (6.0,\yC+0.15) {drag};
        \node[above] at (7,\yC+0.1) {lift, drag};

        \draw[->] (0,\yD) -- (11.25,\yD) node[below] at (11.25, \yD-0.2) {\small$\mathrm{Re}$};
        \foreach \x/\val in {0/0, 4.25/48, 7/315} {
            \draw[thick] (\x,\yD-0.1) -- (\x,\yD+0.1);
            \node[below] at (\x,\yD-0.2) {\val};
        }
        \node[above] at (2.125,\yD+0.1) {linear\\stability};
        \node[above] at (4.25,\yD+0.1) {unstable complex\\eigenmode};
        \node[above] at (7,\yD+0.1) {purely real eigenvalue\\approaches zero};
    \end{tikzpicture}
    \caption{%
      The upper row illustrates the major flow behavior governed by the
      time-dependent Navier--Stokes equations in the Sch\"{a}fer--Turek
      benchmark as the Reynolds number increases. The middle row marks the
      major corresponding qualitative changes as well as minor transitions in the lift or drag profiles of the
      steady solution; mostly on the upstream face of the cylinder.
      The bottom row
      marks the important events happening in the LSA operator spectrum around
      the steady states. In our definition, $\mathrm{Re} = 0$ corresponds to zero velocity, implying no drag and marking the onset of drag for $\mathrm{Re} > 0$. At $\mathrm{Re} = 78$ the lift profile reflects a minor
      transition when the wake width of the decaying vortex street saturates and
      wall-attached vortices occur. Moreover, at similar $\mathrm{Re}$ the steady 3D instability (Mode~E) emerges. Mode A, a 3D unsteady instability, is reported in the literature to occur around $\mathrm{Re} = 177$.
    }
    \label{fig:final_result}
\end{figure}

A~central insight of our study is that the pointwise drag and lift profiles on
the cylinder surface offer a sensitive diagnostic of changes in the long-time flow response.
Non-monotonic features in these local force
distributions arise sharply at Reynolds numbers corresponding to qualitative
changes in long-term dynamics. These features correspond to folds
\eqref{eq:turning_points} of the implicit curve $\partial_{\mathrm{Re}}
t_{\text{c}}(\theta, \mathrm{Re}) = 0$, for a~traction component either
$t_\text{c}=t_\text{drag}$ or $t_\text{c}=t_\text{lift}$.
Interestingly, these changes are primarily located on the upstream face of the obstacle. Yet, the bifurcation mechanisms are known to be global, and the present observations are reported purely a posteriori. This correspondence is synthesized in \cref{fig:final_result}, which integrates flow regime observations, pointwise traction data, and LSA spectral information across the full Reynolds number range.
The most pronounced example occurs near $\mathrm{Re}=315$, where the steady problem exhibits several unstable solutions and the already periodic wake changes its character. In the same Reynolds-number range, several turning points appear in both the drag and lift profiles, making this transition one of the strongest traction signatures in the study.

The LSA spectrum provides complementary insight, with the appearance of an unstable mode coinciding with the primary Hopf bifurcation. While it does not reliably detect other types of attractor transitions, it robustly captures the initial global instability and the onset of oscillations in the vortex wake, see \cref{fig:sfig3:critical_re_eigenmode}, and aligns closely with this major transition. In particular, the detected eigenmode reconstructs the velocity field structure of the emerging attractor well. However, as expected, this capability deteriorates as the Reynolds number increases, and the steady state loses dynamical relevance (i.e., becomes unstable). On the other hand, the traction profiles are also computed from steady solutions that are unstable, yet their turning points still reflect changes in the observed time-dependent flow.

In summary, steady traction profiles computed along the steady continuation provide a compact diagnostic of changes in the flow response across Reynolds numbers. In the present benchmark, their structure corresponds a posteriori to several changes in the underlying dynamics, as classified here by detailed time-dependent computations. This behavior is further illustrated for additional geometries in \cref{ap:C}.

\section*{Data availability}
The supporting code to reproduce the results of this study is available
\citep*{zenodo-v1.1}.

\section*{Acknowledgment}
J.C., S.S., and K.T.\ have been supported by the ERC-CZ Grant LL2105 CONTACT. S.S.\ was further supported by the VR-Grant 2022-03862 of the Swedish Research Council. All authors thank for the support of the Charles University Research program No.\ UNCE/24/SCI/005. J.C.\ also thanks the Charles University Grant Agency for support, Grant No.\ 131124. S.S.\ and K.T.\ are members of the Nečas Center for Mathematical Modeling.

\appendix
\makeatletter%
\@addtoreset{figure}{section}%
\@addtoreset{table}{section}%
\makeatother%
\crefalias{section}{appendix}
\section{Approximation of normal derivative in the finite element method}
\label{ap:A}
For incompressible flow, the traction is essentially given by the normal derivative of the velocity field, or by a more involved expression that depends on it. In this section, we review literature related to the evaluation of the normal derivative using the finite element method, including results on superconvergence properties. Our focus is on corner domains, more precisely, on polytopal domains, that is, domains which can be partitioned into simplices.

Let $\Omega \subset \mathbb{R}^d$ be a bounded Lipschitz domain, and let $f\colon \Omega \to \mathbb{R}$. Consider the Dirichlet problem: Find $u\colon \Omega \to \mathbb{R}$ such that
\begin{subequations}
  \label{eq:laplace}
  \begin{alignat}{2}
    -\Delta u &= f
    \qquad && \text{in } \Omega, \\
    u &= 0
    \qquad && \text{on } \partial\Omega.
  \end{alignat}
\end{subequations}
Our goal is to study the normal derivative $\partial_\mathbf{n} u$. For any $v \in H^{3/2}(\Omega)$, the normal derivative $\partial_\mathbf{n} v$ belongs to $L^2(\partial\Omega)$ and is given by the normal trace of $\nabla v \in H^{1/2}(\Omega)$. However, for a general $v \in H^1(\Omega)$, the normal derivative $\partial_\mathbf{n} v$ is not defined in the classical sense. Nevertheless, for the weak solution $u \in H_0^1(\Omega)$ of~\eqref{eq:laplace}, the normal derivative can be defined in a weak (dual) sense:
\begin{equation}
  \label{eq:vnd}
  \left< \partial_\mathbf{n} u, v \right>_{\partial\Omega}
  \coloneqq \int_\Omega \nabla u \cdot \nabla v \, \mathrm{d}x - \left< f, v \right>,
  \qquad \text{for all } v \in H^1(\Omega).
\end{equation}
If the weak solution is sufficiently regular, i.e., $u \in H^{3/2}(\Omega)$, then the distributional normal derivative coincides with the normal trace of the gradient.

Let $\mathcal{T}_h$ be a regular quasi-uniform family of triangulations of $\Omega$ into simplices, with mesh size $h$. Let the polynomial degree $k \in \mathbb{N}$ be given, and define
\begin{align*}
  V^h
  &\coloneqq
  \{ v \in H^1(\Omega) \mid v|_K \in \mathcal{P}_k \text{ for each } K \in \mathcal{T}_h \}, \\
  V^h_0
  &\coloneqq
  V^h \cap H_0^1(\Omega),
\end{align*}
the space of continuous piecewise polynomials of degree at most $k$, and its subspace with zero trace. Let $u_h \in V^h_0$ denote the finite element solution of~\eqref{eq:laplace}, that is,
\begin{equation}
  \label{eq:fem}
  \int_\Omega \nabla u_h \cdot \nabla v_h \,\mathrm{d}x
  = \left< f, v_h \right>
  \qquad \text{for all } v_h \in V^h_0.
\end{equation}
The space $V^h$ is regular enough for the normal derivative $\partial_\mathbf{n} v_h = \mathbf{n} \cdot \nabla v_h$ to be defined in the usual sense, as the normal trace of the gradient, for any $v_h \in V^h$. On the other hand, one might be tempted to replace both occurrences of $u$ in~\eqref{eq:vnd} with $u_h$, but this would not yield $\partial_\mathbf{n} u_h$. Formula~\eqref{eq:vnd} assumes that $u \in H^1_0(\Omega)$ solves~\eqref{eq:laplace}, which is not the case for $u_h$.

Since $u_h \in V^h_0$ solves~\eqref{eq:fem}, a discrete duality formulation is available for the normal derivative. Consider the trace space
\begin{align*}
  W^h \coloneqq \operatorname{trace} V^h =
  \{ v \in H^{1/2}(\partial\Omega) \mid
     v|_f \in \mathcal{P}_k \text{ for each facet } f \in \mathcal{F}_h^\mathrm{ext} \},
\end{align*}
where $\mathcal{F}_h^\mathrm{ext}$ denotes the set of exterior facets in $\mathcal{T}_h$. We define the \emph{discrete variational normal derivative} (DVND) as $\partial_\mathbf{n}^h u_h \in W^h$ satisfying
\begin{equation}
  \label{eq:dvnd}
  \int_{\partial\Omega} \partial_\mathbf{n}^h u_h \, v_h \,\mathrm{d}S
  \coloneqq \int_\Omega \nabla u_h \cdot \nabla v_h \,\mathrm{d}x - \left< f, v_h \right>
  \qquad \text{for all } v_h \in V^h.
\end{equation}

We now have two distinct quantities:
the normal derivative $\partial_\mathbf{n} u_h$ and the discrete variational normal derivative $\partial_\mathbf{n}^h u_h$.

\begin{theorem*}[{\citet*[Corollary~5.3]{HorgerMelenkWohlmuth2013}}]
  Suppose that \eqref{eq:laplace} admits $H^{3/2+\varepsilon}$ regularity,
  i.e., there exist $\varepsilon>0$ and $c>0$ such that the solution operator
  $T\colon f\mapsto u$ for~\eqref{eq:laplace} satisfies
  \begin{equation*}
    \|Tf\|_{H^{3/2+\varepsilon}(\Omega)}
    \leq c\|f\|_{(H^{1/2-\varepsilon}(\Omega))'}
    \qquad \text{for all $f \in (H^{1/2-\varepsilon}(\Omega))'$.}
  \end{equation*}
  Then there exists $C>0$ such that
  \begin{equation*}
    \|\partial_\mathbf{n} u - \partial_\mathbf{n} u_h\|_{L^2(\partial\Omega)}
    \leq C (1+\delta_{k,1} |\log h|) \begin{cases}
      h^k \, \|u\|_{B^{k+3/2}_{2,1}(\Omega)}, & \\
      h^s \, \|u\|_{B^{s+3/2}_{2,\infty}(\Omega)}, & s\in(0,k), \\
      \|u\|_{B^{3/2}_{2,1}(\Omega)}. &
    \end{cases}
  \end{equation*}
  where $\delta_{k,1}$ denotes the Kronecker delta and $B^s_{p,q}(\Omega)$ are
  the Besov spaces.
\end{theorem*}
\begin{remark*}[Superconvergence of DVND]
  \citet*[Theorem~4.1]{ApelMateosPfeffererRosch2018}, in the context of optimal control, show that on certain superconvergence meshes, one can achieve an additional half-order of convergence for
  $\|\partial_\mathbf{n} u - \partial_\mathbf{n}^h u_h\|_{L^2(\partial\Omega)}$.
  Their analysis is restricted to polygonal domains ($d = 2$) and the lowest-order case ($k = 1$).
  On general quasi-uniform meshes $\mathcal{T}_h$, DVND $\partial_\mathbf{n}^h
  u_h$ exhibits the same convergence rate as the classical normal derivative
  $\partial_\mathbf{n} u_h$.

  However, on meshes where most adjacent triangle pairs form approximate
  parallelograms, a superconvergent rate, up to an additional half-order, can
  be achieved. Specifically, if $\mathcal{T}_h$ is $O(h^2)$-irregular (see
  \citep{ApelMateosPfeffererRosch2018,BankXu2003} for the precise definition),
  then there exists a constant $C > 0$ (depending on~$u$) such that
  \begin{equation*}
    \|\partial_\mathbf{n} u - \partial_\mathbf{n}^h u_h\|_{L^2(\partial\Omega)} \leq C h^{3/2},
  \end{equation*}
  if $u$ is smooth enough and the problem admits sufficient regularity.
\end{remark*}

The DVND exhibits superconvergence on the aforementioned highly regular meshes, but, to the best of our knowledge, the theory is currently available only for the case $d=2$, $k=1$. Nevertheless, numerical experiments show the extra half order also for $k=2,3,\ldots$

The convergence order can be further improved for both
$\|\partial_\mathbf{n} u - \partial_\mathbf{n} u_h\|_{L^2(\partial\Omega)}$
and
$\|\partial_\mathbf{n} u - \partial_\mathbf{n}^h u_h\|_{L^2(\partial\Omega)}$
up to $h^{k+1}$ by using mesh grading (\citet*{pfefferer-winkler-2019}, proved
therein for $d=2$ and $k=1$), or perhaps via adaptive mesh refinement.

Non-polygonal domains $\Omega$ further highlight the advantage of the DVND.
Suppose that $\mathcal{T}_h$ triangulates (possibly with curved elements)
a~domain $\Omega_h$ that approximates $\Omega$. Note that the definition of the
DVND in \eqref{eq:dvnd} does not explicitly involve the normal vector
$\mathbf{n}$. By contrast, when using classical directional derivatives one
must choose between several options,
\begin{equation*}
  \mathbf{n}_{\Omega_h} \cdot \nabla u_h,
  \qquad
  \widehat{\mathbf{n}_{\Omega}} \cdot \nabla u_h;
\end{equation*}
here $\mathbf{n}_{\Omega_h}$ is the exact normal of~$\Omega_h$ and
$\widehat{\mathbf{n}_{\Omega}}$ is a~certain (perhaps isoparametric framework)
mapping of~$\mathbf{n}_{\Omega}$, the exact normal of~$\Omega$, to~$\partial\Omega_h$.
Further options are possible, e.g., a~postprocessing (smoothing)
of~$\mathbf{n}_{\Omega_h}$.

On the other hand, to evaluate DVND $\partial_\mathbf{n}^h u_h$, one does not
explicitly need a~normal; see~\eqref{eq:dvnd}.
See \cref{ap:B} for numerical experiments in the context of incompressible flow.

\section{Traction computation methods}
\label{ap:B}
Suppose that $\Gamma$ is a~compact component of~$\partial\Omega$, where
$\Omega\subset\mathbb{R}^d$, $d=2$ or~$3$, is a~Lipschitz domain.
Traction $\mathbf{t}\colon\Gamma\to\mathbb{R}^d$ is formally given as
\begin{equation}
  \label{eq:traction}
  \mathbf{t} = \mathbb{T}\mathbf{n},
  \qquad
  \mathbb{T} = -p\mathbb{I} + \mu\bigl(\nabla\mathbf{v}+(\nabla\mathbf{v})^\top\bigr).
\end{equation}
Suppose that the Cauchy stress $\mathbb{T}$ satisfies the balance of momentum
\begin{equation*}
  \rho \left(\frac{\partial\mathbf{v}}{\partial t} + (\mathbf{v}\cdot\nabla)\mathbf{v}\right) = \operatorname{div} \mathbb{T}
  \qquad \text{in $\Omega$.}
\end{equation*}
Following \eqref{eq:vnd}, we can define $\mathbf{t}$ by duality as
a~distribution on~$\mathcal{C}_0^\infty(\Gamma)^d$,
\begin{equation}
  \label{eq:variational}
  \left< \mathbf{t}, \boldsymbol\phi \right>_\Gamma
  = \int_\Omega \rho \left(\frac{\partial\mathbf{v}}{\partial t} + (\mathbf{v}\cdot\nabla)\mathbf{v}\right) \cdot \boldsymbol\phi
    + \mathbb{T} \mathbin{:} \nabla\boldsymbol\phi,
  \qquad
  \boldsymbol\phi \in \mathcal{C}_0^\infty(\Omega\cup\Gamma)^d.
\end{equation}
Drag or lift exerted on~$\Gamma$ is obtained by integrating
$\mathbf{t}\cdot\mathbf{e}_x$ or $\mathbf{t}\cdot\mathbf{e}_y$, respectively,
over~$\Gamma$. Equivalently, evaluating~\eqref{eq:variational} with \emph{any}
$\boldsymbol\phi\in\mathcal{C}_0^\infty(\Omega\cup\Gamma)^d$ such that
\begin{equation*}
  \boldsymbol\phi  = \mathbf{e}_x
  \text{ or }
  \mathbf{e}_y, \text{ on~$\Gamma$}
\end{equation*}
gives drag and lift, respectively.
Early use of this method is explicitly described by
\citet*{giles-larson-levenstam-suli-1997},
\citet*[formulas (9.10)]{john-1997},
and \citet*{becker-1998}, who refers to the method as
the \emph{Babu\v{s}ka--Miller trick}
\citep*{babuska-miller-1984a,babuska-miller-1984b}.
See also \citep*[chapter 11]{bangerth-rannacher-2003},
and \citep*[section~2]{gjerde-scott-2024}.

Now we consider velocity--pressure mixed discretization.
Consider a~finite element velocity space
$\mathbf{V}^h\subset\mathbf{W}^{1,\infty}(\Omega)$ which
incorporates homogeneous essential conditions associated
with all essential boundaries but~$\Gamma$, a~subspace
$\mathbf{V}_\Gamma^h\subset\mathbf{V}^h$ satisfying
additionally zero boundary condition on~$\Gamma$, and
a~suitable pressure space $Q^h\subset L^\infty(\Omega)$.
Suppose $(\mathbf{v}_h,p_h)$ is a~discrete solution, i.e.,
$(\mathbf{v}_h-\mathbf{v}^\mathrm{D}_h,p_h)\in \mathbf{V}^h_\Gamma\times Q^h$,
for a~suitable extension~$\mathbf{v}^\mathrm{D}_h$ of the Dirichlet data, and
\begin{equation}
  \label{eq:go}
  \int_\Omega \rho \left(
      \frac{\partial\mathbf{v}_h}{\partial t}
      + (\mathbf{v}_h\cdot\nabla)\mathbf{v}_h
    \right) \cdot \boldsymbol\phi_h
    + \mathbb{T}(\mathbf{v}_h,p_h) \mathbin{:} \nabla\boldsymbol\phi_h
  = 0
  \qquad
  \text{for all $\boldsymbol\phi_h \in \mathbf{V}^h_\Gamma$.}
\end{equation}
We can evaluate traction directly as
\begin{equation}
  \label{eq:th}
  \mathbf{t}(\mathbf{v}_h,p_h,\boldsymbol{\nu})
  = \mathbb{T}(\mathbf{v}_h,p_h)\boldsymbol{\nu},
  \qquad
  \mathbb{T}(\mathbf{v}_h,p_h)
  = -p_h\mathbb{I} + \mu\bigl(\nabla\mathbf{v}_h+(\nabla\mathbf{v}_h)^\top\bigr).
\end{equation}
We leave normal vector $\boldsymbol{\nu}$ so far unspecified.

Suppose that $\mathbf{Z}^h\subset\mathbf{V}^h$ is a~subspace (possibly equal)
and take $\mathbf{W}^h$ as the $\Gamma$-trace space of~$\mathbf{Z}^h$.
For example, if $\mathbf{V}^h\subset\mathbf{W}^{1,\infty}(\Omega)$
is taken as piecewise quadratic polynomials on a~simplicial partition
of~$\Omega$, then $\mathbf{Z}^h\subset\mathbf{W}^{1,\infty}(\Omega)$ and
$\mathbf{W}^h\subset\mathbf{W}^{1,\infty}(\Gamma)$ can be taken as piecewise
affine functions on the simplices and the corresponding facets of~$\Gamma$,
respectively.

Now we can introduce \emph{discrete variational
traction} $\mathbf{t}^\text{dvt}_h\in \mathbf{W}^h$ by $\mathbf{L}^2(\Gamma)$-projection:
\begin{equation}
  \label{eq:dvt}
  \int_{\Gamma} \mathbf{t}^\text{dvt}_h \cdot \boldsymbol{\phi}_h \,\mathrm{d}S
  = \int_\Omega \rho \left(\frac{\partial\mathbf{v}_h}{\partial t} + (\mathbf{v}_h\cdot\nabla)\mathbf{v}_h\right) \cdot \boldsymbol\phi_h
    + \mathbb{T}(\mathbf{v}_h,p_h) \mathbin{:} \nabla\boldsymbol\phi_h
  \qquad
  \text{ for all $\boldsymbol\phi_h \in \mathbf{Z}^h$.}
\end{equation}
The definition \eqref{eq:dvt} is correct provided the Galerkin
orthogonality~\eqref{eq:go} is satisfied.

In the infinite-dimensional case the traction can always be defined as
a~distribution by~\eqref{eq:variational} and it coincides
with~\eqref{eq:traction} as long as a~sufficiently weak notion of normal
derivative is taken (or $\mathbf{v}$ is smooth enough).
On the other hand, in the discrete case the two notions, direct~\eqref{eq:th}
and variational~\eqref{eq:dvt}, do not coincide, in general. Under suitable
conditions they should converge though. In \cref{ap:A} we have seen that
in a~simpler setting of elliptic problems the variational definition might
exhibit superconvergence. In the present context, this has been observed
in the literature for integral quantities such as the lift and drag.
Suppose that we want to approximate $\int_\Gamma \mathbf{t}\cdot\mathbf{e}_x$
and $\int_\Gamma \mathbf{t}\cdot\mathbf{e}_y$. The direct approach
consists of using~\eqref{eq:th} and computing the surface integrals
\begin{equation*}
  \int_\Gamma \mathbf{t}(\mathbf{v}_h,p_h,\mathbf{n})\cdot\mathbf{e}_x,
  \qquad
  \int_\Gamma \mathbf{t}(\mathbf{v}_h,p_h,\mathbf{n})\cdot\mathbf{e}_y.
\end{equation*}
The variational approach consists of making a~choice~$\boldsymbol{\phi}_h$
in~\eqref{eq:dvt} such that $\boldsymbol{\phi}_h=\mathbf{e}_x$ and~$\mathbf{e}_y$,
respectively, on~$\Gamma$. It has been shown in the literature
\citep{giles-larson-levenstam-suli-1997,john-1997,becker-1998,bangerth-rannacher-2003}
that the variational approach exhibits better convergence rate for
the integral quantities. In fact, there is a~lot of degrees of freedom
left for $\boldsymbol{\phi}_h$ to be chosen, which allows to further improve numerical
properties of $\int_\Gamma\mathbf{t}_h^\mathrm{dvt}\cdot\boldsymbol\phi_h$;
this is the basis of adjoint-based methods \citep*{giles-suli-2002}.

An advantage of the variational approach is that it does not explicitly need
the outer normal~$\mathbf{n}$. This seems to provide an advantage especially
when $\Omega$ is approximated by $\Omega_h$ polygonal, or mapped, curvilinear
polygonal. We will see this in the numerical experiments below.

\vspace{1ex}\noindent\textbf{Numerical experiments.}\ 
We consider the setting of the Sch\"{a}fer--Turek benchmark; see
\cref{sec:turek_benchmark,fig:turek_geometry}.
The domain $\Omega$ is curvilinear, non-polynomial due to the circular cylinder $\Gamma$.
We approximate by polygonal (the lowest order, with flat edges) family
$\Omega_h$ and the cylinder boundary $\Gamma_h$;
the only exception is the computation reported in \cref{tab:fem_convergence_b},
where we use isoparametric, piecewise quadratic approximation of~$\Gamma_h$.
The approximate domains $\Omega_h$ are triangulated by regular quasi-uniform $\mathcal{T}_h$.
We use the lowest-order Hood--Taylor pair, i.e.,
$\mathbf{V}^h\times Q^h\subset\mathbf{W}^{1,\infty}(\Omega_h)\times W^{1,\infty}(\Omega_h)$
with $\mathbf{V}^h$ consisting of polynomials of degree~2 on the triangles of~$\mathcal{T}_h$
and $Q^h$ polynomials of degree~1.
We make a~choice for
$\mathbf{Z}^h\subset\mathbf{V}^h$ and
$\mathbf{W}^h$, the trace space of~$\mathbf{Z}^h$ on~$\Gamma_h$,
as affine continuous polynomials on the
triangles of~$\mathcal{T}_h$ and on the edges of~$\Gamma_h$.

\begin{figure}
  \centering
  \begin{subfigure}{0.495\textwidth}
    \caption{$\mathrm{Re}=0$}
    \label{fig:conv_h_1}
    \centering
    \begin{tikzpicture}[scale=0.9]
      \begin{loglogaxis}[
        width=\textwidth,
        ylabel={$\mathbf{L}^2(\Gamma)$ error},
        xlabel={\#DoFs},
        legend style={
          at={(0.5,-0.15)},
          anchor=north,
          legend columns=2,
          legend cell align=left,
        },
        legend to name={mylegend}
      ]
        \def\mydatafile{"%
          Turek_benchmark-convergence_data_S.txt"}
        \addplot+[mark=o, thick, black] table[x=DoF, y=Poincaré-Steklov:CG(1)] {\mydatafile};
        \addlegendentry{$\mathbf{t}_h^\mathrm{dvt}$}

        \addplot+[mark=asterisk, thick, orange] table[x=DoF, y=Directinterpolated:DG(1):analyticaln] {\mydatafile};
        \addlegendentry{$\mathbf{t}(\mathbf{v}_h,p_h,\mathbf{n}_\Omega)$}

        \addplot+[mark=square, thick, blue] table[x=DoF, y=Directprojected:CG(1):analyticaln] {\mydatafile};
        \addlegendentry{$\Pi_1\mathbf{t}(\mathbf{v}_h,p_h,\mathbf{n}_\Omega)$}

        \addplot+[mark=triangle, thick, purple] table[x=DoF, y=Directinterpolated:DG(1):CG(1)n] {\mydatafile};
        \addlegendentry{$\mathbf{t}(\mathbf{v}_h,p_h,\Pi_1\mathbf{n}_{\Omega_h})$}

        \addplot+[mark=x, thick, green] table[x=DoF, y=Directprojected:CG(1):CG(1)n] {\mydatafile};
        \addlegendentry{$\Pi_1\mathbf{t}(\mathbf{v}_h,p_h,\Pi_1\mathbf{n}_{\Omega_h})$}

        \addplot+[mark=diamond, thick, brown] table[x=DoF, y=Directinterpolated:DG(1):FacetNormal] {\mydatafile};
        \addlegendentry{$\mathbf{t}(\mathbf{v}_h,p_h,\mathbf{n}_{\Omega_h})$}

        \slope{bl}{-1}{7.5e4}{1.5e6}{4.5e-6}{$1$}
        \slope{tr}{-0.5}{7.5e4}{1.5e6}{1e-3}{$1/2$}
      \end{loglogaxis}
    \end{tikzpicture}
  \end{subfigure}%
  \hfill
  \begin{subfigure}{0.495\textwidth}
    \caption{$\mathrm{Re}=20$}
    \label{fig:conv_h_2}
    \centering
    \begin{tikzpicture}[scale=0.9]
      \begin{loglogaxis}[
        width=\textwidth,
        ylabel={$\mathbf{L}^2(\Gamma)$ error},
        xlabel={\#DoFs},
      ]
        \def\mydatafile{"%
          Turek_benchmark-convergence_data_NS.txt"}
        \addplot+[mark=o, thick, black] table[x=DoF, y=Poincaré-Steklov:CG(1)] {\mydatafile};
        \addplot+[mark=square, thick, blue] table[x=DoF, y=Directprojected:CG(1):analyticaln] {\mydatafile};
        \addplot+[mark=x, thick, green] table[x=DoF, y=Directprojected:CG(1):CG(1)n] {\mydatafile};
        \addplot+[mark=asterisk, thick, orange] table[x=DoF, y=Directinterpolated:DG(1):analyticaln] {\mydatafile};
        \addplot+[mark=triangle, thick, purple] table[x=DoF, y=Directinterpolated:DG(1):CG(1)n] {\mydatafile};
        \addplot+[mark=diamond, thick, brown] table[x=DoF, y=Directinterpolated:DG(1):FacetNormal] {\mydatafile};

        \slope{bl}{-1}{7.5e4}{1.5e6}{9.0e-6}{$1$}
        \slope{tr}{-0.5}{7.5e4}{1.5e6}{1e-3}{$1/2$}
      \end{loglogaxis}
    \end{tikzpicture}
  \end{subfigure}
  \hfill
  \vskip\floatsep\relax
  \pgfplotslegendfromname{mylegend}
  \caption{%
    Comparison of methods for pointwise traction computation in the
    Sch\"{a}fer--Turek benchmark for the stationary Stokes ($\mathrm{Re}=0$)
    and stationary Navier--Stokes ($\mathrm{Re}=20$) equations.
    Note that slope~$1$ means convergence order~$2$ in terms of mesh
    size~$h\sim(\mathrm{\#DoFs})^{-1/2}$, and similarly slope~$1/2$ means
    order~$1$. Number of degrees of freedom (\#DoFs) refers to the
    dimension of $\mathbf{V}_h\times Q_h$ up to treatment of boundary
    conditions.
  }
  \label{fig:conv_h_combined}
\end{figure}
The finest space $\mathbf{V}^{h_n}\times Q^{h_n}$ we consider has dimension of
around 33~million.
\Cref{fig:conv_h_combined} compares traction computation methods.
Traction on the approximate cylinder~$\Gamma_{h_j}$ is computed using a~given method,
transferred from coarse~$\Gamma_{h_j}$ to the finest~$\Gamma_{h_n}$ (using an ad hoc mapping),
and the $\mathbf{L}^2(\Gamma_{h_n})$-norm of the difference to the finest discrete
variational traction $\mathbf{t}_{h_n}^\mathrm{dvt}$ is computed; these norms
are declared to be the $\mathbf{L}^2(\Gamma)$ errors and are plotted in
\cref{fig:conv_h_combined}. Five of the presented methods are based on the
direct evaluation~\eqref{eq:th}. Here, $\mathbf{n}_\Omega$ is defined in the
whole~$\mathbb{R}^2$ by
\begin{equation*}
  \mathbf{n}_\Omega = -\frac{(x-x_0,y-y_0)}{R},
\end{equation*}
where $(x_0,y_0)$ and $R>0$ are the center and the radius of the exact cylinder~$\Gamma$.
This is an ad hoc extension of the unit normal of~$\Omega$ (which could also be denoted
as~$\mathbf{n}_\Omega$); it is $|\mathbf{n}_\Omega|=1$ only approximately on~$\Gamma_h$.
By $\Pi_1$ we denote the $\mathbf{L}^2(\Gamma_h)$ projection on~$\mathbf{W}^h$, the space
of continuous piecewise affine functions on the edges of~$\Gamma_h$.
\begin{table}
  \captionsetup{position=above}
  \caption{Approximation of drag and lift coefficients \eqref{eq:drag-lift-coeffs} on
    the cylinder in the Sch\"{a}fer--Turek benchmark with $\mathrm{Re}=20$.
    \Cref{tab:fem_convergence_a} and \cref{tab:fem_convergence_b}
    use different geometry approximation on the cylinder.
    All coefficients are approximated using the Babuška--Miller trick, except
    for the reference values from~\cite{Nabh1998}, which are used to estimate
    the experimental order of convergence (EOC, shown in parentheses).
    Number of degrees of freedom (\#DoFs) refers to the dimension
    of $\mathbf{V}^h\times Q^h$ up to treatment of boundary conditions.
    The dimensions do not match for \cref{tab:fem_convergence_a,tab:fem_convergence_b}
    as we used different mesh generation infrastructure.
  }
  \label{tab:fem_convergence}
  \centering
  \begin{minipage}{0.45\textwidth}
  \centering
  \subcaption{Subparametric, piecewise linear boundary~$\Gamma_h$}
  \label{tab:fem_convergence_a}
  \begin{tabular}{@{}rll@{}}
    \#DoFs
      & \multicolumn{1}{c}{$C_\mathrm{D}$ (EOC)}
      & \multicolumn{1}{c}{$C_\mathrm{L}$ (EOC)}
    \\ \hline
    599 & 5.7717427 & \llap-0.3084173\\
    2,206 & 5.5117295 (1.60) & \llap-0.0197592 (3.61)\\
    8,444 & 5.5589576 (1.78) & 0.0075301 (3.41)\\
    33,016 & 5.5744385 (2.05) & 0.0103186 (3.42)\\
    130,544 & 5.5782631 (2.02) & 0.0105893 (3.37)\\
    519,136 & 5.5792152 (2.00) & 0.0106145 (2.76)\\
    2,070,464 & 5.5794548 (2.00) & 0.0106180 (2.27)\\
    8,269,696 & 5.5795151 (2.00) & 0.0106187 (2.08)\\
    33,054,464 & 5.5795302 (2.00) & 0.0106189 (2.03)\\
    reference & 5.5795352 & 0.0106189\\[-5pt]
  \end{tabular}
  \end{minipage}
  \hfill
  \begin{minipage}{0.545\textwidth}
  \centering
  \subcaption{Isoparametric, piecewise quadratic boundary~$\Gamma_h$}
  \label{tab:fem_convergence_b}
  \begin{tabular}{@{}rll@{}}
    \#DoFs
      & \multicolumn{1}{c}{$C_\mathrm{D}$ (EOC)}
      & \multicolumn{1}{c}{$C_\mathrm{L}$ (EOC)}
    \\ \hline
    765       & 5.59108839094 & \llap-0.007350180692\\
    2,835     & 5.57610408144 (1.85) & 0.009791494753 (1.85)\\
    10,890    & 5.57908784213 (3.03) & 0.010598612174 (3.03)\\
    42,660    & 5.57949930985 (3.69) & 0.010619458729 (3.69)\\
    168,840   & 5.57953275640 (3.89) & 0.010619069934 (3.89)\\
    671,760   & 5.57953507204 (3.95) & 0.010618958822 (3.95)\\
    2,679,840 & 5.57953522352 (3.98) & 0.010618948890 (3.98)\\
    10,704,960 & 5.57953523320 (4.01) & 0.010618948177 (3.99)\\
    reference & 5.57953523384 & 0.010618948146\\
    & &\\[-5pt]
  \end{tabular}
  \end{minipage}
  \vskip\floatsep\relax
\end{table}
The experiments show, at least for the present cases, that the discrete
variational traction is superior;
the convergence order in $L^2(\Gamma)$ norm appears to be 2 compared to 1 for
the other methods.
This is consistent with what has been observed for integral drag and lift in the
literature \citep{giles-larson-levenstam-suli-1997,john-1997,becker-1998,bangerth-rannacher-2003}.
We also report integral drag and lift in \cref{tab:fem_convergence}. The results show
that using isoparametric, piecewise quadratic boundary approximation
yields fourth-order convergence of the integral
quantities (see \cref{tab:fem_convergence_b}),
which is in agreement with the theoretical rate $2k$ ($k$ being the polynomial degree
of~$\mathbf{V}^h$) proved in the Stokes case by \citet*{giles-larson-levenstam-suli-1997}.
When piecewise linear boundary is used, the convergence rate drops to~$2$ due to geometric
approximation errors; see \cref{tab:fem_convergence_a}.

We would like to point out that it might make sense to use a~higher-order
space~$\mathbf{Z}^h$ for $\mathbf{t}_h^\mathrm{dvt}$. We have chosen $\mathbf{Z}^h$
(and its trace space~$\mathbf{W}^h$) as piecewise affine functions only because
of technical difficulties in the implementation. If the variational crime
of approximating $\Omega$ by~$\Omega_h$ was not present, one can expect
in the light of \cref{ap:A}, at least for the Stokes case and given sufficient
regularity, that the convergence order of
$\|\mathbf{t}-\mathbf{t}(\mathbf{v}_h,p_h,\mathbf{n})\|_{\mathbf{L}^2(\Gamma)}$
to be the polynomial degree of $\mathbf{Z}^h$ while
$\|\mathbf{t}-\mathbf{t}_h^\mathrm{dvt}\|_{\mathbf{L}^2(\Gamma)}$
might admit superconvergence of up to the extra half order.
Of course, for the present setting of $\Omega_h\not=\Omega$,
the convergence order is limited by the geometry approximation error.

\section{Variations of the Sch\"{a}fer--Turek Benchmark geometry}
\label{ap:C}

We provide a robustness study of the pointwise traction as an indicator under variations of the Sch\"{a}fer--Turek geometry. The same methodology as in the main text is employed, but on one refinement level finer meshes (up to $1.2\times 10^6$ \#DoFs) and with the Reynolds number step of $\Delta \mathrm{Re} = 0.5$, as the computation of a simple continuation in steady states is not prohibitively expensive.

We consider three classes of modifications:
\begin{enumerate}[label=(\alph*)]
    \item symmetric vertical placement of the obstacle within the original geometry;
    \item a small, smooth perturbation of the circular obstacle shape, rotated asymmetrically with respect to the centerline; this configuration is considered both in the original position and in a centered placement;
    \item a channel of double width, with both symmetric and non-symmetric vertical placement of the obstacle (preserving the relative proportions of the original benchmark).
\end{enumerate}

\begin{figure}
    \centering
    \begin{subfigure}{0.440\textwidth}
        \includegraphics[width=\linewidth]{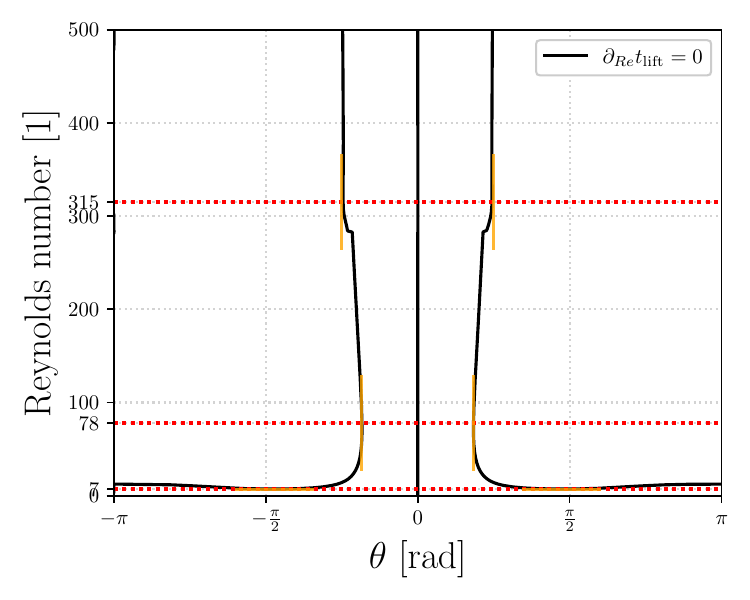}
        \captionsetup{width=0.9\linewidth}
    \caption{Pointwise lift: $\partial_{\mathrm{Re}} t_\text{lift}(\theta, \mathrm{Re}) = 0$ shown~in black}
    \end{subfigure}
    \begin{subfigure}{0.440\textwidth}
        \includegraphics[width=\linewidth]{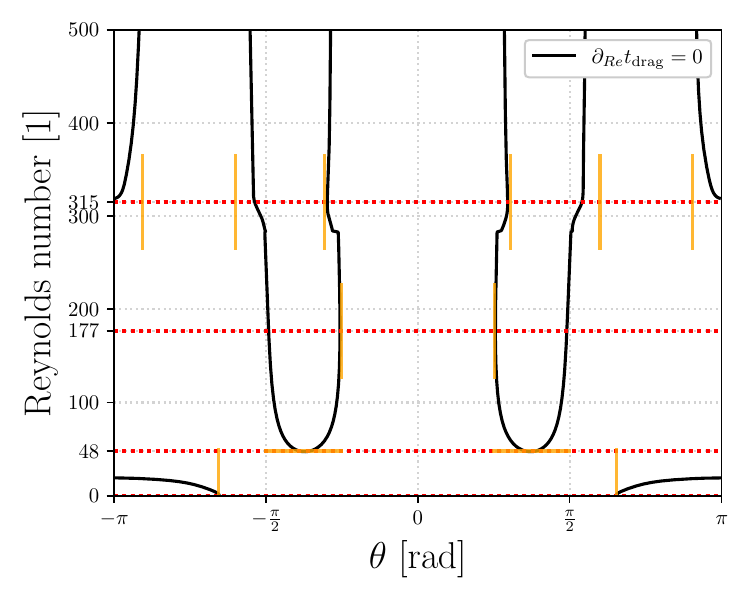}
    \captionsetup{width=0.9\linewidth}
    \caption{Pointwise drag: $\partial_{\mathrm{Re}} t_\text{drag}(\theta, \mathrm{Re}) = 0$ shown in black}
    \end{subfigure}
\caption{Symmetric vertical placement of the cylinder. Implicit curves defined by $\partial_{\mathrm{Re}} t_{\text{drag,lift}}(\theta, \mathrm{Re}) = 0$ are shown in black along the boundary $\theta \in (-\pi, \pi)$ for $\mathrm{Re} \in (0, 500)$. Their folds indicate turning points; orange tangent lines and red horizontal lines are retained to facilitate comparison with the original geometry.}
\label{fig:profiles_cylinder_symm}
\end{figure}

\vspace{1ex}\noindent\textbf{(a).}\  Regarding the symmetric vertical placement of the cylinder, the zero-derivative curves exhibit turning points at Reynolds numbers consistent with those of the original imperfect geometry, see \cref{fig:profiles_cylinder_symm}. The conditions \eqref{eq:turning_points} remain clearly identifiable. As before at $\mathrm{Re} = 315$, a mixed response in both traction components is observed; in addition, both turning-point conditions appear simultaneously, with \eqref{eq:turning_points_a} occurring at the trailing point ($\theta = \pi$). 

In contrast to the original case, where the baseline branch is asymmetric and follows an imperfect pitchfork structure (see \cref{fig:bifurcation_diagram}), the branch obtained here by simple continuation remains symmetric.

\vspace{1ex}\noindent\textbf{(b).}\  Regarding robustness with respect to the obstacle shape, we consider a small, smooth perturbation of the circular geometry. Since the perturbation is mild, one may expect the global unsteady flow behavior to remain qualitatively similar, allowing for comparison of the corresponding turning points.

We define the obstacle boundary as a parametric curve $\theta \mapsto (x(\theta), y(\theta))$, $\theta \in [0, 2\pi)$, given by
\begin{align*}
   r(\theta) &= R \bigl(1 + \varepsilon \sin(k \theta)\bigr),\\
   x(\theta) &= x_0 + r(\theta)\cos\theta, \\
   y(\theta) &= y_0 + r(\theta)\sin\theta,
\end{align*}
where $R = 0.05$ m is the reference radius of the benchmark geometry, $\varepsilon = 0.01$ controls the amplitude of the perturbation, $k=8$ is the number of oscillations, and $(x_0,y_0)$ denotes the center of the obstacle; see \cref{fig:flower_shape}. For $k=8$, the resulting shape is not symmetric with respect to either the horizontal line $y=y_0$ or the vertical line $x=x_0$.

\begin{figure}
  \def\R{1.0}
  \def\eps{0.04}
  \def\k{8}
  \centering
  \begin{tikzpicture}[scale=1.5]
    \draw[domain=0:360, samples=400, smooth, variable=\t]
    plot ({(\R*(1+\eps*sin(\k*\t)))*cos(\t)},
          {(\R*(1+\eps*sin(\k*\t)))*sin(\t)});
    \fill (0,0) circle (0.02);
    \draw[->] (-1.4,0) -- (1.4,0) node[above] {$x$};
    \draw[->] (0,-1.4) -- (0,1.4) node[right] {$y$};
    \draw[dashed] (0,-1.2) -- (0,1.2);
    \draw[dashed] (-1.2,0) -- (1.2,0);
  \end{tikzpicture}
  \caption{%
    Perturbed obstacle boundary defined by $r(\theta)=R\bigl(1+\varepsilon \sin(k\theta)\bigr)$,
    $k=\k$, here shown with $\varepsilon=\eps$ for the purpose of visualization.
    The shape breaks the rotational symmetry and is used in the traction analysis.
  }
  \label{fig:flower_shape}
\end{figure}

\begin{figure}
    \centering
    \begin{subfigure}{0.440\textwidth}
        \includegraphics[width=\linewidth]{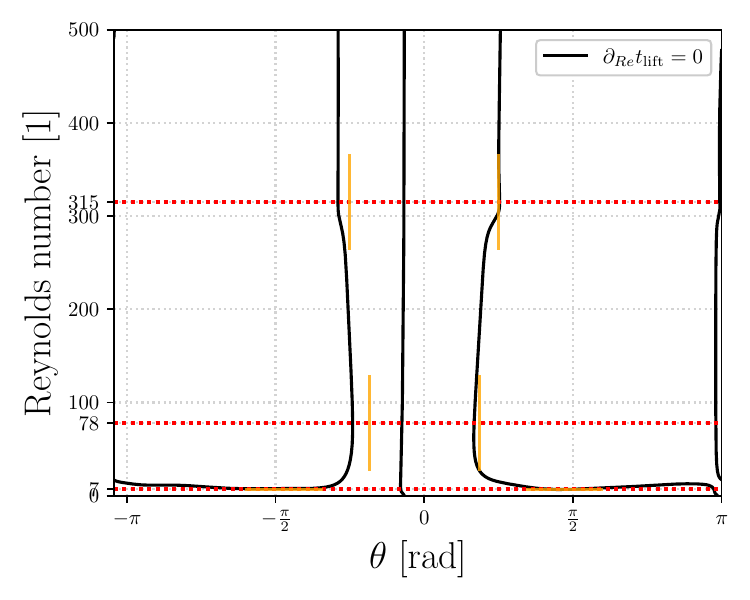}
        \captionsetup{width=0.9\linewidth}
    \caption{Pointwise lift: $\partial_{\mathrm{Re}} t_\text{lift}(\theta, \mathrm{Re}) = 0$ shown~in black}
    \end{subfigure}
    \begin{subfigure}{0.440\textwidth}
        \includegraphics[width=\linewidth]{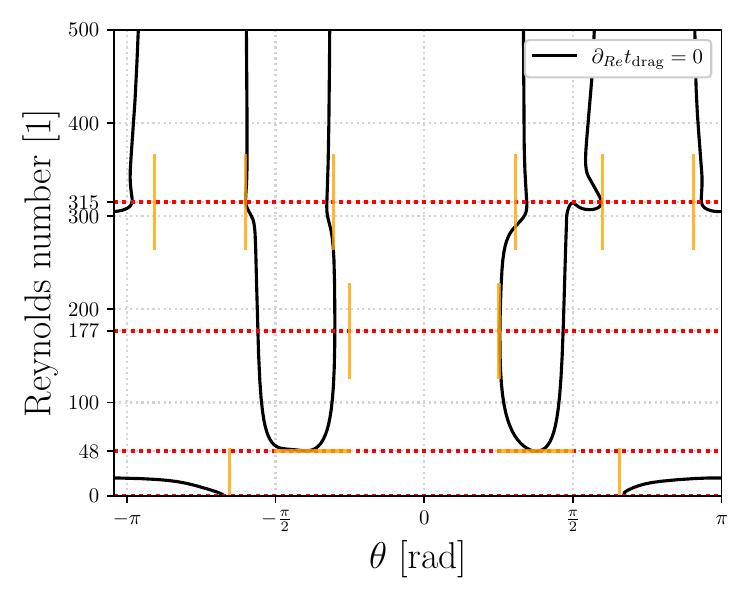}
    \captionsetup{width=0.9\linewidth}
    \caption{Pointwise drag: $\partial_{\mathrm{Re}} t_\text{drag}(\theta, \mathrm{Re}) = 0$ shown in black}
    \end{subfigure}
\caption{Perturbed circular obstacle at original vertical position. Implicit curves defined by $\partial_{\mathrm{Re}} t_{\text{drag,lift}}(\theta, \mathrm{Re}) = 0$ are shown in black along the boundary $\theta \in (-\pi, \pi)$ for $\mathrm{Re} \in (0, 500)$. Their folds indicate turning points; orange tangent lines and red horizontal lines are retained to facilitate comparison with the original geometry.}
\label{fig:profiles_flower}
\end{figure}

\begin{figure}
    \centering
    \begin{subfigure}{0.440\textwidth}
        \includegraphics[width=\linewidth]{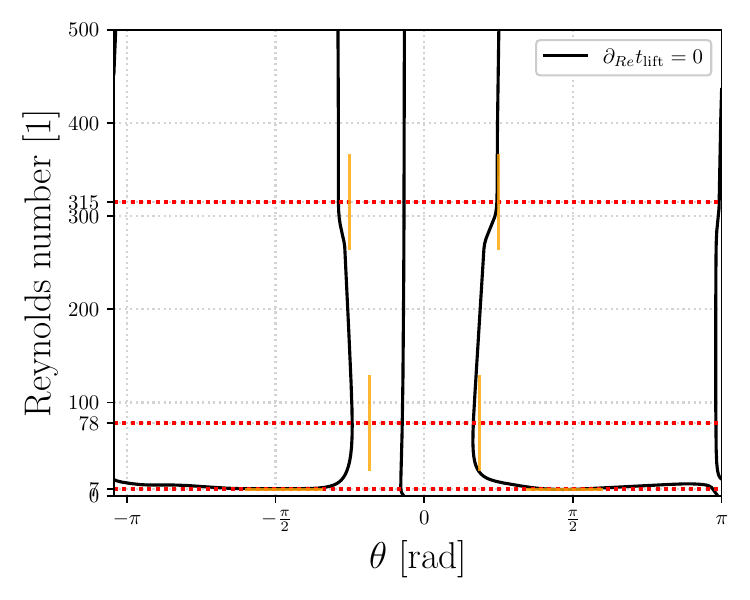}
        \captionsetup{width=0.9\linewidth}
    \caption{Pointwise lift: $\partial_{\mathrm{Re}} t_\text{lift}(\theta, \mathrm{Re}) = 0$ shown~in black}
    \end{subfigure}
    \begin{subfigure}{0.440\textwidth}
        \includegraphics[width=\linewidth]{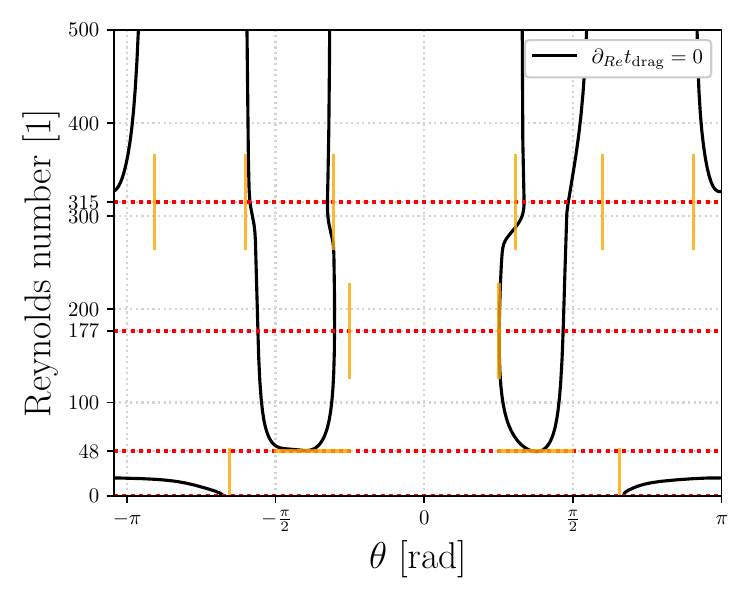}
    \captionsetup{width=0.9\linewidth}
    \caption{Pointwise drag: $\partial_{\mathrm{Re}} t_\text{drag}(\theta, \mathrm{Re}) = 0$ shown in black}
    \end{subfigure}
\caption{Symmetric vertical placement of the perturbed circular obstacle. Implicit curves defined by $\partial_{\mathrm{Re}} t_{\text{drag,lift}}(\theta, \mathrm{Re}) = 0$ are shown in black along the boundary $\theta \in (-\pi, \pi)$ for $\mathrm{Re} \in (0, 500)$. Their folds indicate turning points; orange tangent lines and red horizontal lines are retained to facilitate comparison with the original geometry.}
\label{fig:profiles_flower_symm}
\end{figure}

The perturbed geometry demonstrates insignificant differences to the original counterpart, see the perturbed cylinder shape at original location, \cref{fig:profiles_flower}, and at symmetric vertical placing \cref{fig:profiles_flower_symm}.

\vspace{1ex}\noindent\textbf{(c).}\  Finally, we consider a~channel of double width $H_{2\times} = 0.82\,\mathrm{m}$, with both symmetric and non-symmetric vertical placement of the obstacle, while preserving all other absolute proportions of the original benchmark. The comparison is made with the turning points identified for the original channel width.

\begin{figure}
    \centering
    \begin{subfigure}{0.440\textwidth}
        \includegraphics[width=\linewidth]{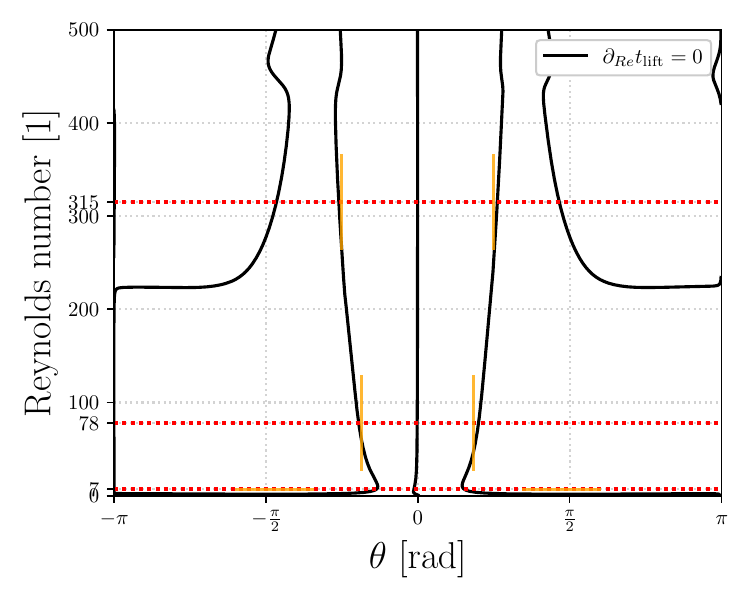}
        \captionsetup{width=0.9\linewidth}
    \caption{Pointwise lift: $\partial_{\mathrm{Re}} t_\text{lift}(\theta, \mathrm{Re}) = 0$ shown~in black}
    \end{subfigure}
    \begin{subfigure}{0.440\textwidth}
        \includegraphics[width=\linewidth]{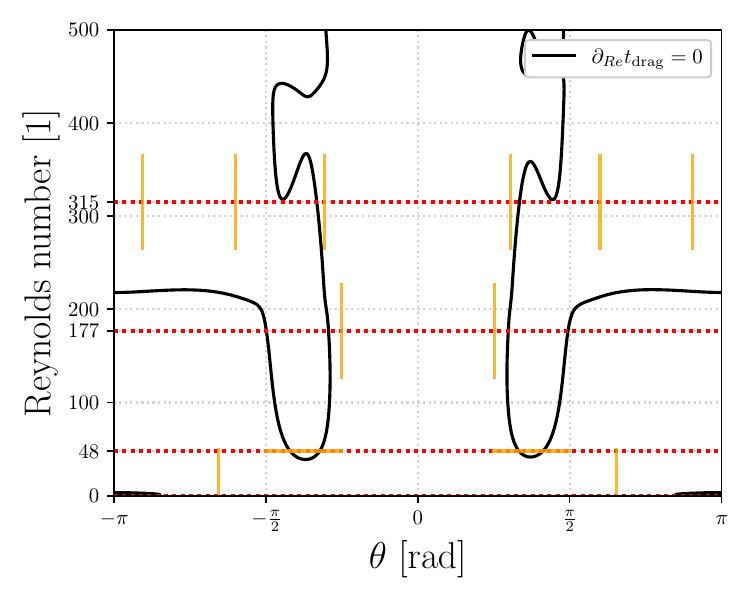}
    \captionsetup{width=0.9\linewidth}
    \caption{Pointwise drag: $\partial_{\mathrm{Re}} t_\text{drag}(\theta, \mathrm{Re}) = 0$ shown in black}
    \end{subfigure}
\caption{Double-width channel with circular obstacle at original relative vertical position. Implicit curves defined by $\partial_{\mathrm{Re}} t_{\text{drag,lift}}(\theta, \mathrm{Re}) = 0$ are shown in black along the boundary $\theta \in (-\pi, \pi)$ for $\mathrm{Re} \in (0, 500)$. Their folds indicate turning points; orange tangent lines and red horizontal lines are retained to facilitate comparison with the original geometry.}
\label{fig:profiles_cylinder_wide}
\end{figure}

For the original asymmetric placement, see \cref{fig:profiles_cylinder_wide}, we observe a shift in the onset of the recirculating wake, which occurs at lower Reynolds numbers. Furthermore, the turning point previously associated with the wall-effect and Mode~E is no longer observed, consistent with the reduced blockage ratio. In the drag response, the turning point tentatively associated with the Mode~A instability shifts to a lower value, around $\mathrm{Re} \approx 130$. 

The turning point related to multiplicity appears to be reflected in the lift response at a higher Reynolds number, around $\mathrm{Re} \approx 465$, where behavior similar to that in the original configuration is observed. The corresponding drag response in this regime is more involved; however, a vertical turning point (see \eqref{eq:turning_points_b}) can also be identified at the same Reynolds number. This is inferred from similar turning-point features in the original geometry and has not been independently verified.

Additionally, new features emerge, such as a pronounced mixed response around $\mathrm{Re} \approx 220$, which may warrant further investigation. Overall, the wider channel exhibits a greater number of turning points in the traction response, consistent with the reduced confinement of the flow by the channel walls. At the same time, the wide-channel configuration mitigates some confinement effects and, while leading to shifts in critical Reynolds numbers and the disappearance of some wall-induced responses, largely preserves the overall structure of the traction-based description. The comparison between the centered and off-centered obstacle placement in the double-width channel further suggests that, in the present tests, the effect of vertical placement is weaker than the effect of changing the channel width.
Overall, the double-width channel reduces the influence of the walls and changes the traction response quantitatively. The critical Reynolds numbers are shifted, some wall-induced features disappear, and the set of observed turning points changes. Nevertheless, the main structure of the traction-based description remains present, which suggests that the observed relation between traction profiles and flow-regime transitions is not specific only to the original channel width.

\begin{figure}
    \centering
    \begin{subfigure}{0.440\textwidth}
        \includegraphics[width=\linewidth]{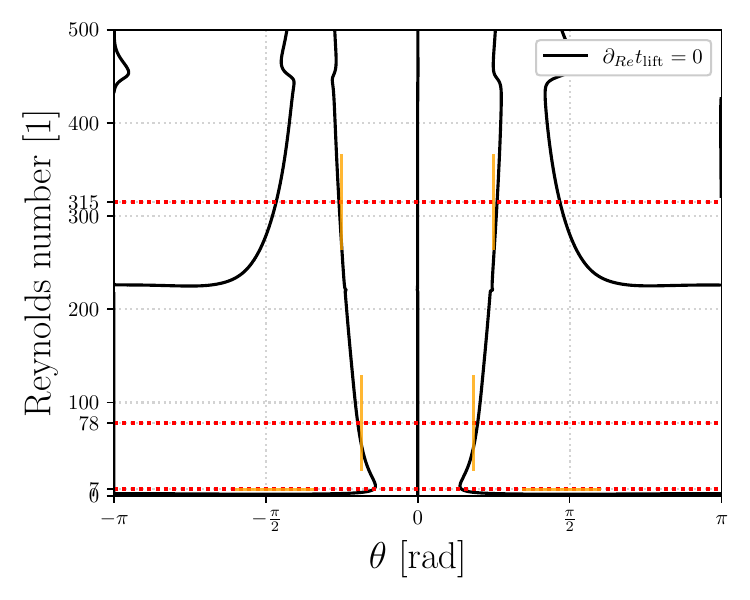}
        \captionsetup{width=0.9\linewidth}
    \caption{Pointwise lift: $\partial_{\mathrm{Re}} t_\text{lift}(\theta, \mathrm{Re}) = 0$ shown~in black}
    \end{subfigure}
    \begin{subfigure}{0.440\textwidth}
        \includegraphics[width=\linewidth]{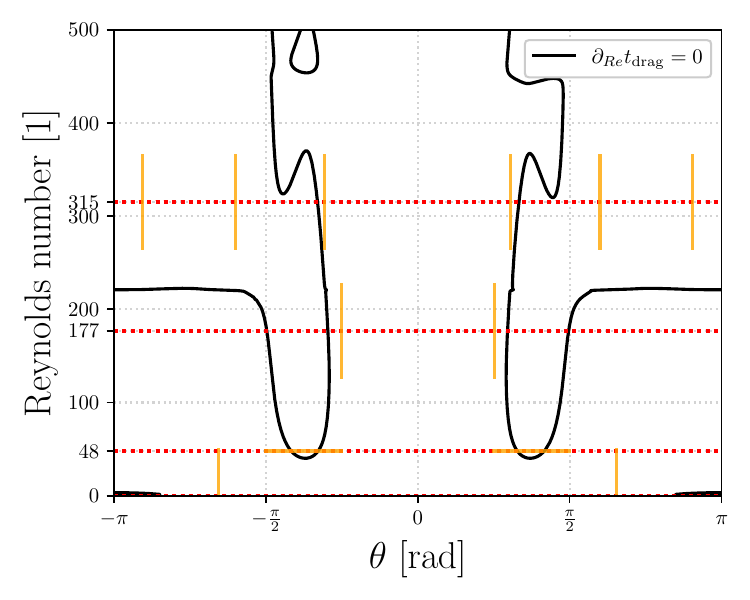}
    \captionsetup{width=0.9\linewidth}
    \caption{Pointwise drag: $\partial_{\mathrm{Re}} t_\text{drag}(\theta, \mathrm{Re}) = 0$ shown in black}
    \end{subfigure}
\caption{Double-width channel with circular obstacle at symmetric vertical position. Implicit curves defined by $\partial_{\mathrm{Re}} t_{\text{drag,lift}}(\theta, \mathrm{Re}) = 0$ are shown in black along the boundary $\theta \in (-\pi, \pi)$ for $\mathrm{Re} \in (0, 500)$. Their folds indicate turning points; orange tangent lines and red horizontal lines are retained to facilitate comparison with the original geometry.}
\label{fig:profiles_cylinder_symm_wide}
\end{figure}

The comparison between the symmetric and non-symmetric obstacle placements in the double-width channel shows only small differences; see \cref{fig:profiles_cylinder_symm_wide}. Thus, in the present tests, the vertical position of the obstacle appears to be less important than the channel width. The main practical difference is that, even with the same mesh density as in the original-width channel, we do not follow the symmetric branch beyond the possible multiplicity bifurcation at $\mathrm{Re} \approx 465$.

\FloatBarrier

\begingroup
\phantomsection
\addcontentsline{toc}{section}{References}
\raggedbottom
\interlinepenalty=10000

\bibliographystyle{elsarticle-num-names}
\endgroup

\end{document}